\documentclass[a4paper]{article}

\usepackage{ae}
\usepackage[colorlinks=true, pdfstartview=FitV, linkcolor=blue, citecolor=blue, urlcolor=blue]{hyperref}
\usepackage{ifpdf}
\ifpdf
\usepackage[pdftex]{graphicx}
\graphicspath{{all_png/}}
\else
\usepackage[dvips]{graphicx}
\graphicspath{{all_eps/}}
\fi

\ifpdf
\fi

\title{\textsc{Hector}, a fast simulator for the transport of particles in beamlines}
\author{J. de Favereau,  X. Rouby and K. Piotrzkowski \footnote{Email addresses: jdf@fynu.ucl.ac.be and rouby@fynu.ucl.ac.be and krzysztof.piotrzkowski@cern.ch} \\ 
	\small \textit{Center for Particle Physics and Phenomenology (CP3), Universit\'e catholique de Louvain} \\ 
	\small \textit{B-1348 Louvain-la-Neuve, Belgium}}

\begin{document}

\maketitle
\section*{Abstract}
Computing the trajectories of particles in generic beamlines is an important ingredient of experimental
particle physics, in particular regarding near-beam detectors. A new tool, \textsc{Hector}, has been built for such 
calculations, using the transfer matrix approach and energy corrections. The limiting aperture effects are 
also taken into account. As an illustration, the tool was used to simulate the LHC beamlines, in particular 
around the high luminosity interaction points (IPs), and validated with results of the \textsc{Mad-X} simulator. The 
LHC beam profiles, trajectories and beta functions are presented. Assuming certain forward proton detector 
scenarios around the IP5, acceptance plots, irradiation doses and chromaticity grids are produced. Furthermore, 
the reconstruction of proton kinematic variables at the IP (energy and angle) is studied as well as 
the impact of the misalignment of beamline elements.

\textit{PACS:} 29.27.-a ; 41.85.Ja\\

\textit{Keywords:} Particle transport, transfer matrix, LHC beamline, forward detectors, misalignment \\

\textit{Preprint:} CP3-07-13

\section{Introduction}

In the context of the LHC beamlines and in particular of very forward, near beam detectors, we introduce
a new fast simulator, called\footnote{The user guide is available as an appendix of this document.}$^{,}$\footnote{An homonymic program is used in the DIS community, for the calculation of QED, QCD and electroweak
corrections to e p and lepton+- N deep inelastic neutral and charged current scattering.} \textsc{Hector}, for the transport of single particles through generic
beamlines, although primarily dedicated to the LHC. The simulator is based on a linear approach of the beamline 
optics, implementing transport matrices from the optical element magnetic effective length, and with 
correction factors on magnetic strength for particles with non nominal energy. \textsc{Hector} deals 
with the computation of the position and angle of beam particles, and the limiting aperture of the 
optical elements. It has been designed to be fast, light and user friendly and its object oriented 
structure, using the \textsc{Root} framework \cite{ROOT}, helps its usage, its maintenance and the 
future improvements.

In forward physics, the measurement of the position and the angle of particles in dedicated detectors 
(as so-called \emph{roman pots} for example) hundreds of meters away from the interaction point (IP), 
helps in reconstructing the kinematics of the event at the related central detector. \textsc{Hector} 
links the information from very forward detectors (VFDs) to the one from the central detector, by 
precise calculation of the particle trajectory.

We present here the physics content and the simulation techniques used in \textsc{Hector}, as well as 
its overall performance. The validation of \textsc{Hector} output comes from direct comparisons with 
the \textsc{Mad-X} simulator \cite{MADX}, for the beta functions and relative positions of both LHC 
beams. Some physics results, like beam profiles and irradiation levels due to the forward  proton 
from inelastic $\mathrm{pp} \rightarrow \mathrm{pX}$ interactions are plotted for two reasonable 
scenarios for the very forward detectors. Finally, first studies of the 
forward-scattered proton kinematics reconstruction are performed, resulting in the discussion
of limiting factors and best strategies for the VFD designs.

\section{The LHC beamlines}

Even if \textsc{Hector}'s class structure is compatible with the most frequent beamline 
elements, the physics motivation on backstage is driven by the forward physics at the LHC. This includes 
detection of protons which have for example been quasi-elastically scattered by exchange of a photon. The 
reconstruction of the proton momentum using information from a VFD requires a good knowledge of the 
LHC beams and related beamlines. Such a beamline is described here below ; note that both LHC beams are basically identical. 

\subsection{Beamline elements and beam parameters}
\label{LHC_param}

The position and lateral size of the LHC beams are mainly controlled by two types of elements :
\begin{itemize}
\item focusing (quadrupoles) or deflecting (dipoles, kickers) magnets;
\item collimators and optical element apertures, limiting the lateral beam extension.
\end{itemize}

Quadrupole magnets can focus the beam either vertically or horizontally. Dipole magnets also come in 
two kinds : rectangular dipoles with a straight shape, and sector dipoles which are bent to match 
the beam curved trajectory. The former are mainly used in the long and short straight sections (LSS, 
SSS) of the beamline, while the latter are placed in the dispersion suppressors (DSs) and the bending 
sections (ARCs) \cite{LHC_EDMS}, \cite{LHC_TDR}. The kickers are dipole magnets dedicated to produce 
the crossing angle at the IP.

Some parameters are of direct interest when describing the nominal 7 TeV proton beams:
\begin{itemize}
\item the horizontal and vertical emittances 
$\epsilon_{x,y} \sim 500 \times 10^{-6} ~ \mu \mathrm{m}$, which are constant along the orbit
according to Liouville's theorem ;
\item the lateral beam size at the IP 
$\sigma = 16.6 ~ \mu \mathrm{m} = \sqrt{\epsilon\beta^*}$ , where 
$\beta^*$ is value (= 0.55 m) at the IP of the so-called beta-function ;
\item the lateral vertex size (or the transverse size of collision region) 
$\sigma^v = 11.76 ~ \mu \mathrm{m} = \sigma/\sqrt{2}$ ;
\item the angular divergence at the IP $\sigma_{\theta} = 30.2 ~ \mu \mathrm{rad} = 
\sqrt{\epsilon/\beta^*}$ ;
\item the crossing angle is equal to $2 \times 142.5 ~ \mu \mathrm{rad}$, in vertical 
(horizontal) plane at the IP1 (IP5) ;
\item the energy spread $\sigma_E = 1.129 \times 10^{-4} E$.
\end{itemize}

\subsection{Beamline map}

From here onwards, focus will be put on interaction region 5 (IR5). The elements affecting the 
beam will be described starting at interaction point 5 (IP5), going eastward when looking from the 
center. This is the propagation direction of beam 1. The other side of the beamline is very similar. 
As the region of interest for the forward physics related detectors extends up to approximately 
$500 ~ \mathrm{m}$ from the IP, our description of the beamline stops there. Within this part of the 
beamline, all aperture shapes are \textit{rectellipses} \cite{LHC_TDR}.

\subsubsection*{Before $85 ~ \mathrm{m}$ from IP5}

After the IP, beam 1 gets separated from the incoming beam 2 because of the crossing angle. The two 
beam paths stay close ($\sim 4 ~ \mathrm{mm}$) for the first $85 ~ \mathrm{m}$. They encounter the 
first three collimators at $19 ~ \mathrm{m}$, $45 ~ \mathrm{m}$ and $55 ~ \mathrm{m}$ from the IP
(the so-called low-{$\beta$} inner triplet), 
then a kicker ($30 ~ \mathrm{m}$) and then quadrupoles (H: $23$ and $47$ m ; V: $32$ and $39 ~ 
\mathrm{m}$) and consecutive dipoles (R-Bend) from $60$ to $81 ~ \mathrm{m}$, which deviate the 
two beam paths to lead them to their final separation of $194 ~ \mathrm{mm}$. The dipoles are positioned
symmetrically with respect to the two beams.

\subsubsection*{From $85 ~ \mathrm{m}$ to $269 ~ \mathrm{m}$ from IP5}

At about $160 ~ \mathrm{m}$ from the IP big R-Bend dipoles are used to make the beam orbits parallel,
so from there on the beams get to their nominal relative positions. These dipoles are positioned in 
such a way that the beam 1 (2) enters (exits) it perpendicularly. Four more collimators are located 
respectively at $141$, $149$, $184$ and $256 ~ \mathrm{m}$ from the IP5. Two kickers, at $165$ and 
$199 ~ \mathrm{m}$, lead them to the ideal path that they would have followed if there were not any 
crossing angle, due to the so-called closed bump orbit. The beam is kept under control with five more 
quadrupoles (H: $194$, $260$, $264$ ; V: $168$ and $226 ~ \mathrm{m}$ ) and one dipole ($153 ~ 
\mathrm{m}$). This is the end of the straight section.

\subsubsection*{From $269 ~ \mathrm{m}$ onwards}

There the beam path starts its bending. The beamline is composed in alternations of quadrupoles (> 6) 
to keep the beam focused and sector dipoles (> 8 S-dipoles) to bend it. The first sextupoles with 
non-zero field occur only after $439 ~ \mathrm{m}$ from the IP.

\section{Simulation techniques}

\subsection{Physical description}

The simulator relies on a linear approach to single particle propagation. 
A beamline consists of a set of optical elements, amongst dipoles, quadrupoles, drifts, collimators, 
kickers and VFDs. The optical elements are described by their magnetic field, their length and their 
aperture. In turn, a set of particles, with all possible smearings of initial positions, angles or 
energies, is propagated by \textsc{Hector} through the beamline, particle by particle. In 
electromagnetism, the influence of an external magnetic field is given by the Lorentz force. 
Let's consider the Taylor expansion of the vertical component of magnetic field $B_y$, around its 
central value :
$$\frac{e}{p} B_y(x) = \frac{e}{p} B_y + \frac{e}{p} \frac{\partial B_y}{\partial x} x + \frac{1}{2} \frac{e}{p} \frac{\partial^2 B_y}{\partial x^2} x^2 + \ldots$$
where $p$ is the momentum of the particle and $e$ its electric charge. The terms of this sum are 
interpreted as respectively dipolar ($\frac{1}{R} = \frac{e}{p} B_y$), quadrupolar 
($k = \frac{e}{p} \frac{\partial B_y}{\partial x} $) and sextupolar fields. In the co-moving coordinate 
system, neglecting small deviations ($x \ll R$, $y  \ll R$) and small momentum loss ($\Delta p \ll p$), 
this leads to the following equation of motions for a particle traveling along path length $s$ through 
a magnetic element \cite{Klaus_Wille} :

\begin{equation} 
\label{eq_motion}
\Bigg\{ \begin{array}{l} x''(s) + \left( \frac{1}{R^2(s)} - k(s) \right) x(s) = \frac{1}{R(s)} \frac{\Delta p}{p} \\
	y''(s) + k(s) y(s) = 0.
\end{array}
\end{equation}

The solution $(x(s), x'(s), y(s), y'(s))$ to these equations can be expressed as a linear combination 
of the initial values $(x_0, x'_0, y_0, y'_0)$, where the rotation matrices are defined by the 
properties of the optical element (length and magnetic field). The typical values 
$R \sim 200 ~ \mathrm{m}$ and $x < 0.01 ~ \mathrm{m}$ match the first assumption ($x \ll R$). 
As $\Delta p \ll p$ is not always verified, we apply a further correction to the magnetic field 
(for mode details, see section \ref{E_correction}) of each optical element of the beamline.

In other words, for the simulation of the transport of a particle in a beamline, each beam particle is 
represented by a phase space vector and each optical element by a transfer matrix by which the vector is multiplied. The 
propagation of a single particle is thus the \textit{rotation} of the phase space vector by the $n$ 
optical element matrices.

$$ X(s) = X(0) \underbrace{M_1 M_2 ... M_n}_{M_{\mathrm{beamline}}} $$

The whole beamline is modeled as a single transport matrix acting on each particle phase space vector 
(no intrabeam interactions). The optical element description also refers to its physical aperture. 
When a particle is propagated through an optical element, two tests check whether its path is 
compatible with the element acceptance or not, at its entrance and exit. The fact that a bent path 
could hit a central part of the element is neglected. A particle which does not pass through the 
element acceptance is flagged, allowing to know the stopping element for given particle parameters 
(for the computation of acceptance and irradiation rates for instance).

\subsection{Implementation}
\label{E_correction}

The simulator has an object-oriented design, using the C++ \textsc{Root} framework \cite{ROOT}, 
with dedicated classes describing :
\begin{itemize}
  \item the beam particles. The 6-components phase space vector is $X = (x,x',y,y',E,1)$, with horizontal 
$(x,x')$ and vertical $(y,y')$ coordinates and angles ; $E$ is the particle energy. The sixth 
component is just a factor used to add an angular kick on the particle momentum direction.
  \item the optical elements (dipoles, drifts...) inheriting from a common class.
\end{itemize}

The 6x6 transport matrices can be decomposed into blocks
$$
      \mathbf{M_{units}} =
      \left(
      \begin{array}{cccccc}
	\mathcal{A} & \mathcal{A} & 0 & 0 & 0 & 0 \\
	\mathcal{A} & \mathcal{A} & 0 & 0 & 0 & 0 \\
	0 & 0 & \mathcal{B} & \mathcal{B} & 0 & 0 \\
	0 & 0 & \mathcal{B} & \mathcal{B} & 0 & 0 \\
	\mathcal{D} & \mathcal{D} & 0 & 0 & 1 & 0 \\
	K & K & K & K & 0 & 1 \\
      \end{array}
      \right)
$$ where
\begin{itemize}
\item $\mathcal{A}$ (and $\mathcal{B}$) blocks refer to the action (focusing, defocusing, drift) on 
horizontal (and vertical, resp.) coordinate and angle.
\item $\mathcal{D}$ terms reflect the dispersion effects of the horizontal dipoles on off-momentum 
particles.
\item $K$ factors are the angular action of kickers.
\end{itemize}

A beamline is the implementation of a list of optical elements with such transport matrices. The 
introduction of dispersion terms allow a proper description of off-momentum particles ($\Delta p / p 
\neq 0$). The \textit{dispersion function} can be defined from eq. (\ref{eq_motion}), for horizontal 
dipoles ($k=0$), taking $\Delta p / p = 1$ :
\begin{equation}
D''(s) + \frac{1}{R^2} D(s) = \frac{1}{R}.
\end{equation}

Developing the solution of this equation leads to a correction term for the deflection of 
off-momentum particles in the dipoles : 
\begin{equation}
x_\mathrm{off-momentum}(s) = x(s) + D(s) \frac{\Delta p}{p}.
\end{equation}

For explicit description of these $\mathcal{D}$ terms, refer to the appendix, Eq. (\ref{s_dip}). 
Besides the dispersion correction which is valid for low $\Delta p$, the actual dependence of the 
particle deflection on its momentum is also taken into account. This is the \textit{chromaticity}, 
or the energy dependence of the transport matrix, implemented by rescaling the magnetic field terms 
($R$, $k$, $K$) with a factor ($\frac{p}{p - \Delta p }$). The propagation of particles different 
from protons is also possible by rescaling these magnetic field terms :
\begin{equation} 
	k_i(\Delta E, m_p, q_p) = k_i \frac{p_0}{p_0 - \Delta p } \frac{q_p}{q_\mathrm{proton}} ~ ~ , ~ ~ ~ ~ k_i = R,k,K ;
\end{equation}
where $q_p$ and $m_p$ are the particle charge and mass, respectively.
The \textsc{Root} framework easily interfaces \textsc{Hector} with the output of the high energy 
physics event generator \textsc{Pythia} \cite{PYTHIA}. Forward particles from the final state can be 
then propagated through the beamline via \textsc{Hector}, for example in photon induced interactions.

As \textsc{Hector} has been designed for the LHC beamlines, one class can parse the official LHC optics 
tables, interfacing them directly into the code at run-time. For a better compatibility with different 
table layout, \textsc{Hector} can recognize column headers and is thus compliant with different table
types \cite{LHC_Optics}. All the default parameters are hard coded (in \texttt{H\_Parameters.h}), 
like nominal beam energy, position and angle, and their divergences, as well as the nominal crossing 
angle, but these default value can obviously be rewritten at run time. The effects of the kickers or 
the bending of the sector magnets can also be easily switched on or off. For the following analysis, 
LHC optics version 6.500 has been used, corresponding to $\beta^* = 0.55$~m at the IP1 and IP5.

Limitations : sextupoles and magnets of higher order (multipoles) are neglected in \textsc{Hector} 
(harmless the first $430$ m after the IP). The energy dependence of the transfer matrices is taken 
into account, but neither beam-beam interactions, nor the field non-uniformities. The beam particle 
mass is also neglected compared to its energy.

\section{Validation and performances}
\subsection{Performances}
\begin{figure} [tb]
\centering
\includegraphics[width=0.8\textwidth]{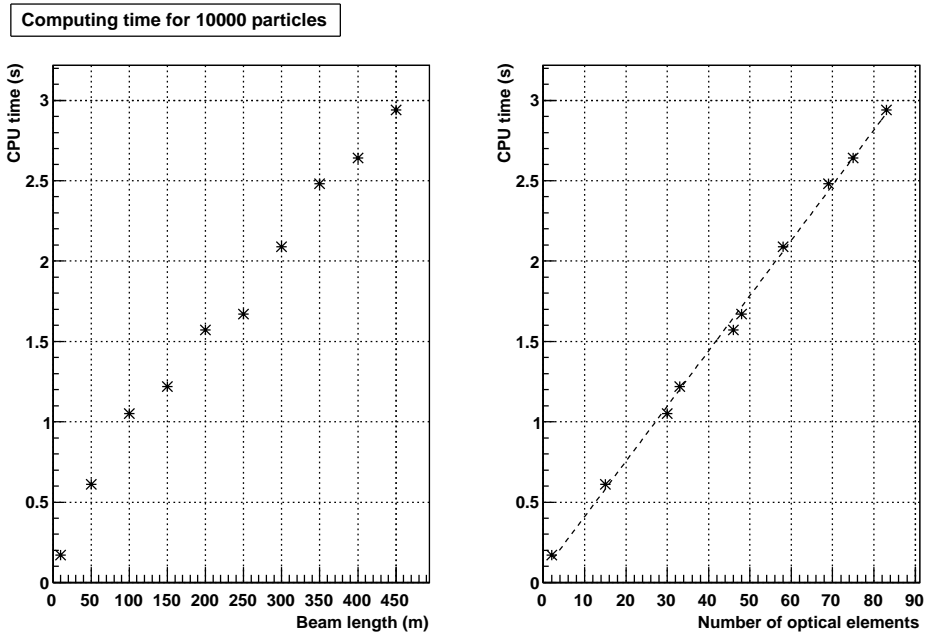}
\caption[\textsc{Hector} performance]{\textsc{Hector} speed performance : CPU time with respect to the 
propagation distance (\emph{left}) and the number of optical elements (\emph{right}) in the beamline, 
computed for a $10^4$ particle beam around the IR5. As the propagation of particles corresponds to the 
multiplication of matrices, the computation time is directly proportional to the amount of optical 
elements in the beamline, as well as the number of beam particles. For a time estimation for large 
number of particles and long beamlines, a good approximation is $4 ~ \mu s$  per optical element 
(including drifts) per particle. For this test ($3.43 ~ \mu \mathrm{s}$ per element per particle), 
\textsc{Hector} is run on a regular laptop with $1.7 ~ \mathrm{GHz}$ Intel Pentium Centrino processor, 
Linux rel. Ubuntu.} \label{hector_performance}
\end{figure}

\textsc{Hector} has been designed to be fast and light. All the code and its library altogether weight 
less than $2.5 ~ \mathrm{MB}$ for about $12000$ lines of code. The figure \ref{hector_performance} 
depicts its performance, leading for a first estimate to a CPU time of $4 ~ \mu s$ per beam particle 
per beamline optical element at the LHC. 
As expected, the CPU time is linear with the number of optical elements and the number of protons. 

\subsection{Validating \textsc{Hector} with \textsc{Mad-X}}
\begin{figure}[p]
\centering
\includegraphics[width=0.85\textwidth]{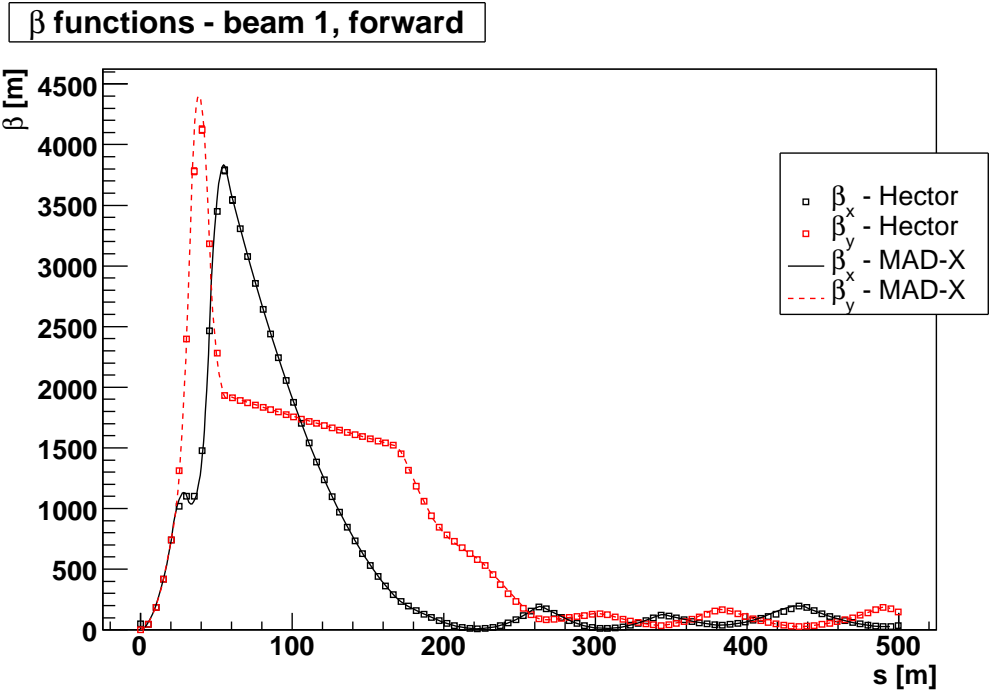}
\includegraphics[width=0.85\textwidth]{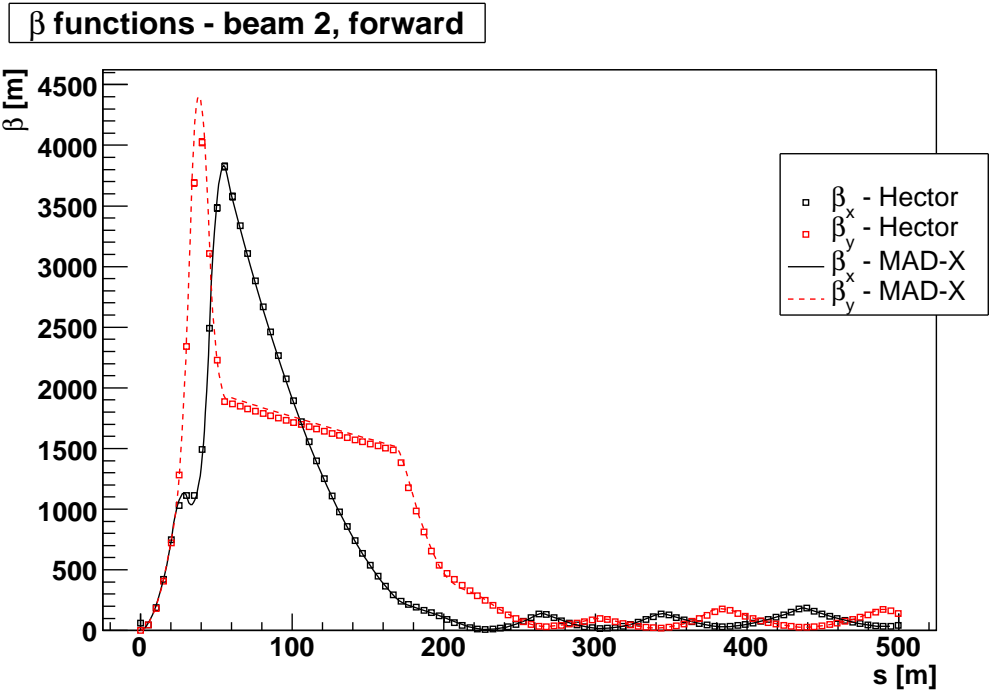}
\caption[\textsc{Hector} validation : Beta functions]{Beta functions $\beta_x$ and $\beta _y$ for 
beams 1 (\emph{above}) and 2 (\emph{below}) near the IP5, going forward: these functions depict the 
beam size variation, and are computed from the beam emittance and the beam lateral profiles at 
successive positions. Indexes $x$ and $y$ are corresponding to respectively the horizontal and 
vertical directions. The plain and dashed curves, coming from \textsc{Mad-X}, match exactly 
\textsc{Hector}'s output (\emph{squares}).}
\label{beam_beta}
\end{figure}

\begin{figure}[p]
\centering
\includegraphics[width=0.83\textwidth]{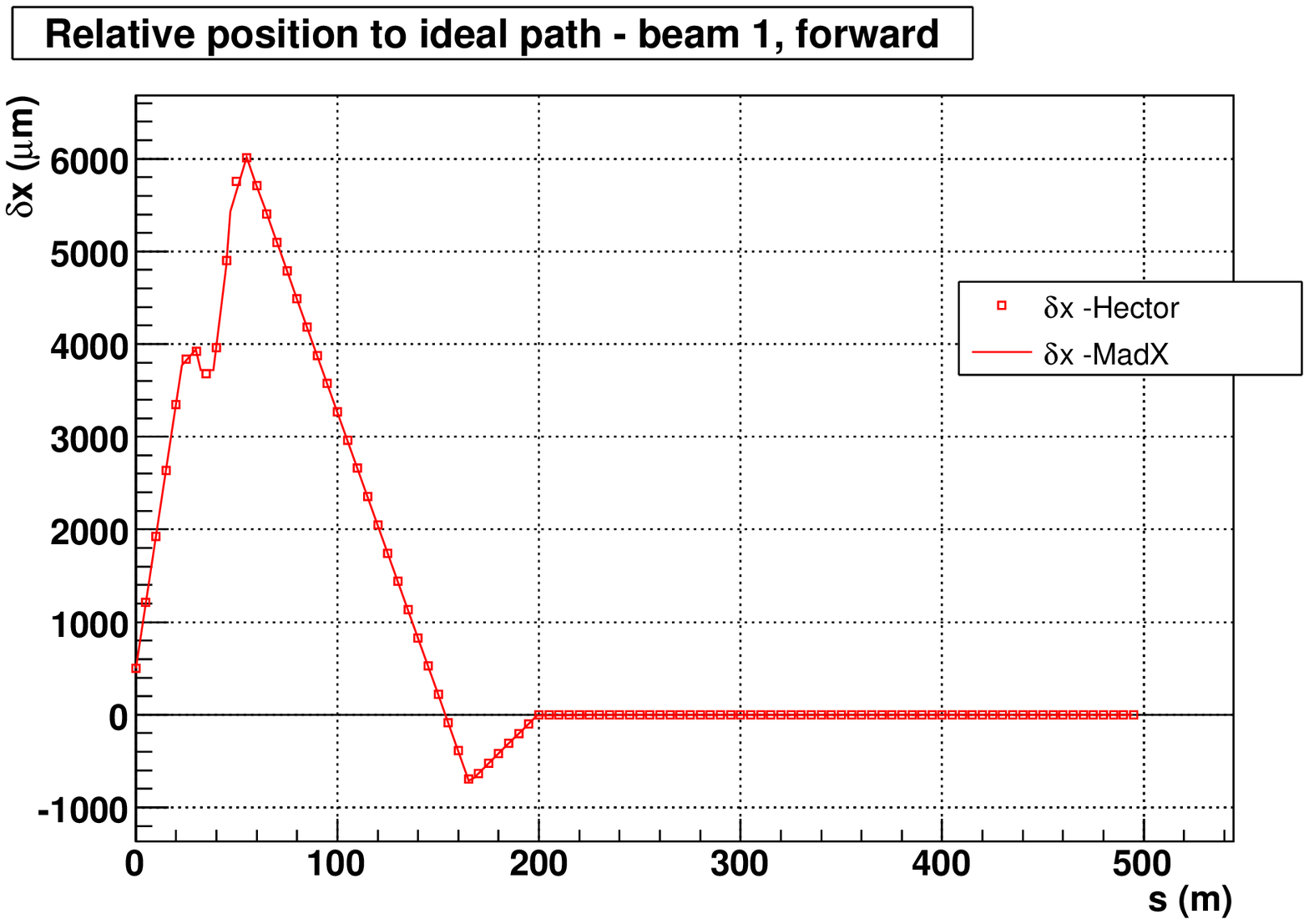}
\includegraphics[width=0.83\textwidth]{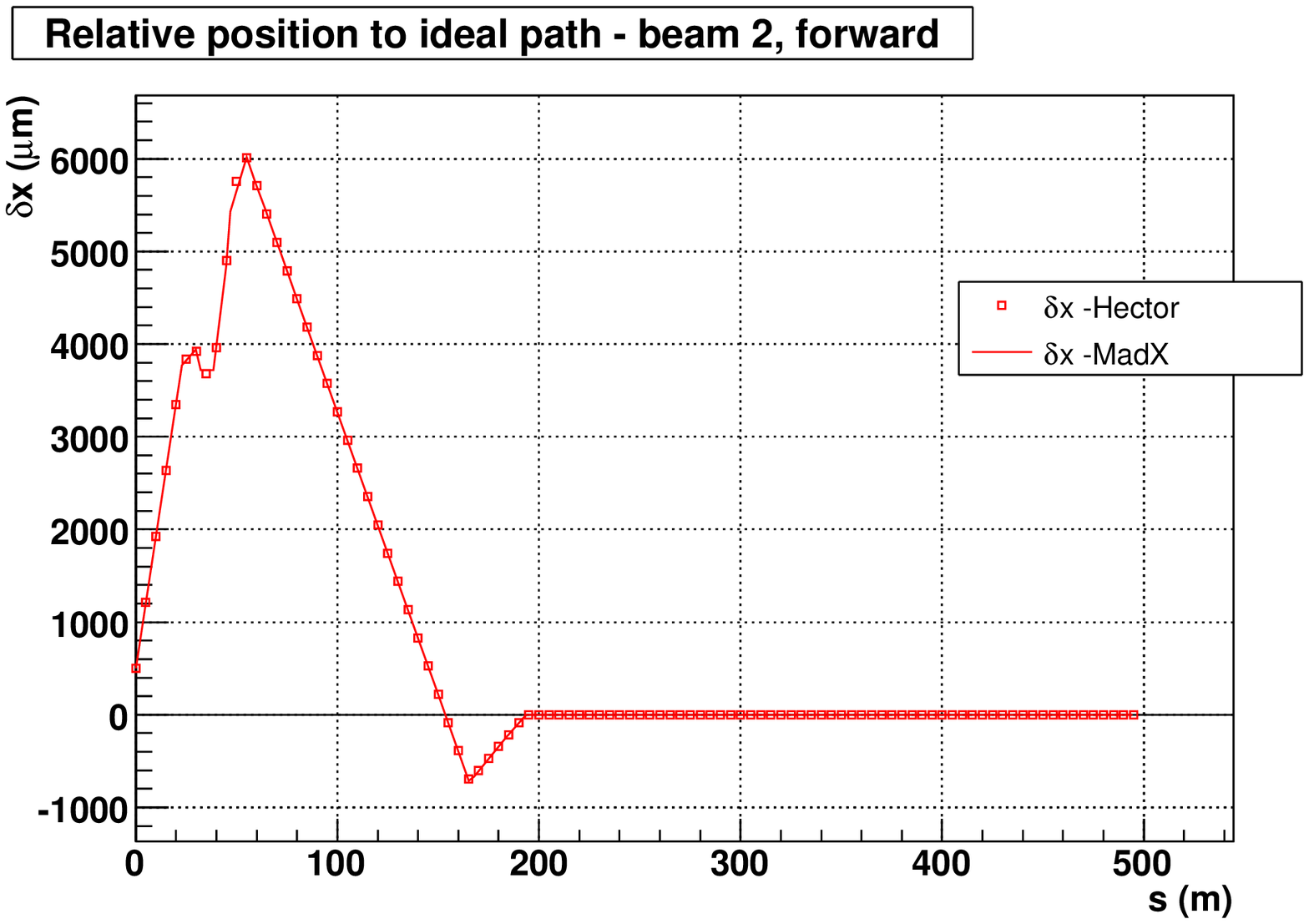}
\caption[\textsc{Hector} validation : Relative positions]{Relative horizontal positions for beams 1 
(\emph{above}) and 2 (\emph{below}), around the IP5, in the forward direction. The relative position 
is defined as the difference along the ideal orbit between the beam lateral positions with and 
without the crossing angle. 
This relative position graph (\emph{squares}) matches \textsc{Mad-X} predictions (\emph{plain lines}). 
One clearly sees the crossing angle at the IP5 ($s = 0 ~ \mathrm{m}$) and the effects of the kickers 
that move the beams back to the ideal orbit.}
\label{beam_position}
\end{figure}

In order to validate \textsc{Hector}, cross-checks have been made with the output of \textsc{Mad-X} 
\cite{MADX}, such as the beam beta functions (Fig. \ref{beam_beta}) and the beam relative positions 
(Fig. \ref{beam_position}). For each point, the beta functions are computed from the beam emittance 
and the RMS of its profile in the transverse plane. The beam beta functions, which describe the lateral
beam size variation along the orbit, as output by \textsc{Hector}, match perfectly the results from 
\textsc{Mad-X}. The lateral beam position shown in Fig. \ref{beam_position} is relative to its ideal 
path, defined for disabled kickers (and no crossing angle). These graphs (Fig. \ref{beam_beta} and 
\ref{beam_position}), and their equivalents in backward direction, are the strongest tests for 
\textsc{Hector}, where smearings of position, angle and energy are applied as previously explained.

\subsection{Validating off-beam energy simulation}
The comparisons of \textsc{Hector} and \textsc{Mad-X} results directly validate only the simulation of a
7~TeV proton propagation, and additional checks are necessary to validate simulations of the off-beam 
energy particles. This has been done by comparing numerically the exact calculations and the 
\textsc{Hector} predictions for a couple of proton trajectories at energies between 10 and 100\% of the 
beam energy. It was found that for all the energies of interest for the VFDs, i.e. for energies 
above 80\% of the beam energy, \textsc{Hector} remains very accurate, providing the particle lateral 
positions with precision at least at a few micron level.


\clearpage

\section{Beamline simulation}
By using the \textsc{Hector} classes and library physics studies can be performed by means of the
particle propagation inside the beamlines, away from the interaction point.

\subsection{Trajectories}
\begin{figure}[p]
\vspace{-2cm}
\centering
\includegraphics[width=0.85\textwidth]{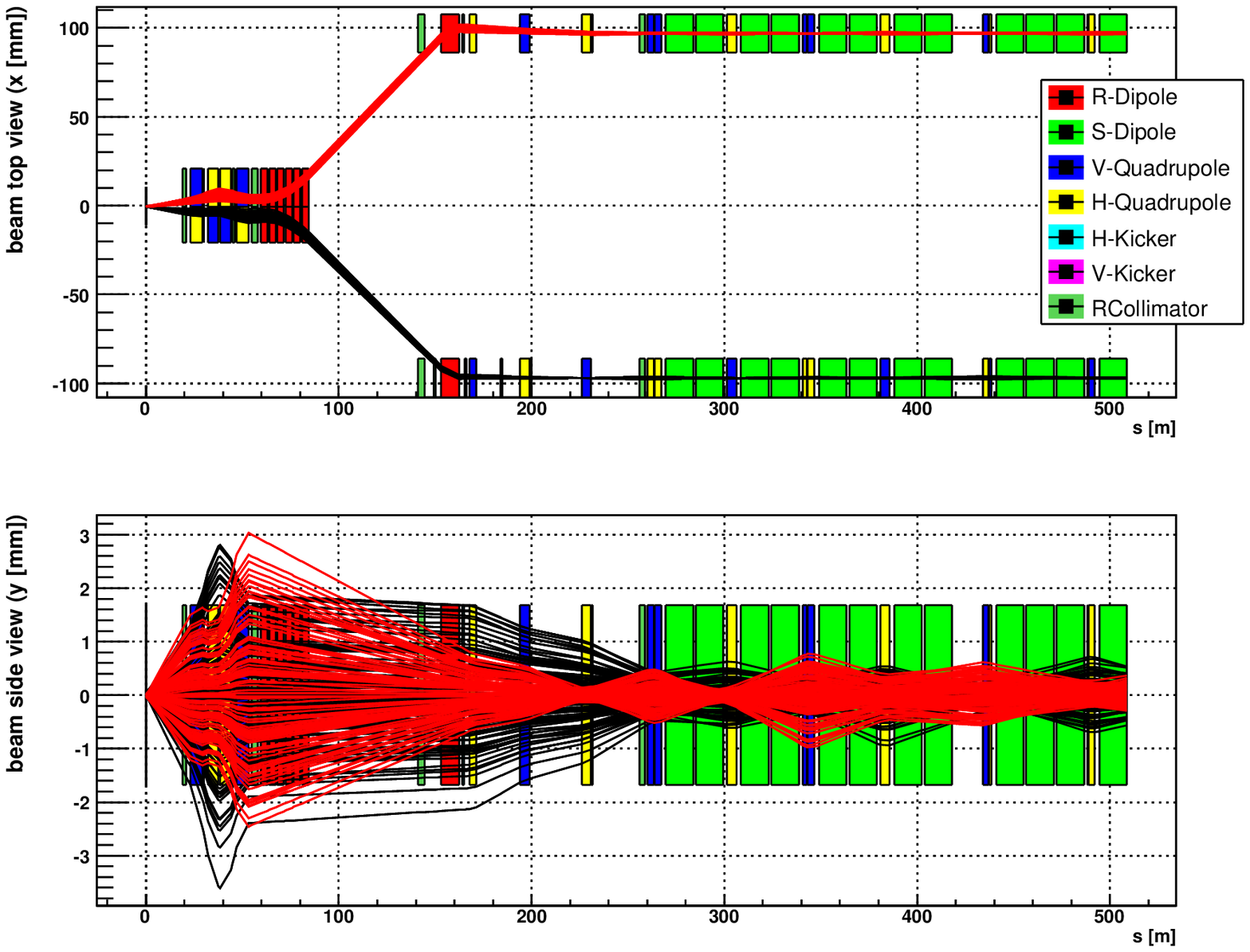}
\includegraphics[width=0.85\textwidth]{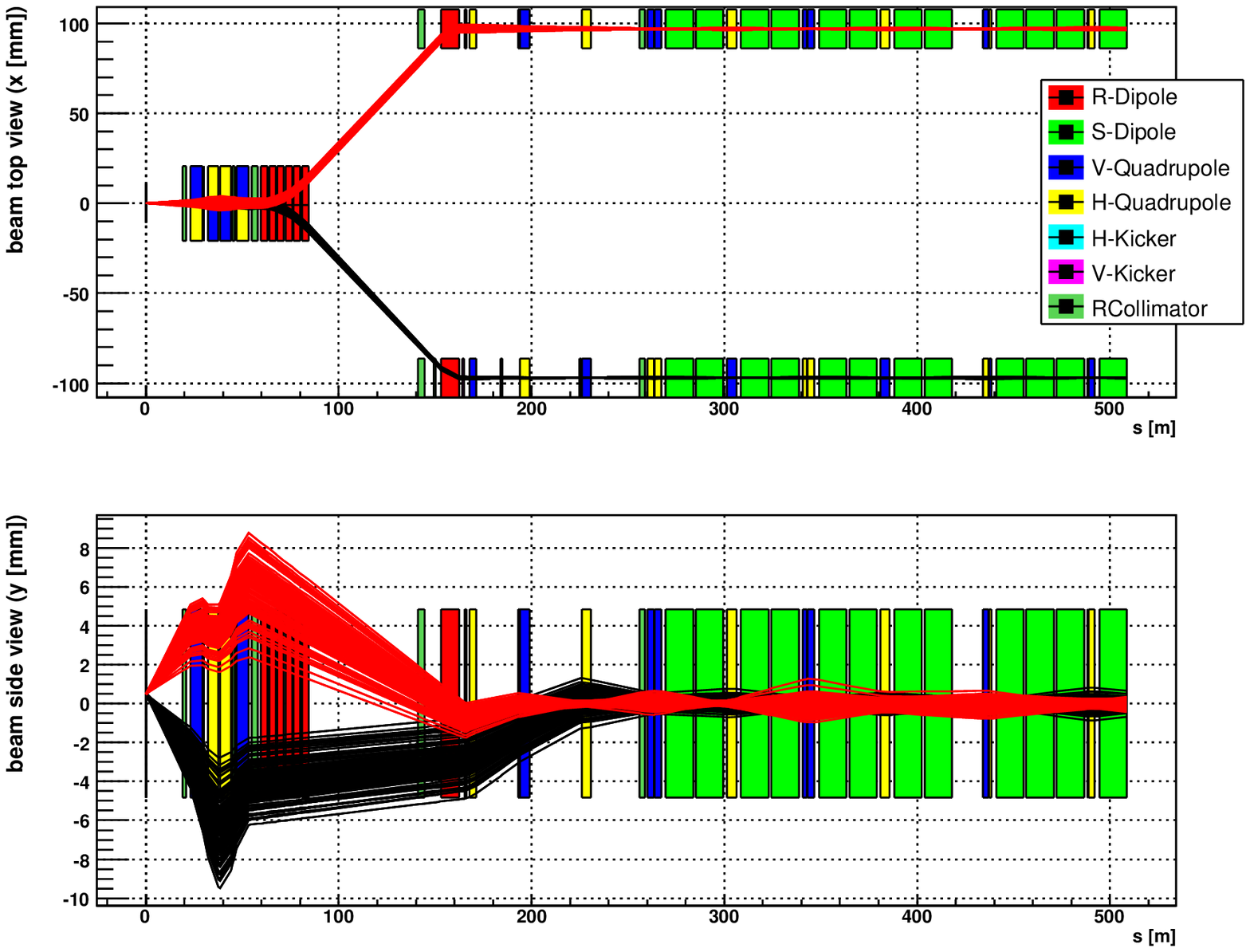}
\caption[LHC beams : top/side views]{Top and side views of both LHC beams around the IP5 
(\emph{first two graphs}) and at the IP1 (\emph{last two graphs}). The IP is located at $s=0$ m. 
Beam 2 (red) flows from the right to the left and is seen before its passage at the IP. In turn, beam 1 
(black) is seen downstream, passing from the left to the right after the crossing. In this graph, the 
bending effect of the sector magnets (S-Dipoles) has been switched off, thus rectifying the beam path 
to a straight line after $250 ~ \mathrm{m}$. The difference between the IP1 and IP5 is clearly seen, as the crossing planes are perpendicular to each other. 
Note: In the LHC absolute frame, the position of the IP1 is at $s = 0 ~ \mathrm{m}$ and the IP5 is at 
$s = 13329 ~ \mathrm{m}$.}
\label{2beams}
\end{figure}

Knowing the optics tables for both LHC beams, their trajectories can be compared simultaneously, in 
both top and side views, for the two LHC beams aside (incoming beam 2 and outgoing beam 1, at the IP5 
and IP1: Fig. \ref{2beams}). The top view shows the beams on the horizontal plane, clearly depicting 
the crossing angle at the IP5, and the beam separation after $70 ~ \mathrm{m}$ away from the 
interaction point. 
The bending of the sector dipoles has been switched off in order to makes graphics more clear -- 
this is why both beams are straight and parallel after $250 ~ \mathrm{m}$. However, the optical 
elements have been shifted (without tilt) in 
the horizontal plane by the half of the beam separation distance, from $180 ~ \mathrm{m}$ away from 
the IP, in order to match the ideal beam path: a proton with nominal energy and on the ideal orbit 
should travel through optical elements in their geometrical center. The side view in turn shows the 
beams in the vertical plane, emphasizing the difference between IP1 and IP5. In addition, the major 
optical elements have been drawn: rectangular dipoles in red, sector dipoles in light green, and
quadrupoles in yellow and blue.

\subsection{Beam profiles}
\begin{figure}[p]
\centering
\begin{tabular}{cc}
\includegraphics[width=0.4\textwidth]{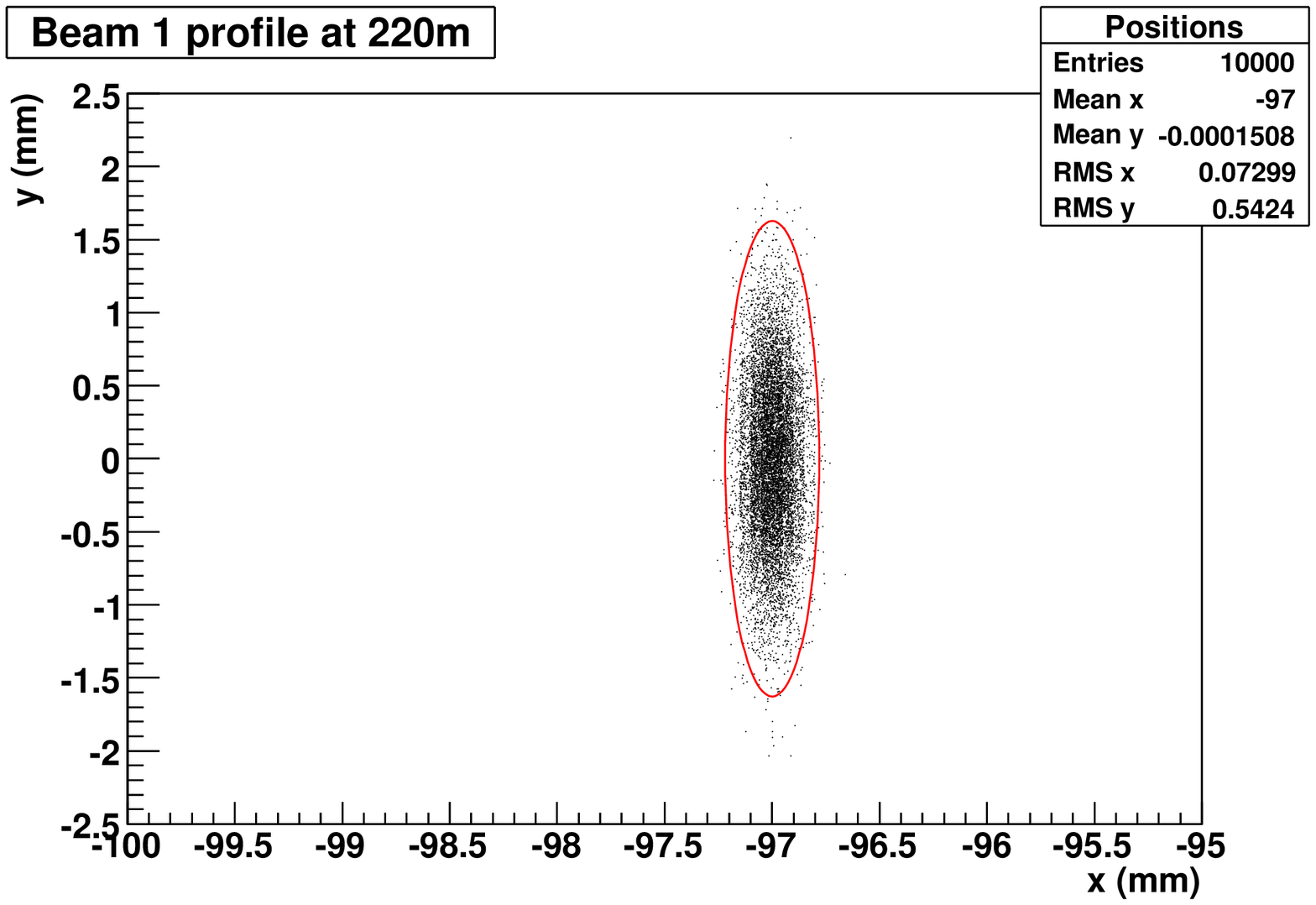} &
\includegraphics[width=0.4\textwidth]{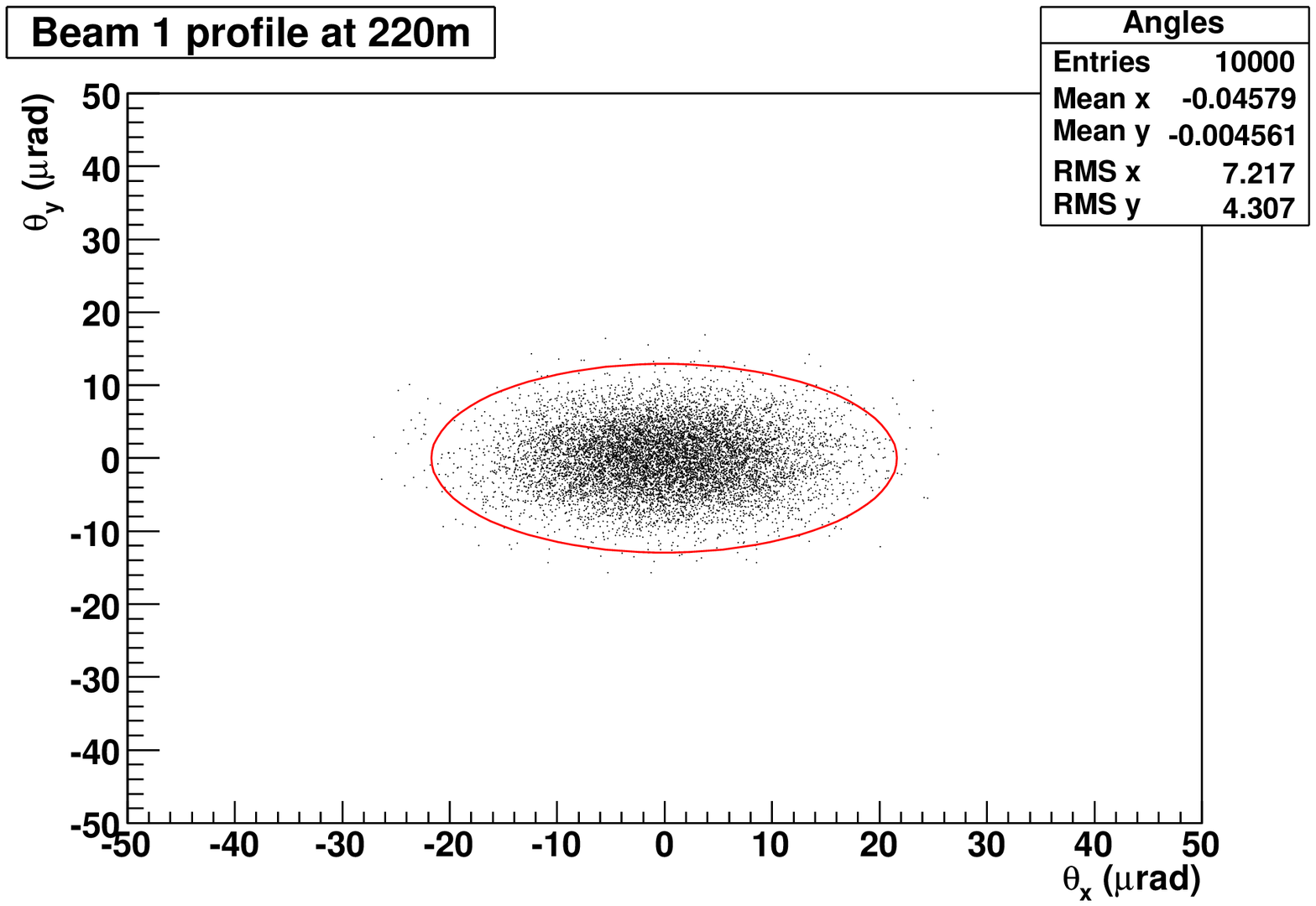} \\
\includegraphics[width=0.4\textwidth]{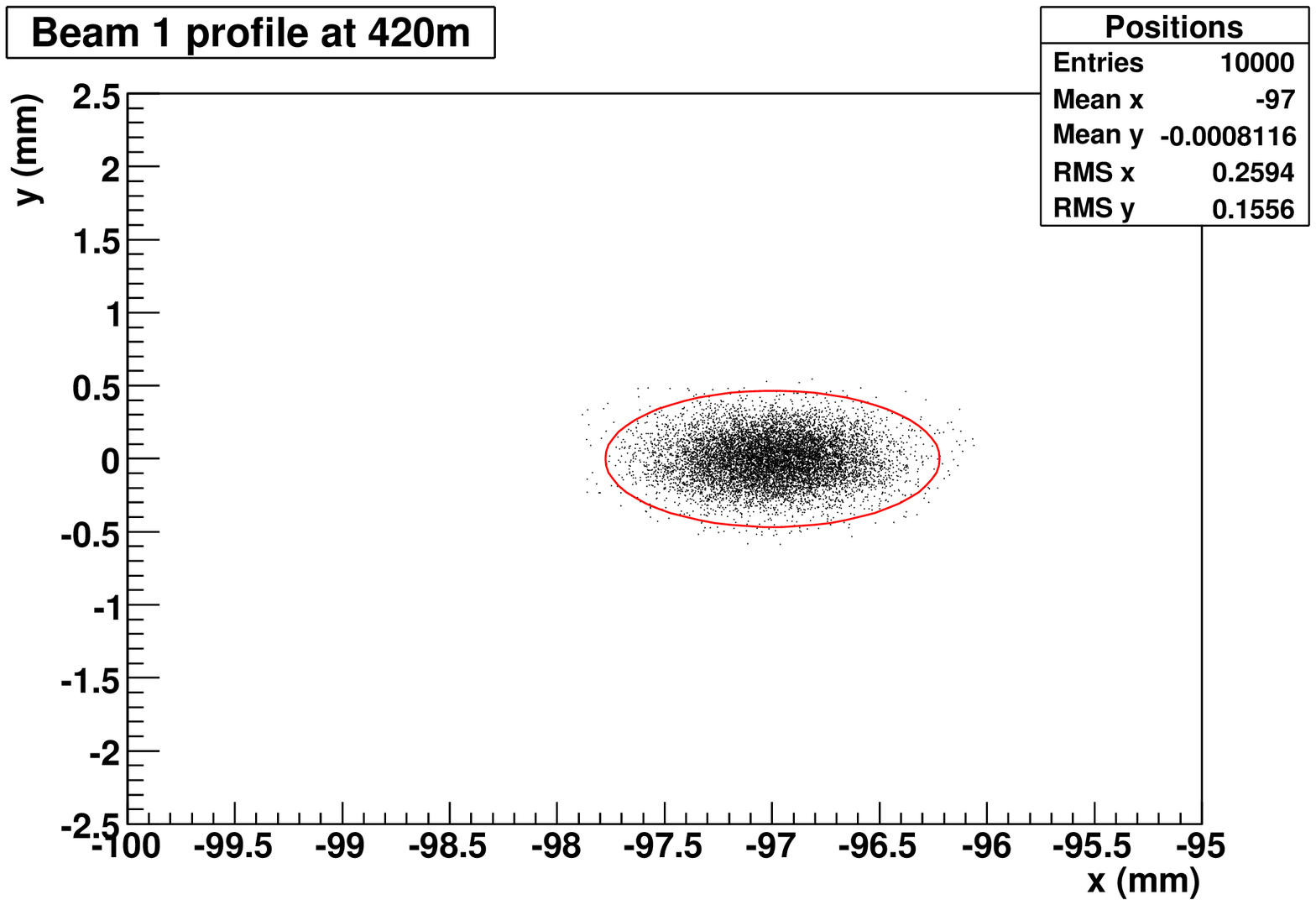} &
\includegraphics[width=0.4\textwidth]{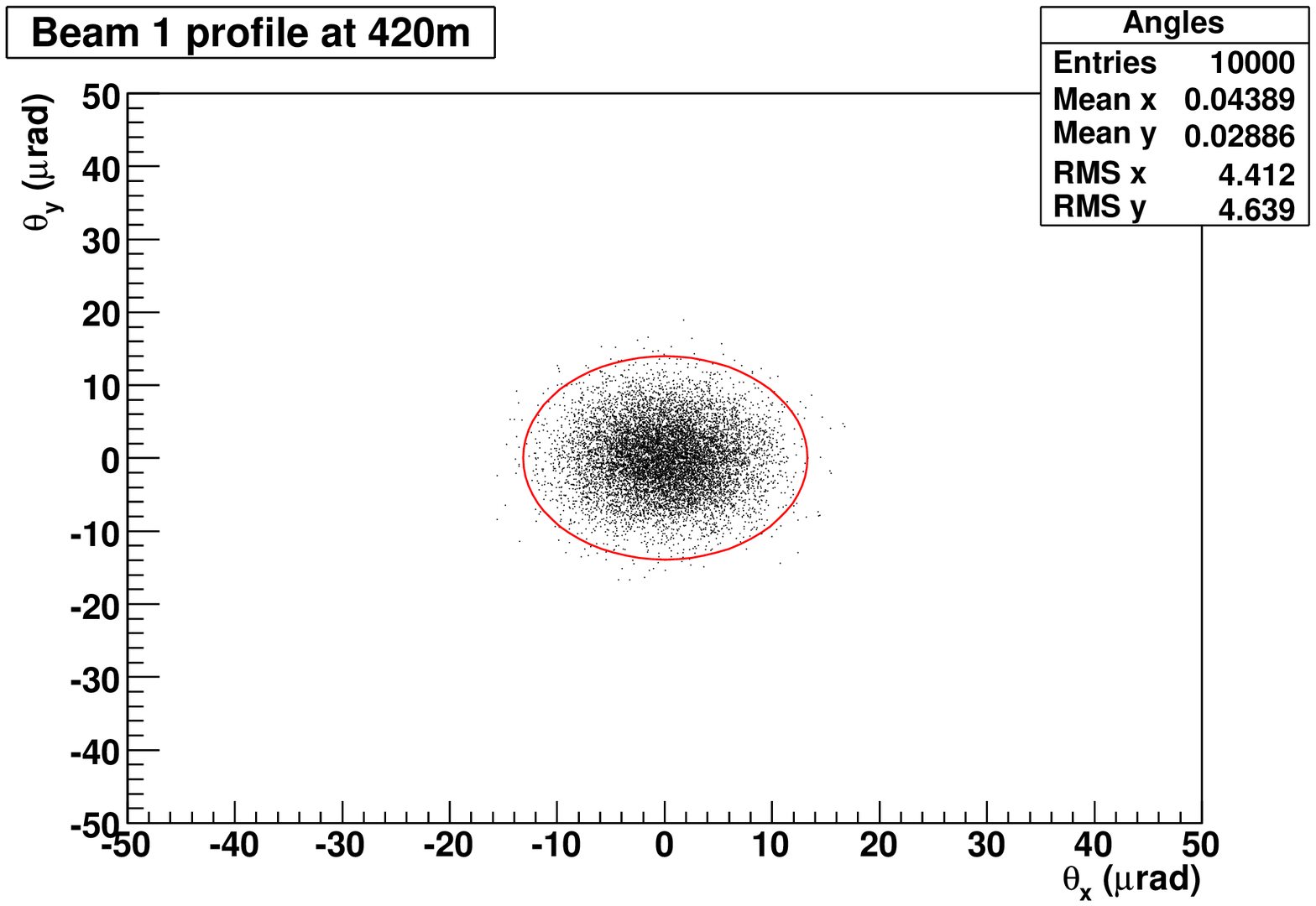} \\
\includegraphics[width=0.4\textwidth]{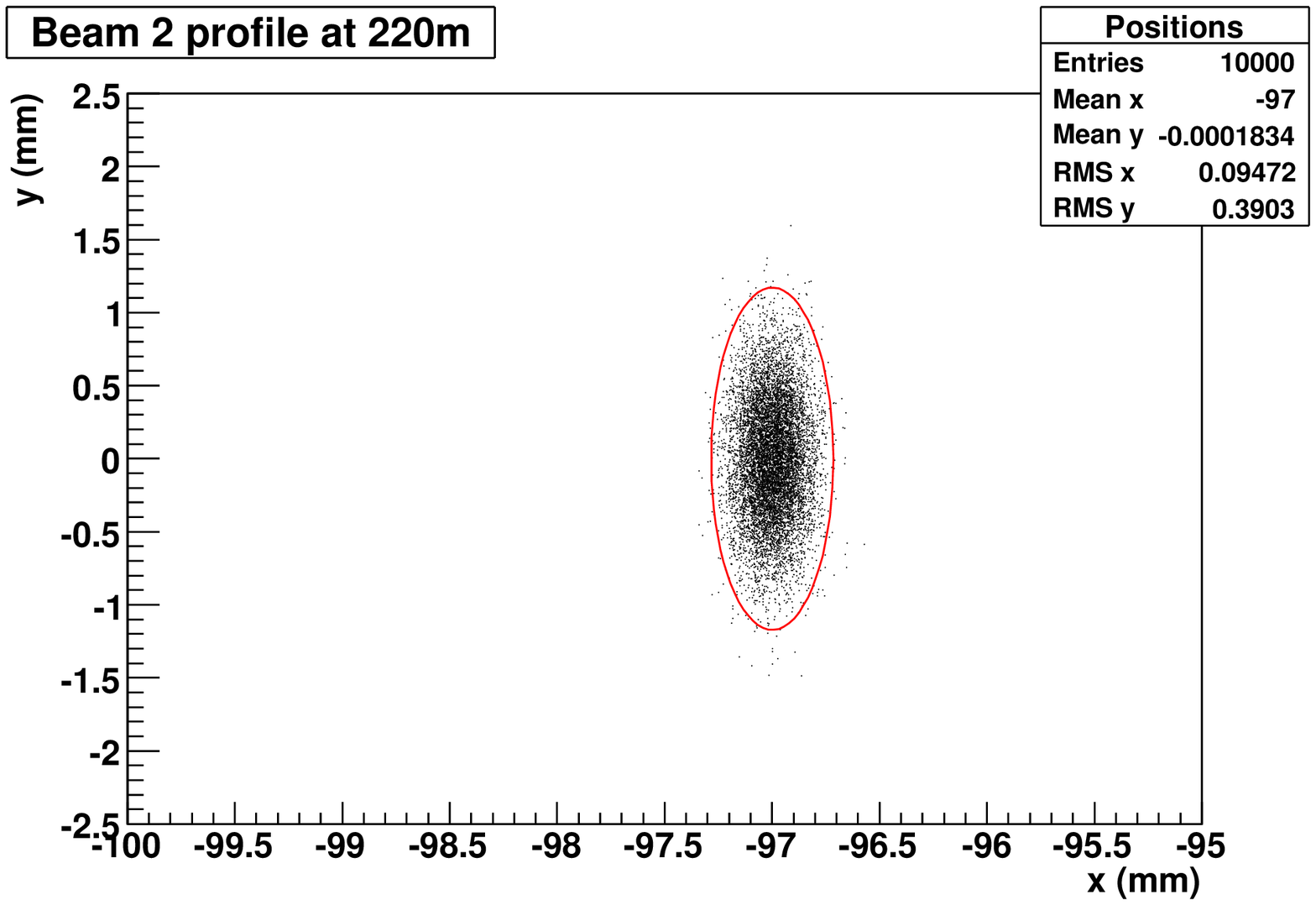} &
\includegraphics[width=0.4\textwidth]{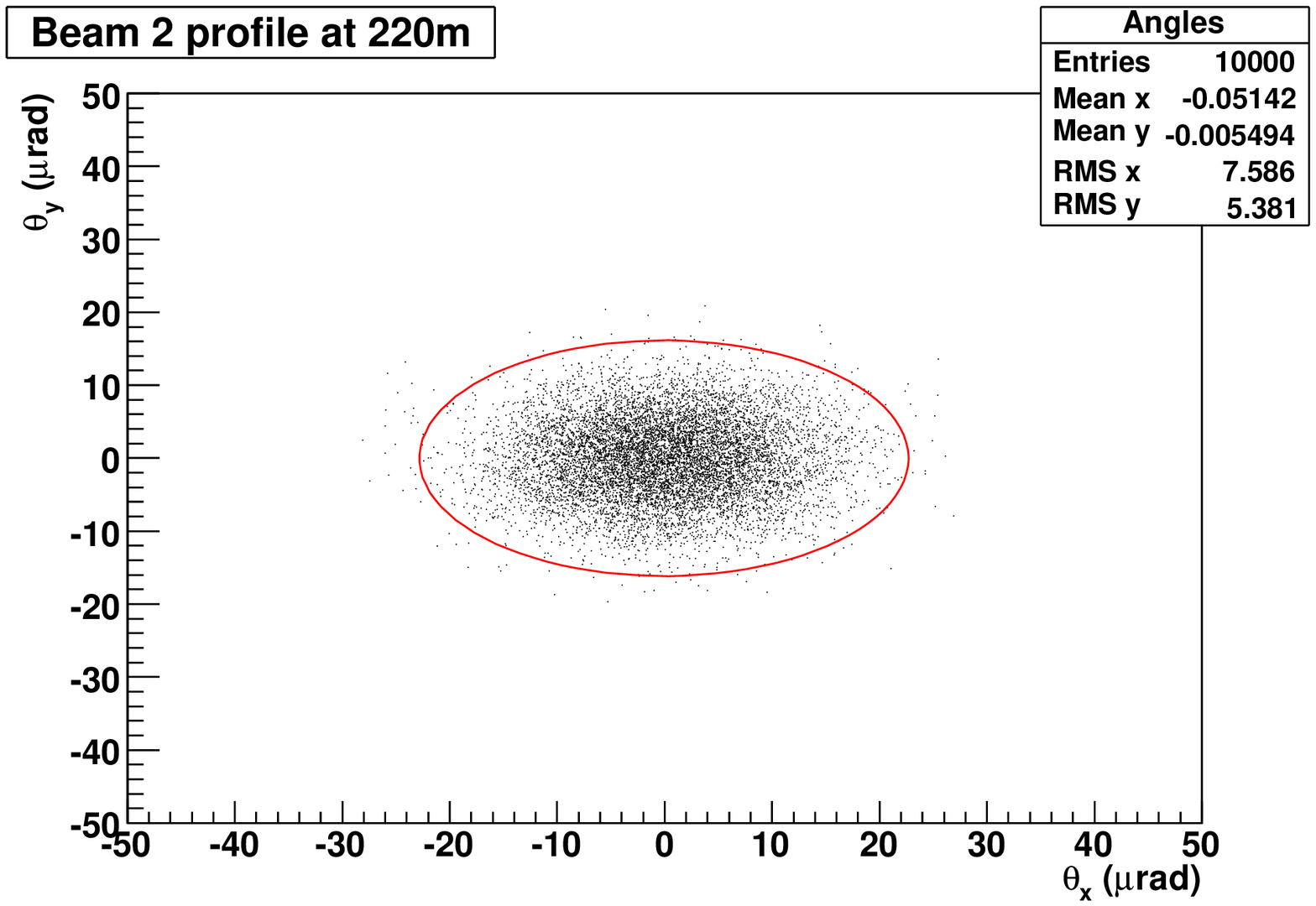} \\
\includegraphics[width=0.4\textwidth]{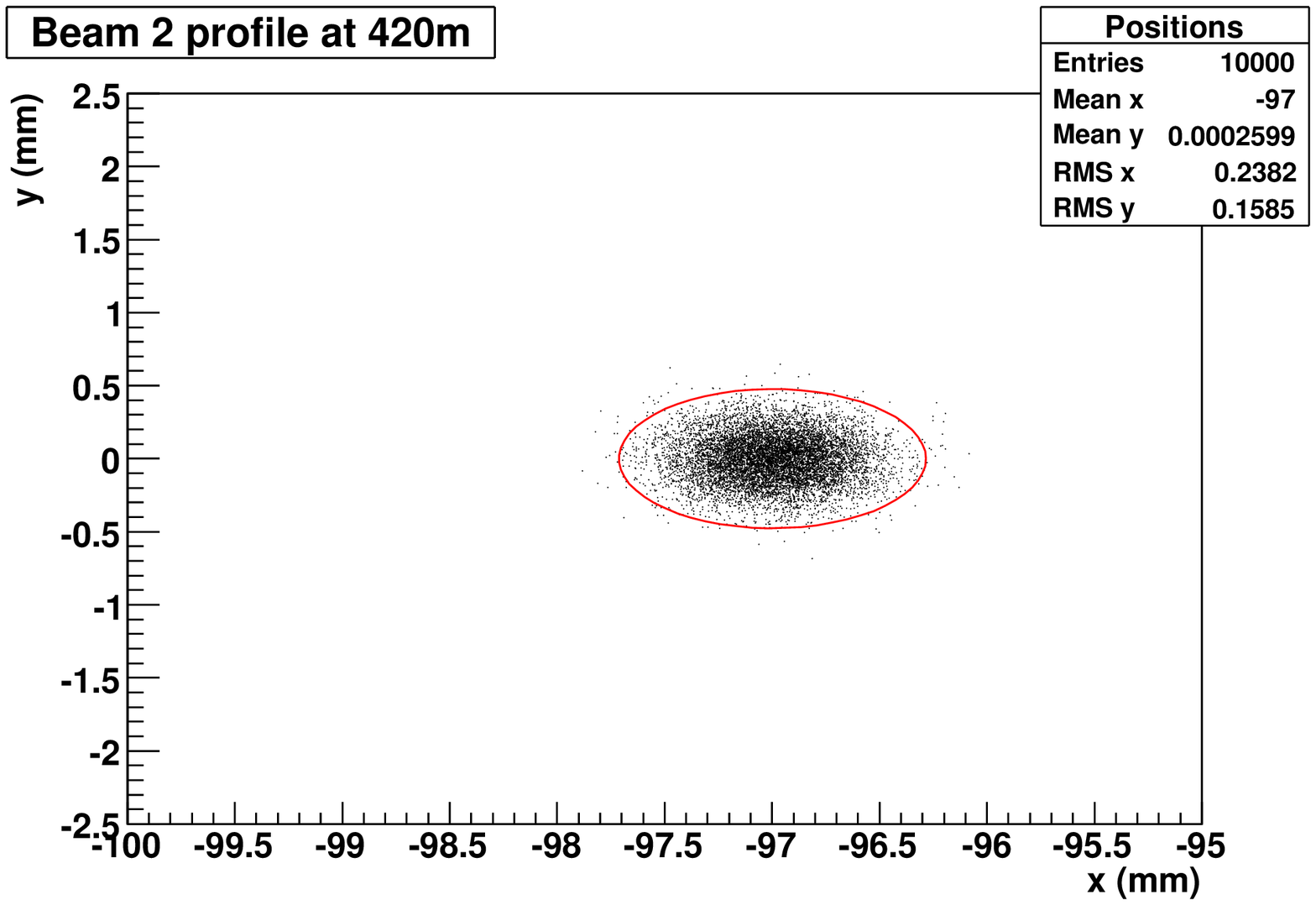} &
\includegraphics[width=0.4\textwidth]{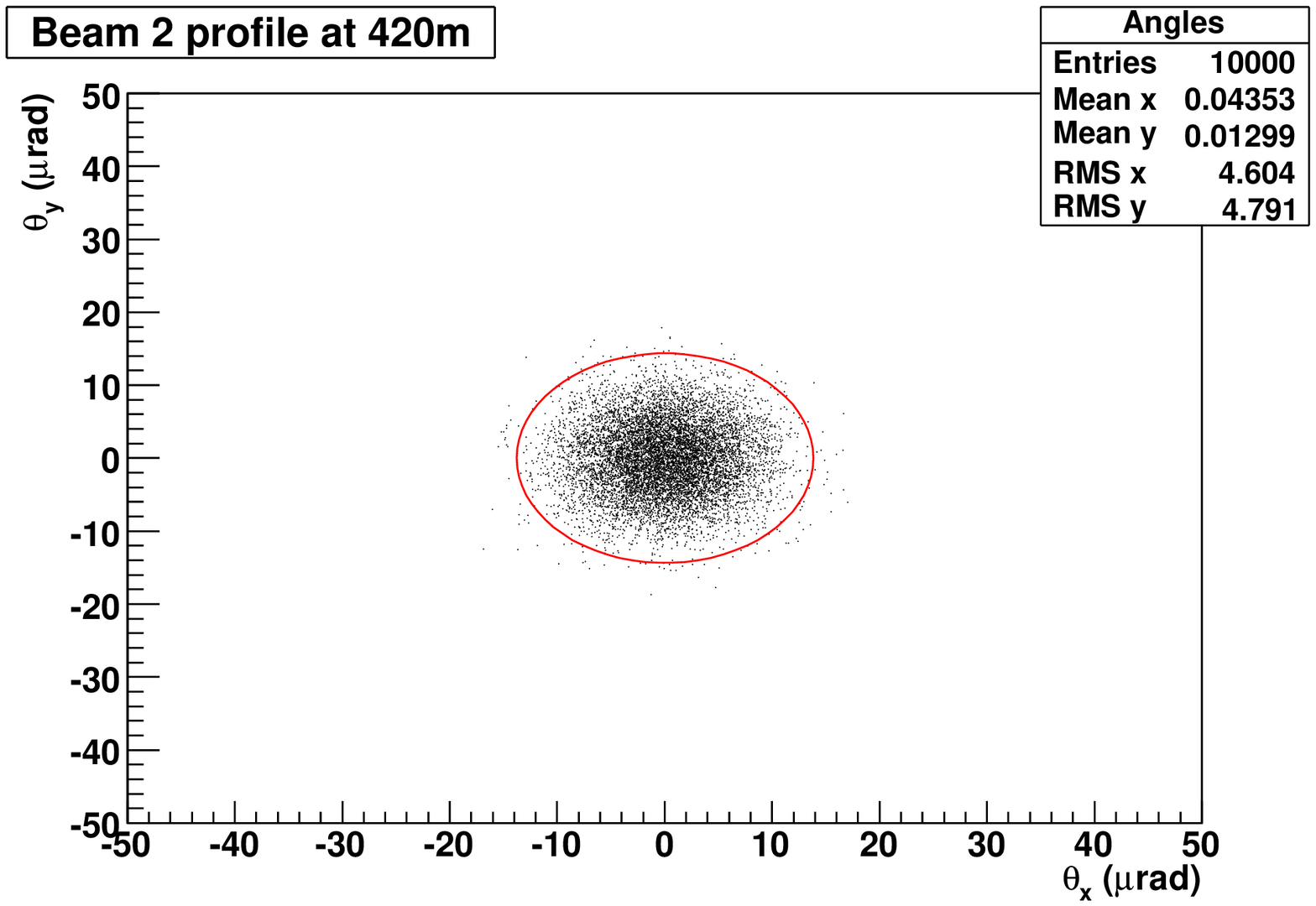} \\
\end{tabular}
\caption[LHC beam profiles in ($x$,$y$) plane]{LHC beam profiles in the transverse plane at 
$220 ~ \mathrm{m} $ and $420 ~ \mathrm{m}$ from the IP5, for both beam 1 (\emph{above}) and 2 
(\emph{below}). The graphs in the left column plot the horizontal and vertical position profiles. 
The graphs in the right column show the horizontal and vertical angular profiles. These plots are 
obtained by propagating ten thousand 7 TeV protons through the LHC beamlines, in the forward direction, 
from the IP5. The mean horizontal position $\bar x =-97 ~ \mathrm{mm}$ matches the half of the 
separation distance between both beams. The red ellipse delimits the $3 \sigma$ contour for 
each plots.} \label{beam_prof}
\end{figure}

\begin{figure}[p]
\centering
\begin{tabular}{cc}
\includegraphics[width=0.4\textwidth]{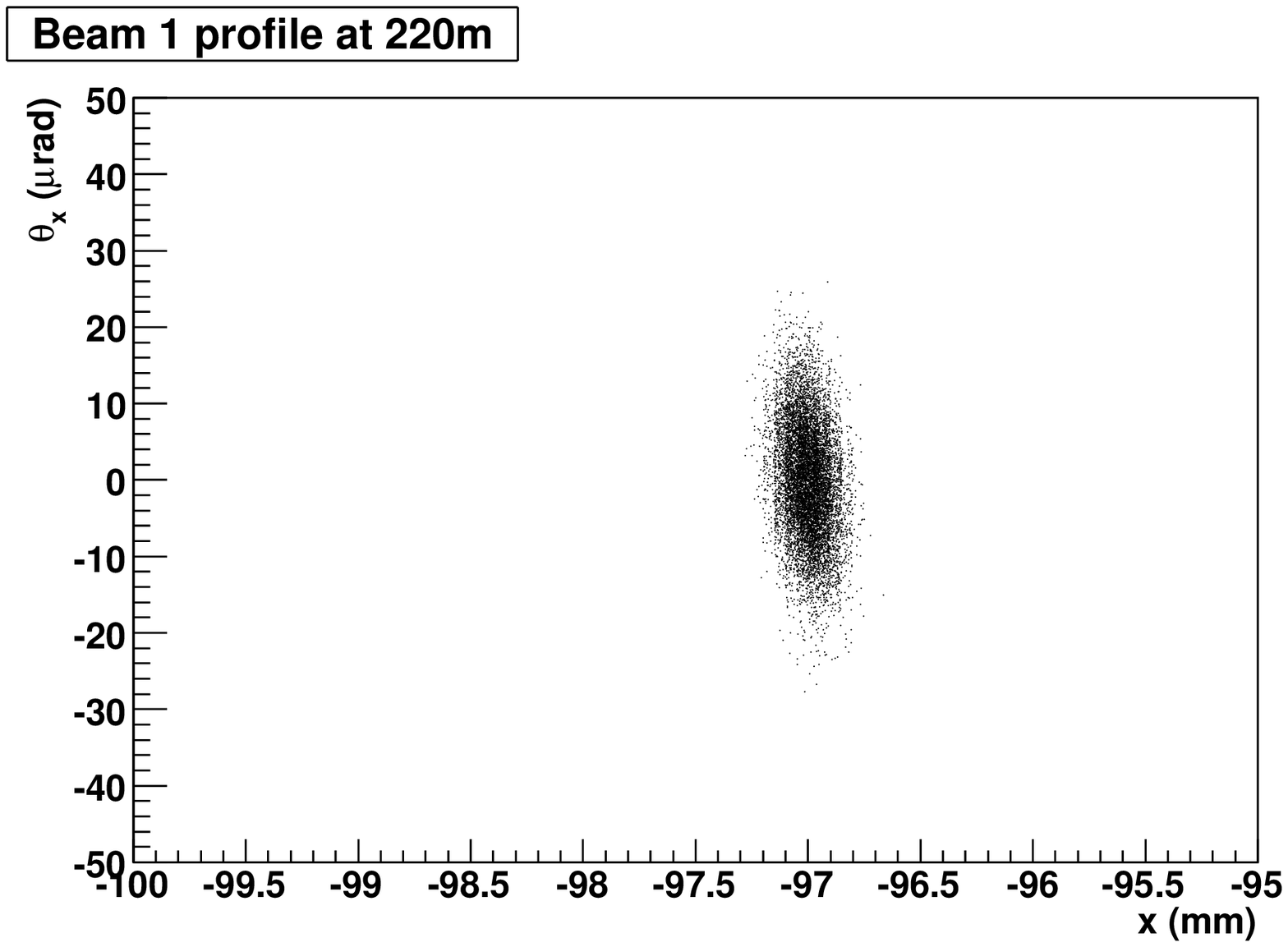} &
\includegraphics[width=0.4\textwidth]{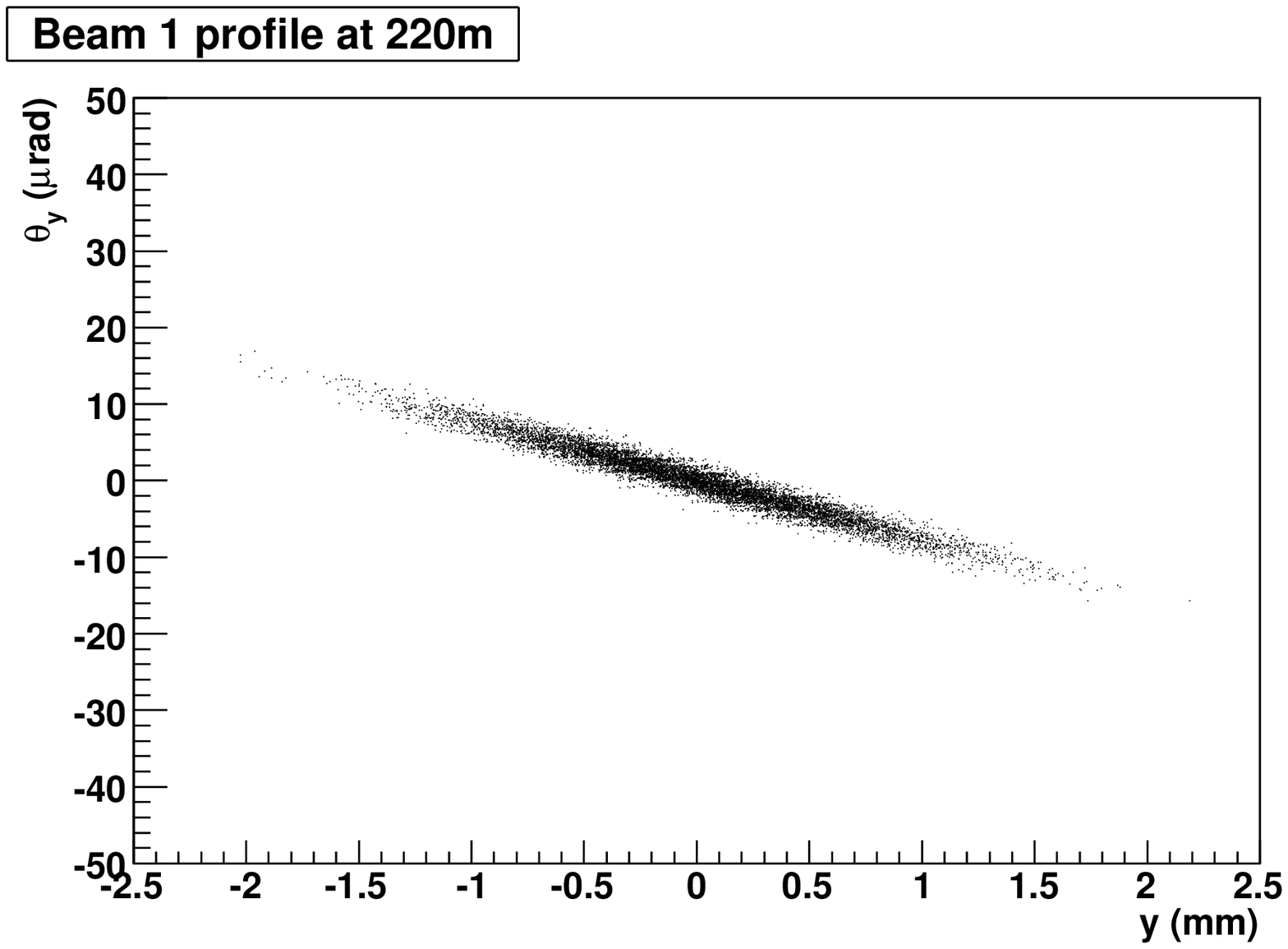} \\
\includegraphics[width=0.4\textwidth]{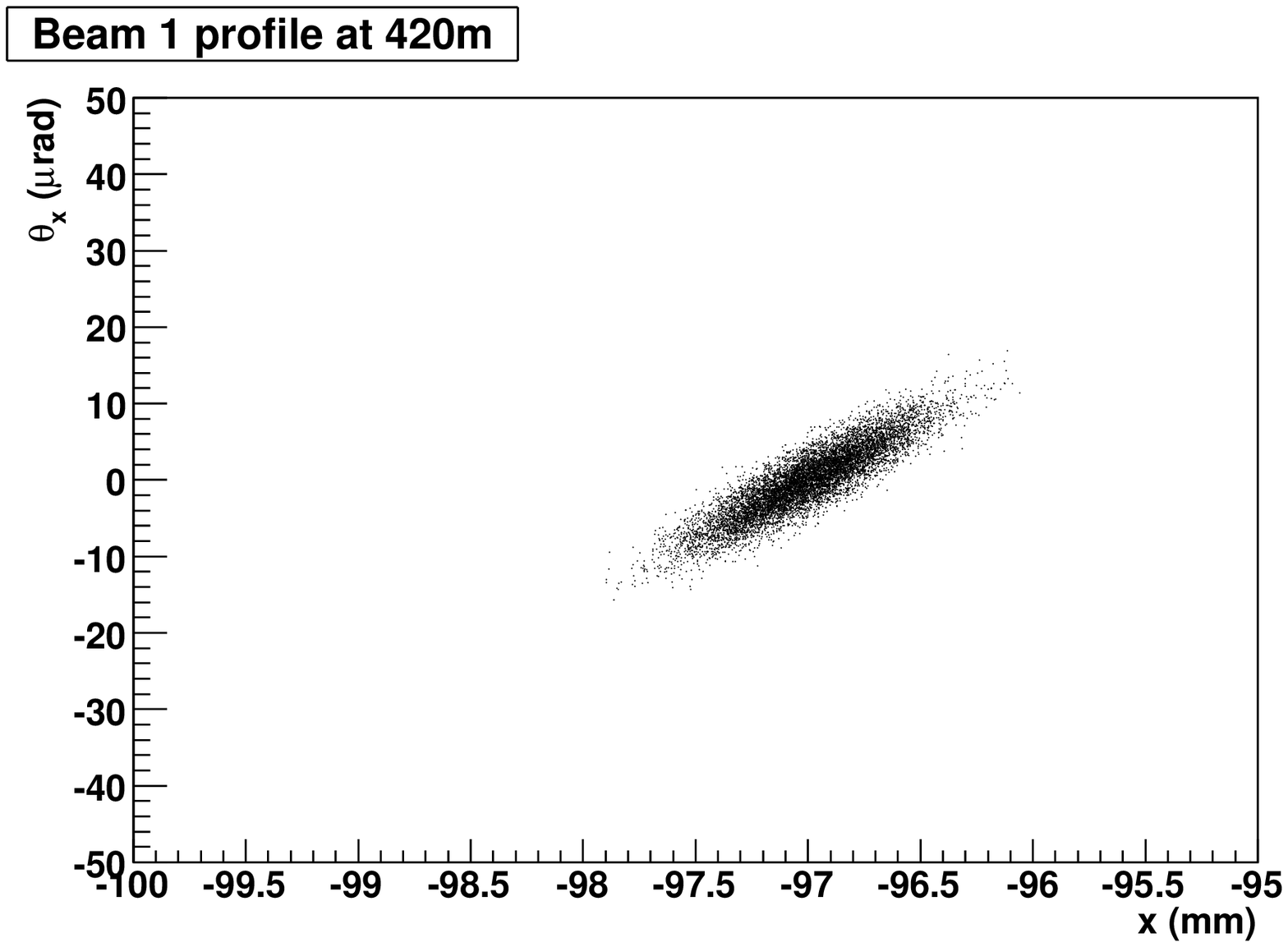} &
\includegraphics[width=0.4\textwidth]{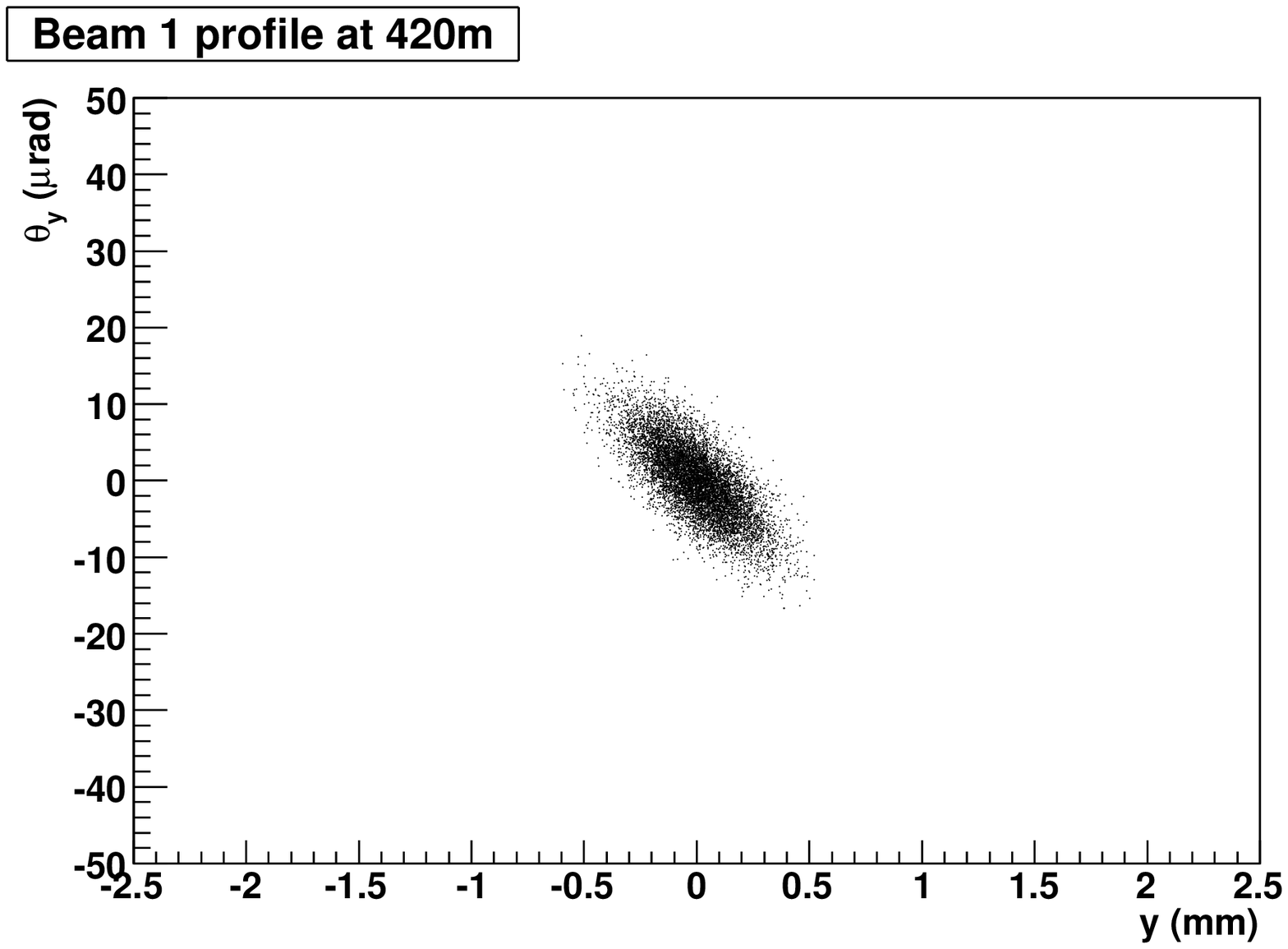} \\
\includegraphics[width=0.4\textwidth]{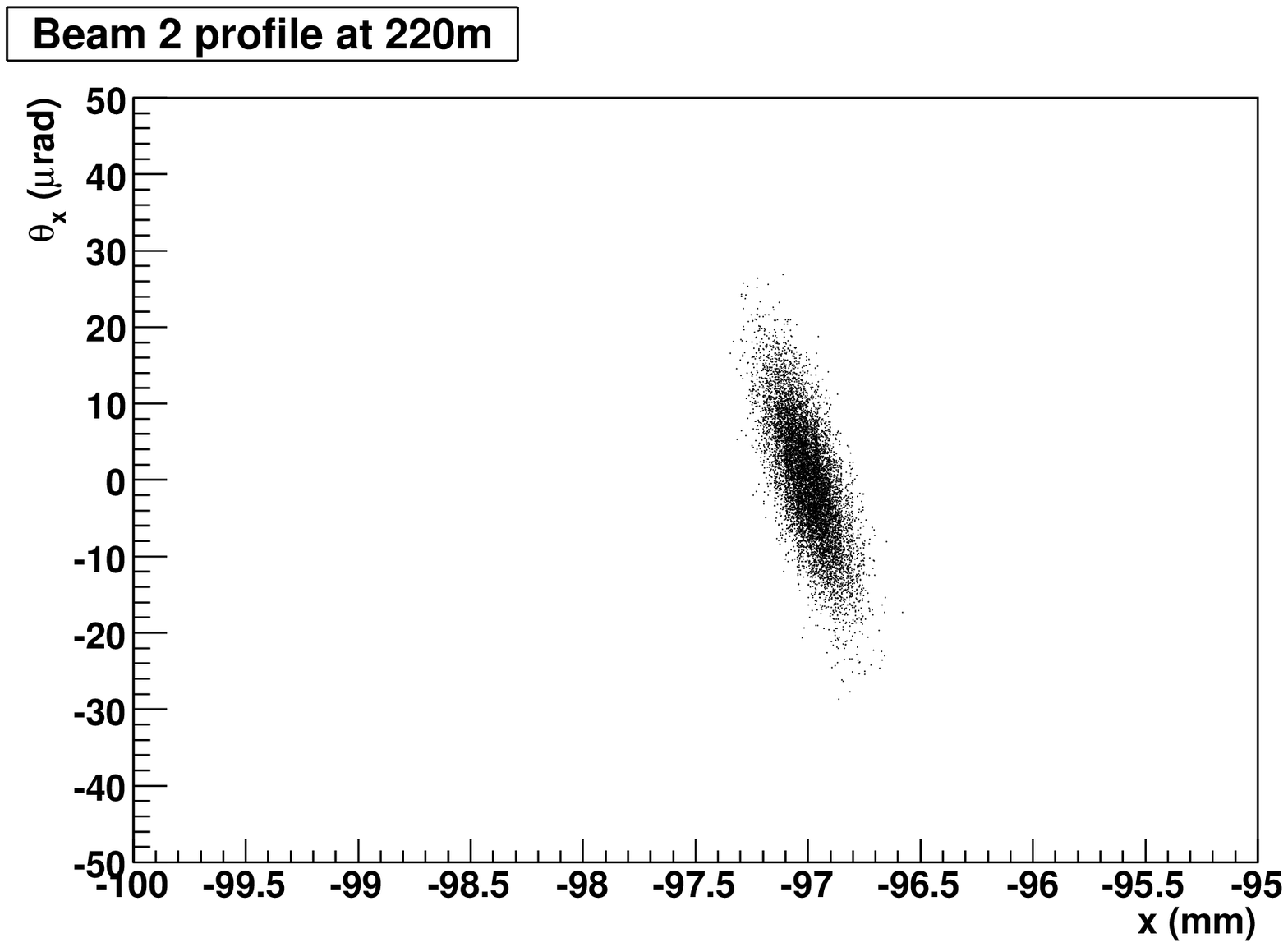} &
\includegraphics[width=0.4\textwidth]{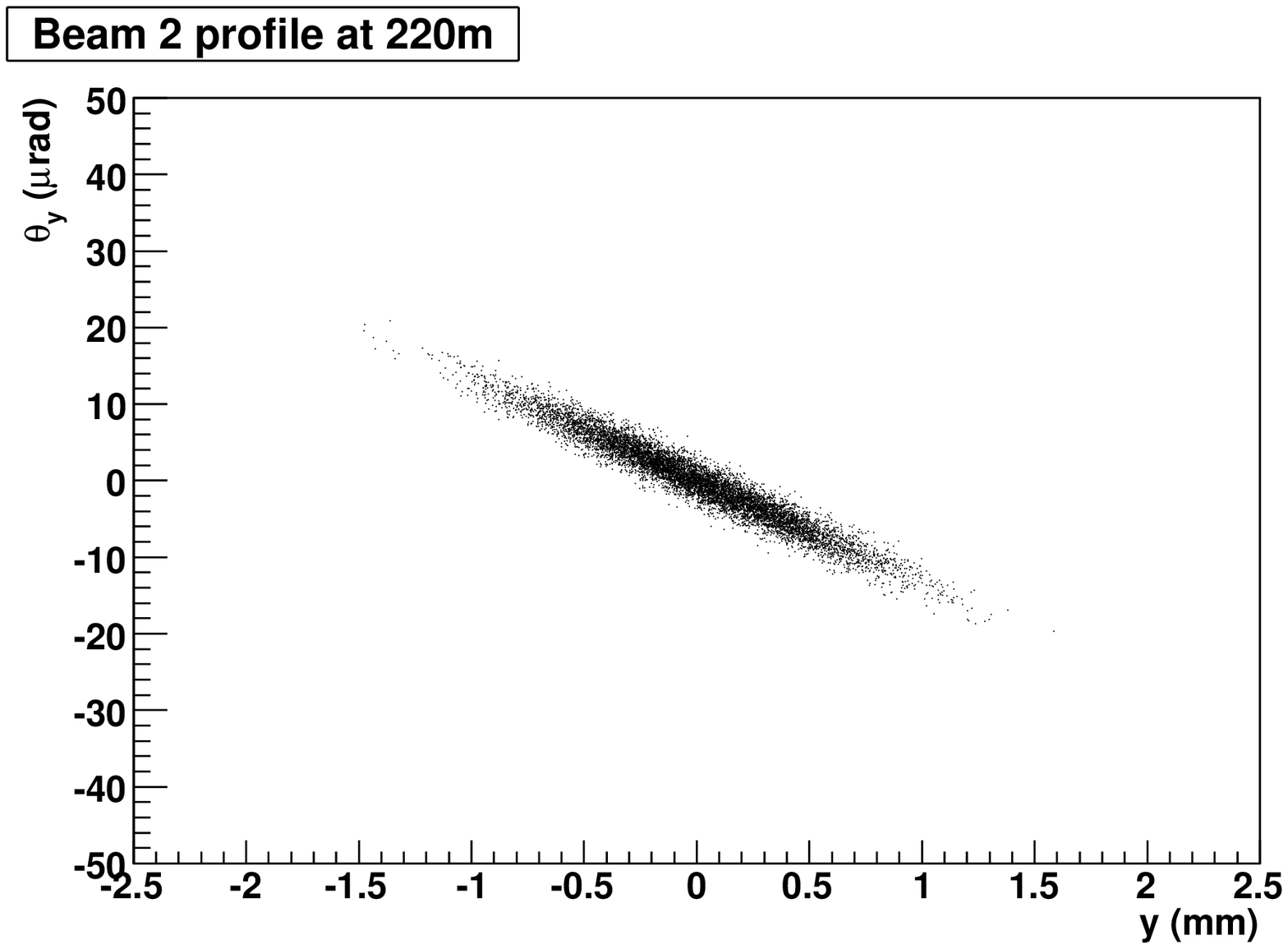} \\
\includegraphics[width=0.4\textwidth]{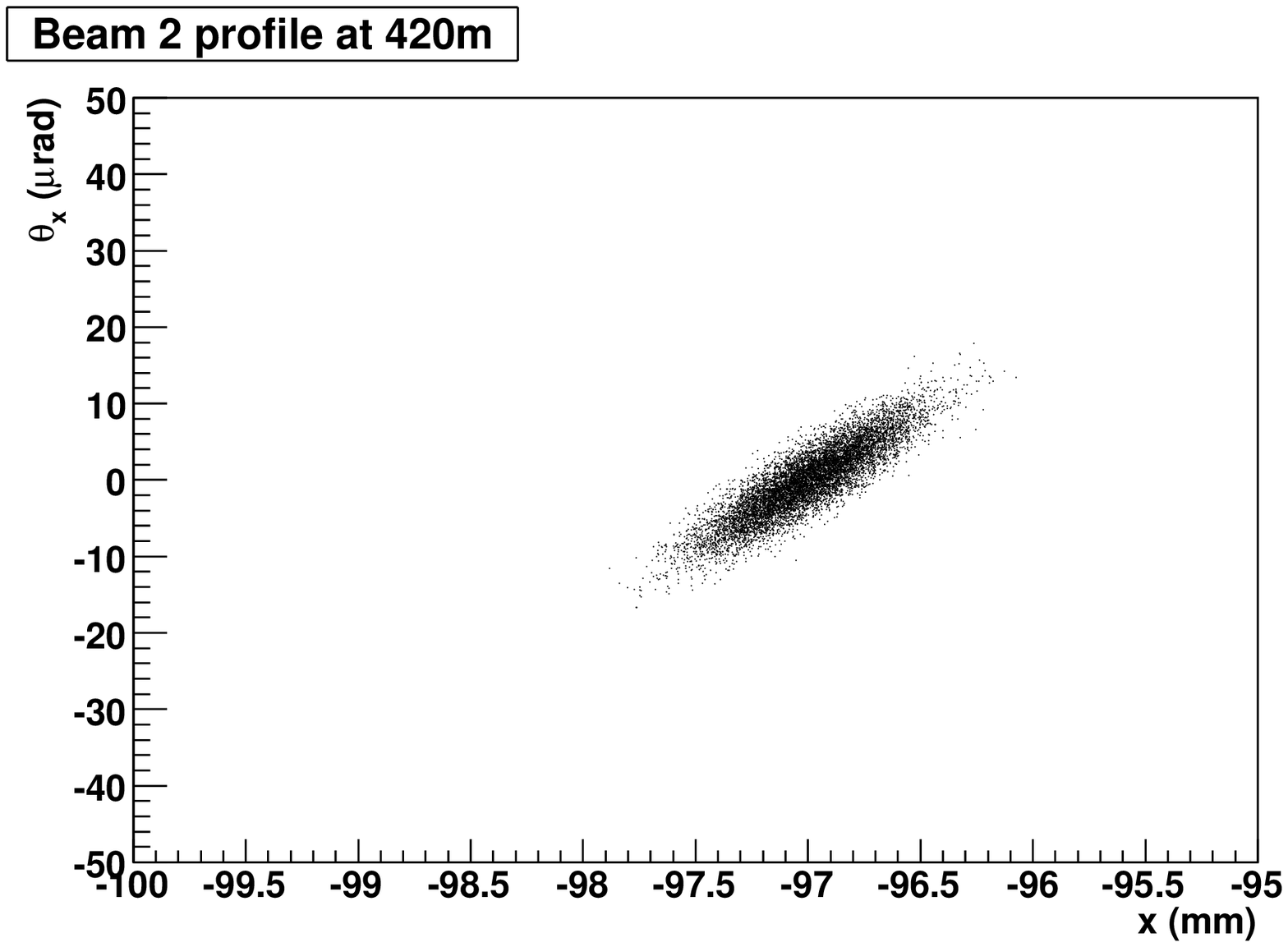} &
\includegraphics[width=0.4\textwidth]{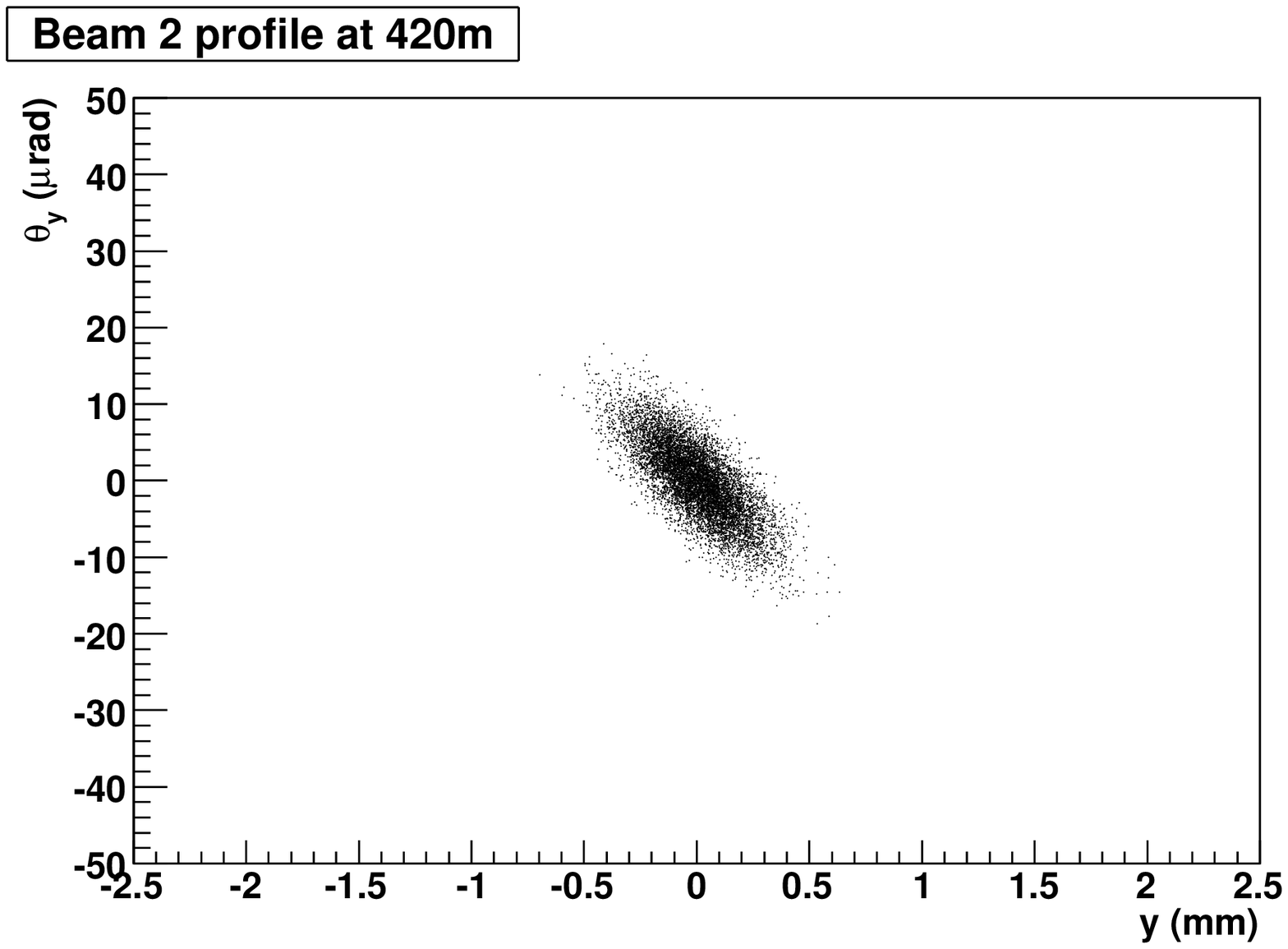} \\
\end{tabular}
\caption[LHC beam profiles in ($x$,$x'$) plane]{LHC beam profiles in the transverse phase space at 
$220 ~ \mathrm{m}$ and $420 ~ \mathrm{m}$ from the IP5, for both beam 1 (\emph{above}) 
and 2 (\emph{below}). These graphs depict the horizontal $(x,\theta_x)$ and vertical $(y,\theta_y)$ 
distributions obtained by propagating $10^4$ 7 TeV protons through the LHC beamlines, in the forward 
direction from the IP5.} 
\label{beam_prof_phase}
\end{figure}

Lateral beam profiles along the beamline are of interest, for instance for studies of very 
forward detectors (Fig. \ref{beam_prof}). The beam size and shape are obtained after propagation 
of beam particles using initial dispersion in position, angle and energy at the interaction point. 
Assuming a Gaussian distribution for these variables, the evolution of both beams can be seen in the 
transverse plane for example at $220 ~ \mathrm{m}$ and $420 ~ \mathrm{m}$ away from the IP. As expected, 
the beam particles are distributed symmetrically on the vertical plane ($\bar y = 0 \mathrm{mm}$), 
and are already separated in the horizontal  plane ($\bar x = \pm 97 \mathrm{mm}$). The symmetry between 
beam 1 and 2 is striking. The $3 \sigma$ contours (in red) guides the eye for the lateral beam 
extension. In the angular $(\theta_x , \theta_y )$ plane, the beams evolve according to the focusings 
and the defocusings they undergo, even if the mean value of these angles remains effectively equal to $0 ~\mu$rad. 
The evolution of the beams in the phase space has also been drawn (Fig. \ref{beam_prof_phase}). Once 
again, the similitude between both beams is clear, even if they differ from each other by slight 
details. All these profiles where obtained by simulation of $10^4$ 7 TeV beam protons.

\begin{figure}[tp]
\centering
\includegraphics[width=0.8\textwidth]{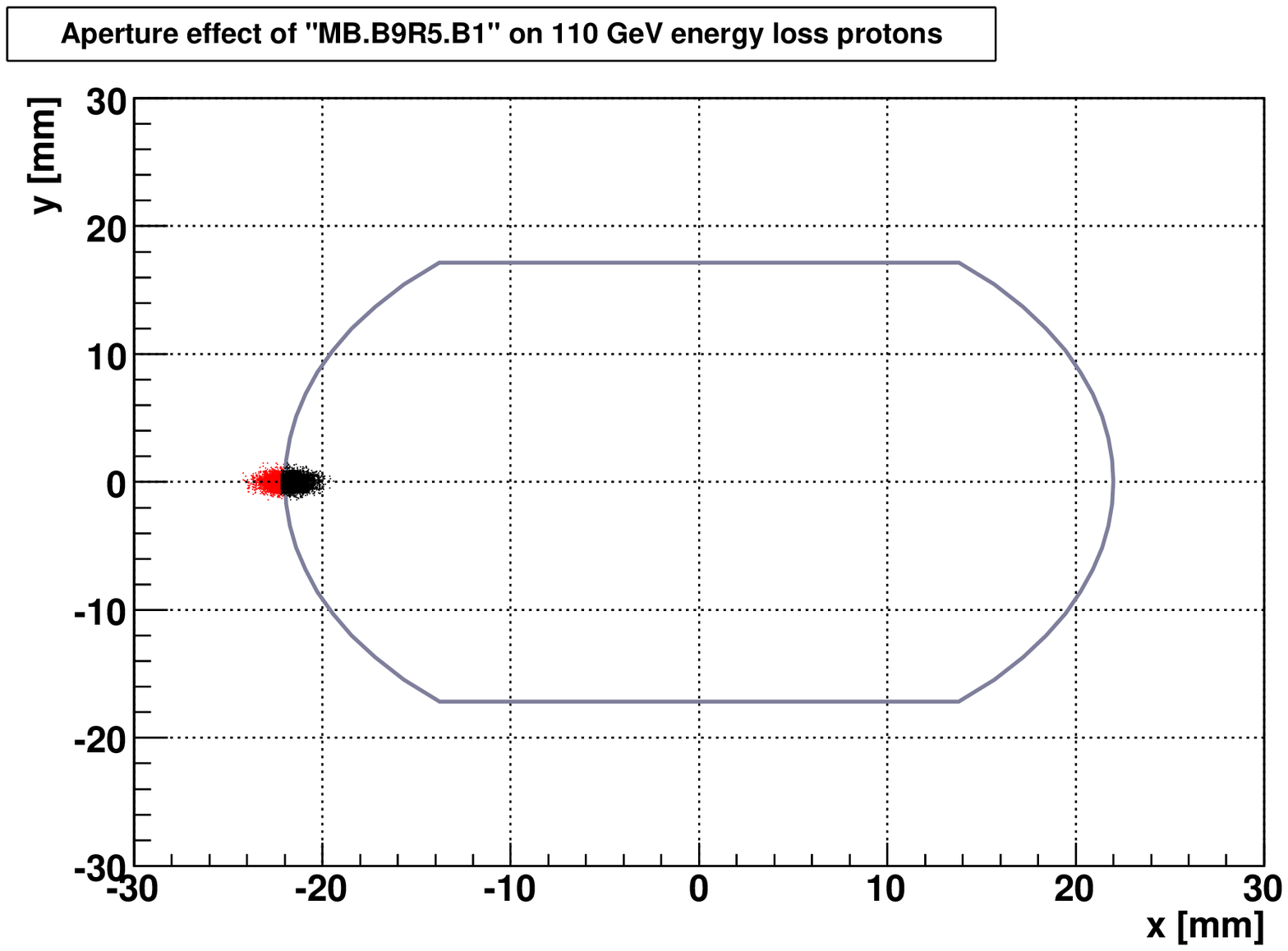}
\includegraphics[width=0.8\textwidth]{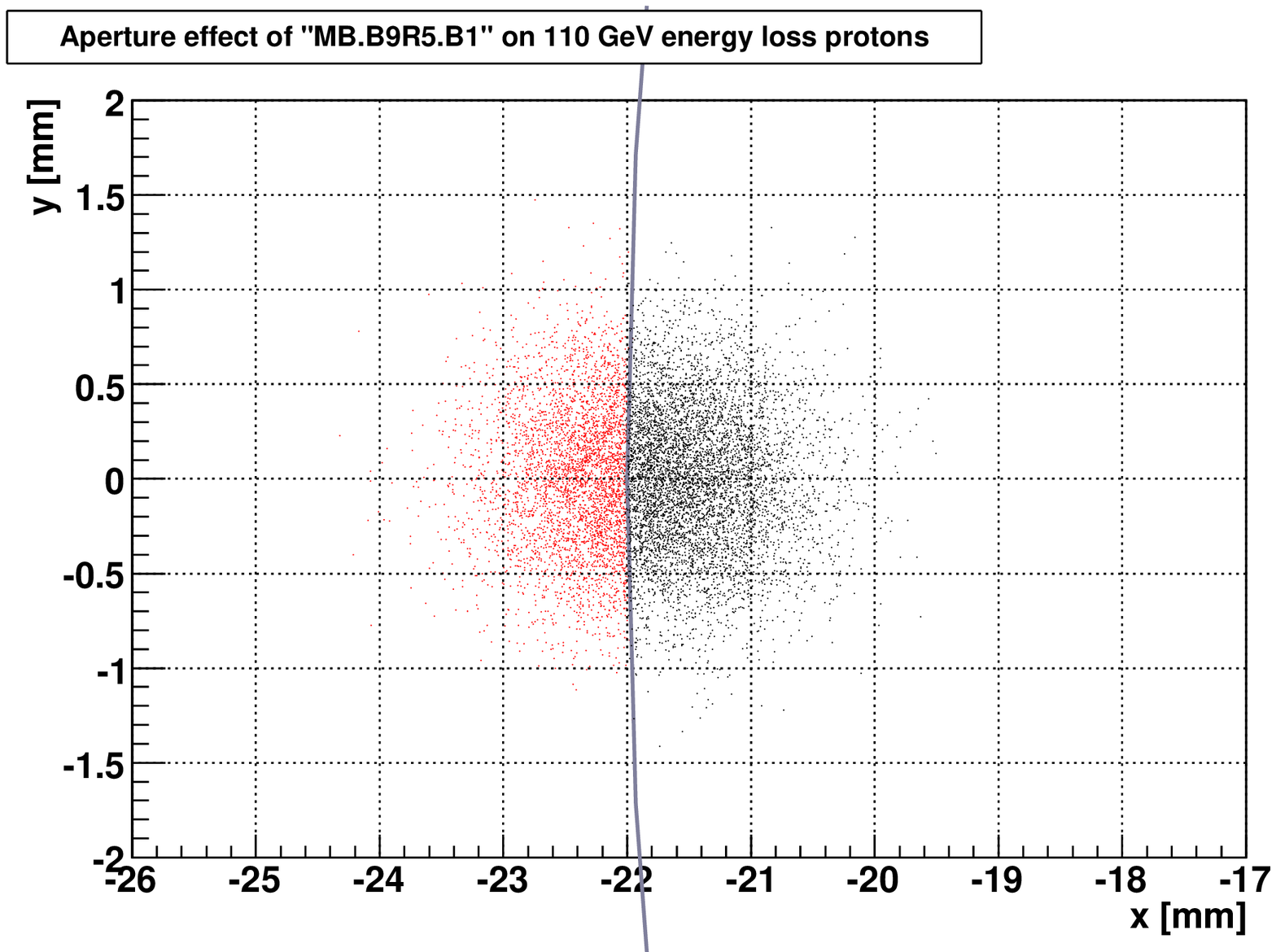}
\caption[Example of aperture check]{Example of aperture check for the \textit{MB.B9R5.B1} dipole ($s=338$ m), 
drawn in the $(x,y)$ transverse plane, at the exit of this optical element. A set of protons, with a 
mean energy loss around $110 ~ \mathrm{GeV}$, has been propagated from the IP5 to the 
\textit{MB.B9R5.B1} dipole using the LHC beam 1 optics. The beam 1 is centered in $(0,0)$ but is not 
shown. The closed blue area corresponds to the \textit{rectellipse} aperture shape of the optical element. 
Protons that passed through the dipole are the black ones, while protons hitting the walls are tagged 
in red. The first graph is scaled for the full aperture shape, the second one has been zoomed in 
around the region of interest.}
\label{Aperture_check}
\end{figure}

\subsection{Apertures}
The physical aperture of the real optical elements has also to be taken into account. If a particle 
hits a collimator or a wall of the primary vacuum system, it is assumed to be lost. Each time a particle 
enters or leaves an optical element, a test is performed to determine if its position matches the element acceptance. 
This is illustrated in Fig. \ref{Aperture_check}. Protons from the LHC beam 1, around the IP5 have lost 
some energy, which leads to a deflection with respect to the nominal beam position in $(0,0)$ in the 
transverse plane. The aperture shape of the \textit{MB.B9R5.B1} optical element is the rectelliptical 
hatched area. The proton energy has been chosen to $110 ~ \textrm{GeV}$, in order to match the element 
limiting energy acceptance. Due to the initial smearings, some protons (in black) still pass through the 
optical element without being stopped. On the other hand, some of them hit the limiting aperture shape 
and are tagged as stopped protons (in red). In such a case, the particles are considered as lost 
from the beam, and not propagated any further.

\clearpage

\section{Very Forward Detectors}
The physics program at the LHC has been extended to processes in which isolated particles at very large 
(pseudo-)rapidities are produced \cite{opus} in diffractive and photon induced interactions \cite{KPgamma}. To study those processes, 
several very forward detectors are being prepared for installation along the beamline. This section
will discuss in particular two scenarios, differing by their distance from the IP ($s$) and their minimal
horizontal distance ($x$) from the nominal beam position. The assumed location of the first detector is 
$(s=$220--224$\textrm{ m,} ~ x=2000 ~ \mu\textrm{m})$, and of the second one at 
$(s=$420--428$\textrm{ m,} ~ x=4000 ~ \mu\textrm{m})$. No hypothesis is made on their detection 
efficiency or their resolution, unless quoted. We consider here VFDs providing 2D-measurement 
($x$ and $y$ coordinates), each consisting in fact of two stations separated by 4 and $8 ~\textrm{m}$ 
as a lever arm for the angle measurement, with no magnetic element in between. In the following several 
examples of studies performed using \textsc{Hector} are given.

\subsection{Acceptance}
\begin{figure}[!bh] 
\centering
\begin{tabular}{cc}
\includegraphics[width=0.45\textwidth]{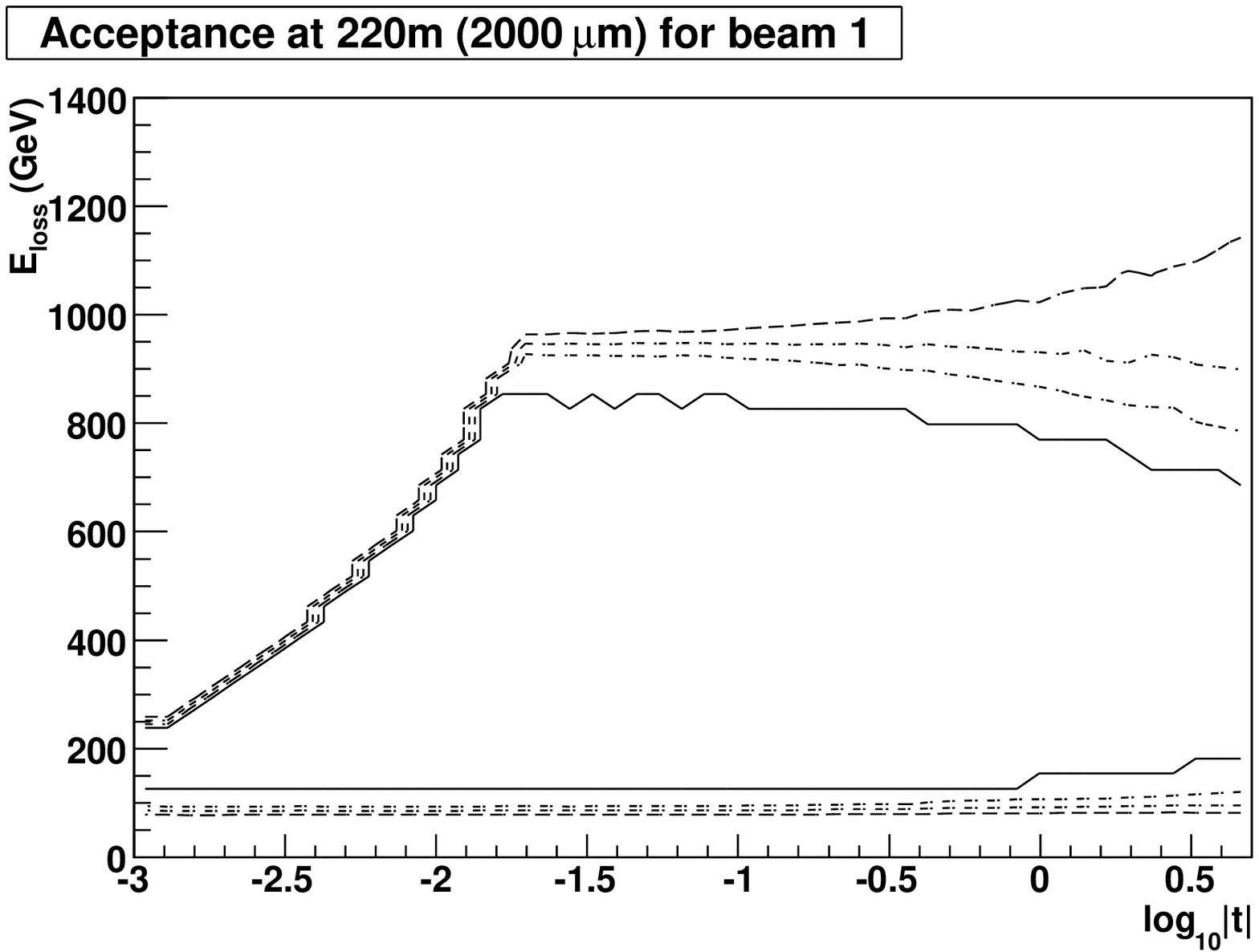}  &
\includegraphics[width=0.45\textwidth]{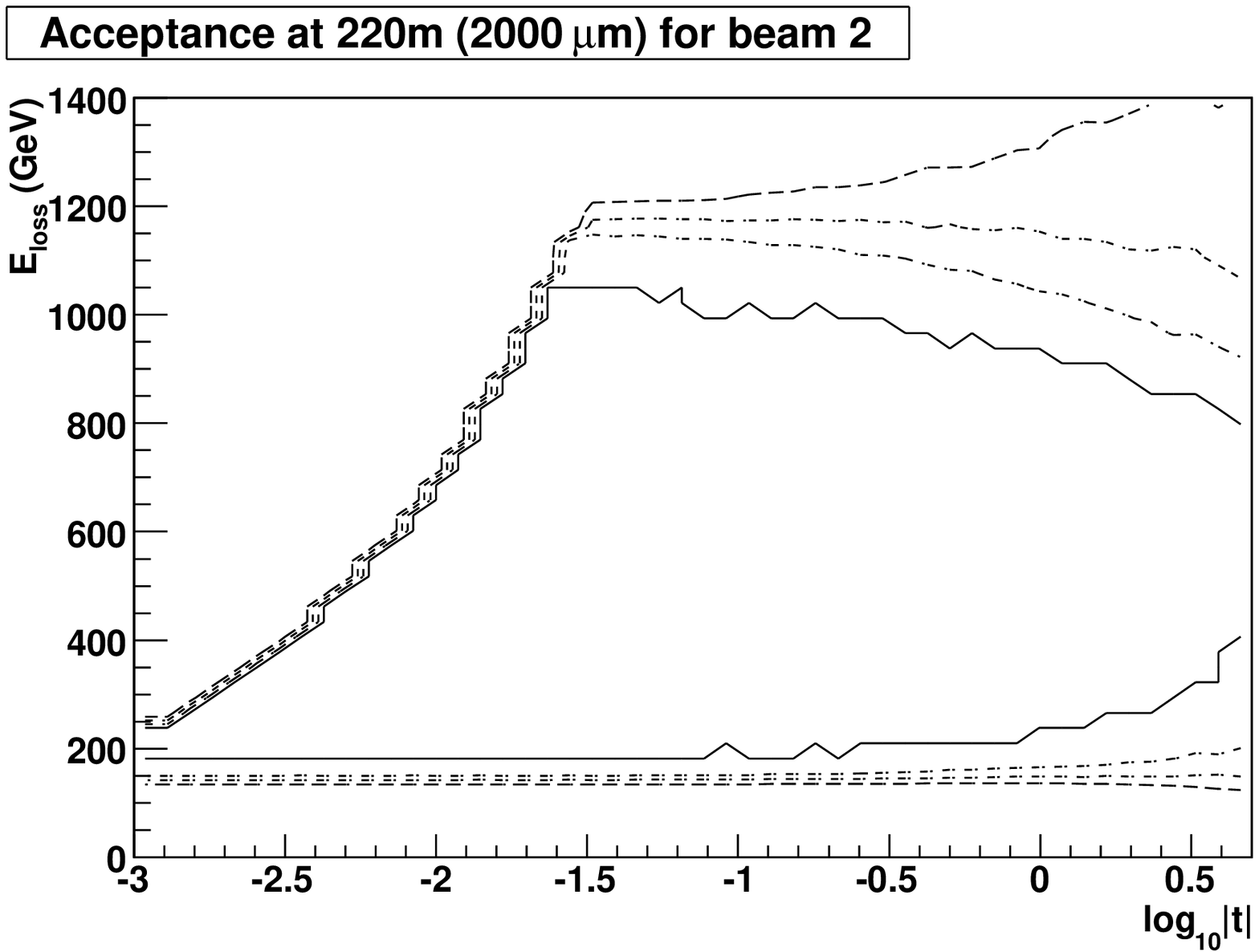}  \\
\includegraphics[width=0.45\textwidth]{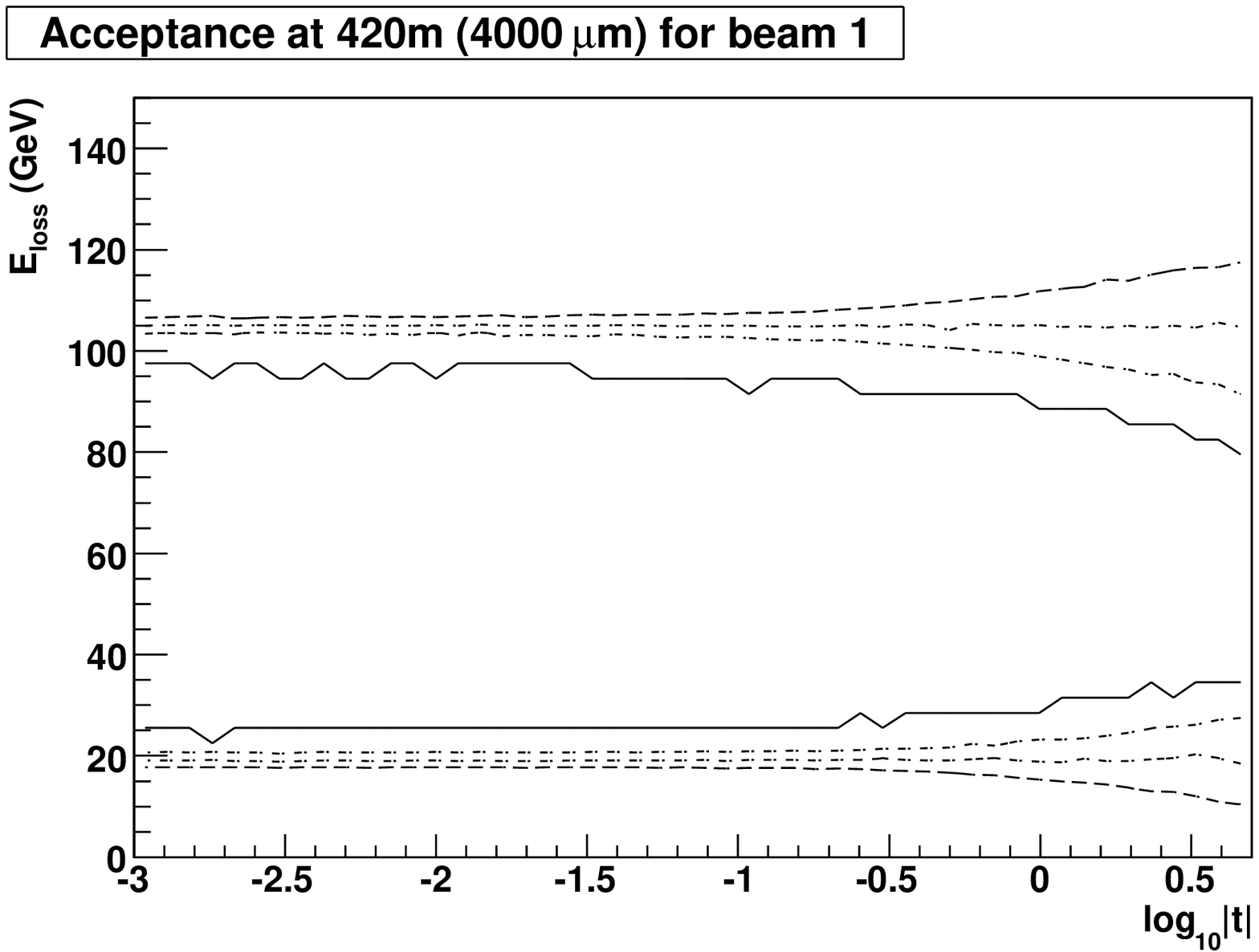}  &
\includegraphics[width=0.45\textwidth]{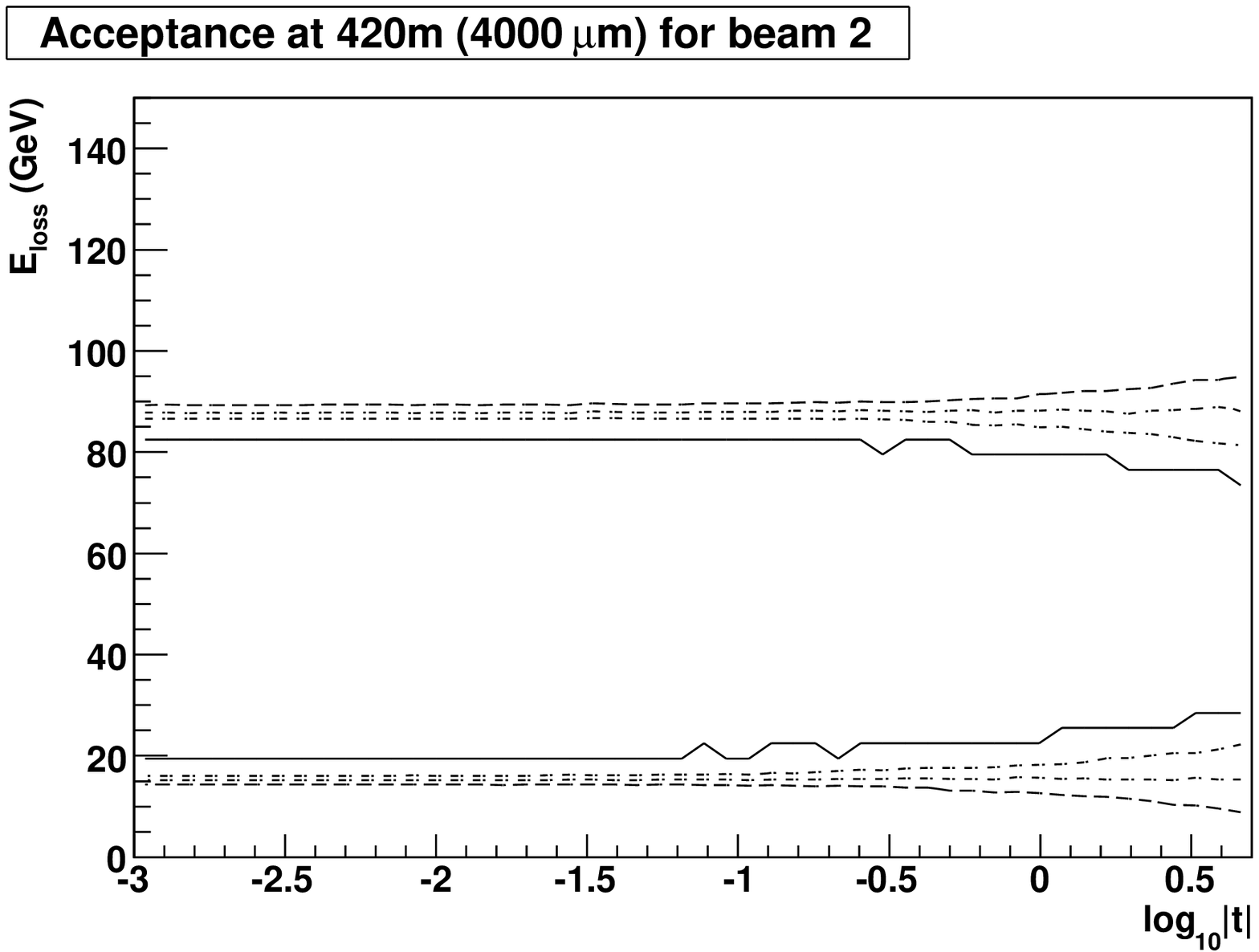}  \\
\end{tabular}
\caption[Roman pot 2D-acceptances ($E$,$Q^2$)]{VFD proton acceptance for the LHC beam 1 
(\emph{left}) and 2 (\emph{right}) around the IP5. These VFDs are located at ($s=220$ m, 
$x=2000 ~ \mu$m horizontal position ; \emph{above}) and ($s=420$ m, $x=4000 ~ \mu$m 
horizontal position ; \emph{below}). This map shows contours of $25\%$, $50\%$, $75\%$ 
and (plain curve) $100\%$ acceptance. The acceptance is roughly rectangular, i.e. independent 
of $t$. At $220$ m, the missing triangle in the acceptance at low $|t|$ and high energy loss is 
prohibited by kinematics. This area corresponds to non-physical protons. The stair-stepping effect 
of the lower border of this triangle is only due the binning of the graph. At bottom right corner 
of each graph, the angular kick coming from the large momentum transfers leads to an 
increasing smearing of the graph lower edge.} \label{RP_acceptance}
\end{figure}

\begin{figure}[!th] 
\centering
\begin{tabular}{cc}
\includegraphics[width=0.45\textwidth]{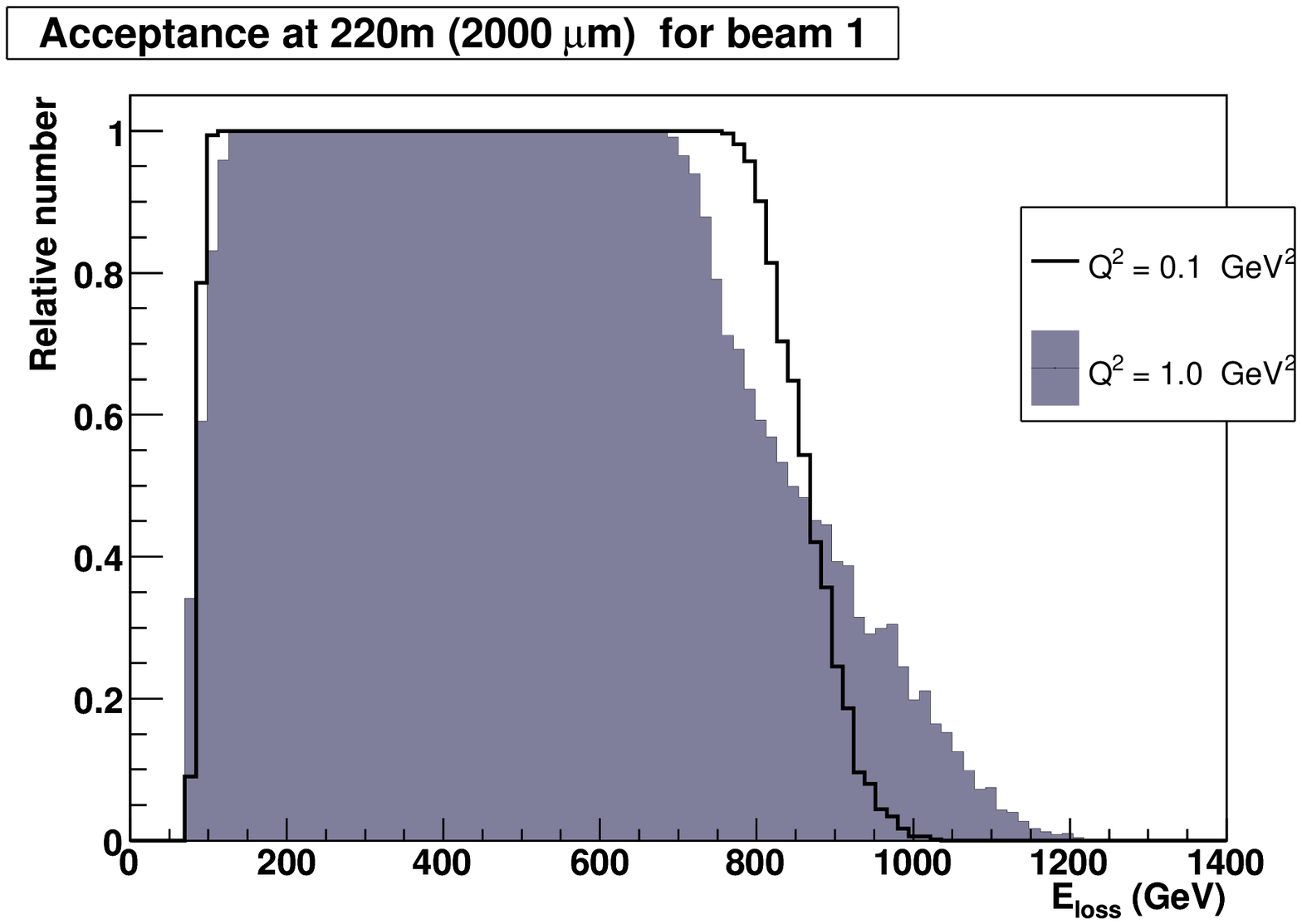} &
\includegraphics[width=0.45\textwidth]{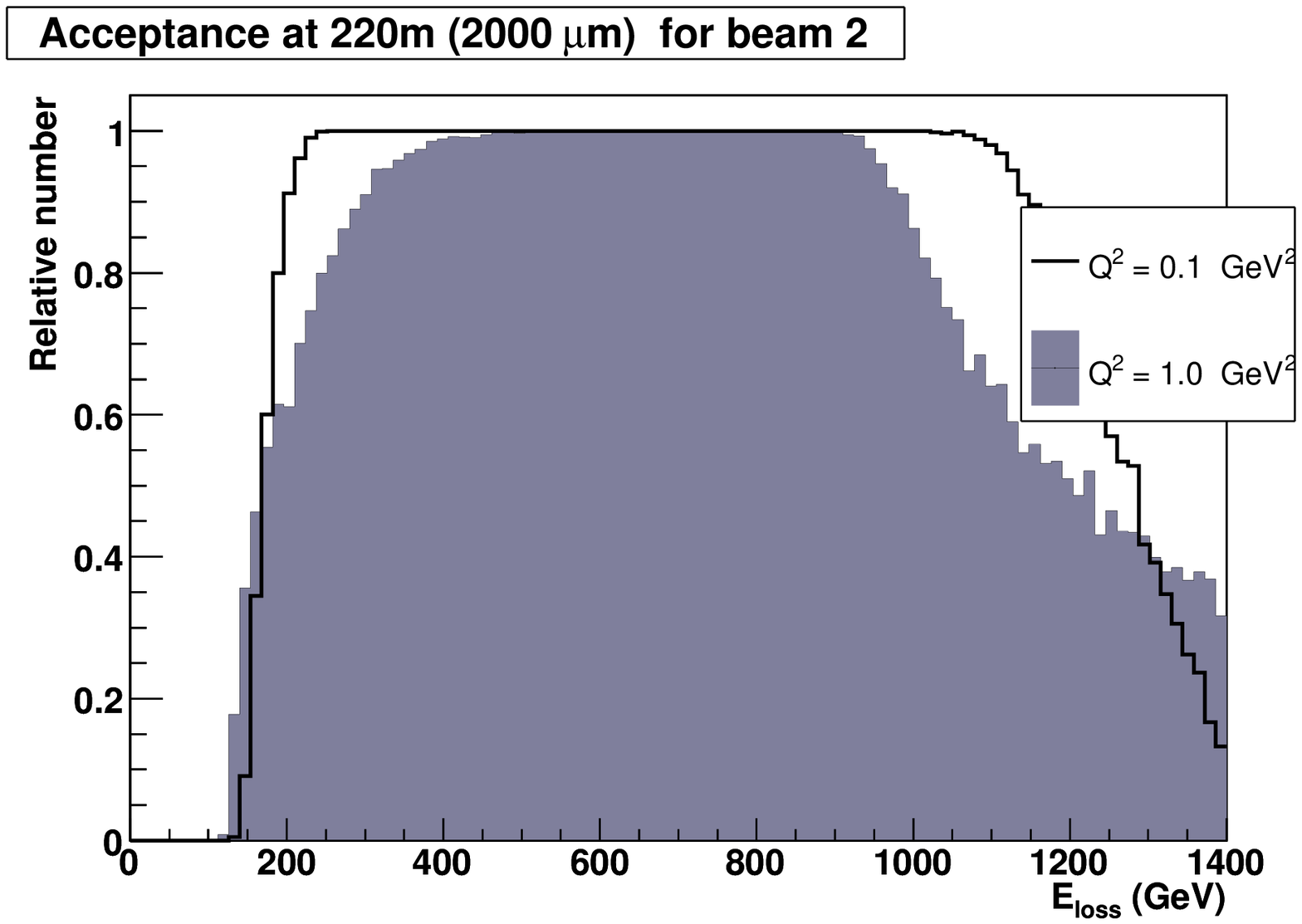} \\
\includegraphics[width=0.45\textwidth]{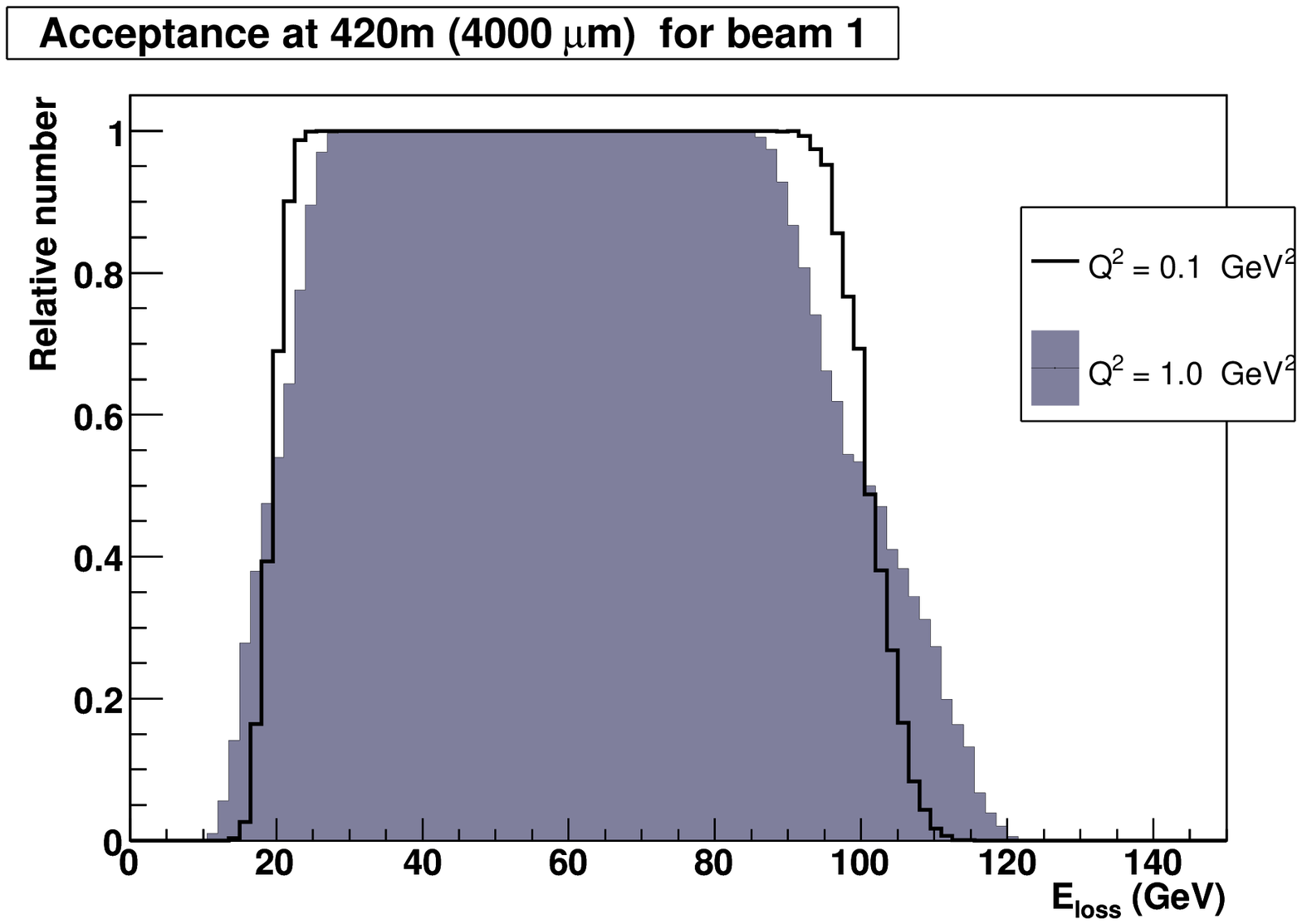} &
\includegraphics[width=0.45\textwidth]{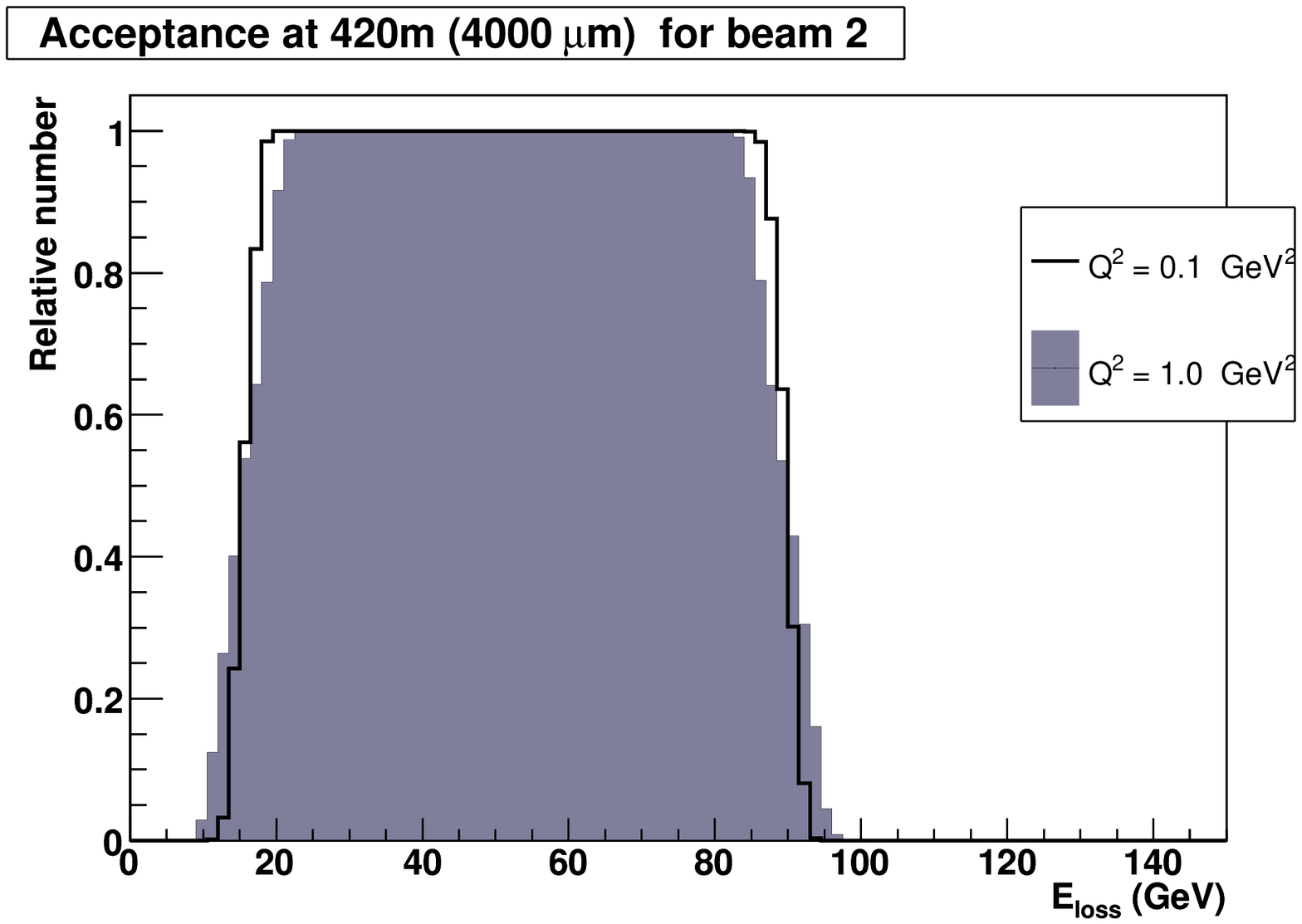} \\
\end{tabular}
\caption[Roman pot 1D-acceptances ($E$)]{VFD acceptance as a function of energy loss, for two fixed 
virtualities (Reminder: $Q^2=-t$). See previous figure for more details.}
\label{RP_acceptance_1D}
\end{figure}

Using \textsc{Hector}'s aperture description, it is possible to identify the characteristics of the 
protons that will hit the VFD. The exchange of a photon or a pomeron, leaving the proton intact, 
results in a proton energy loss ($E_{loss}$) and a scattering angle, directly linked to the
four-momentum transfer squared ($t$). In case of the photon-induced interactions $-t$ is equal to the 
virtuality $Q^2$ of the exchanged photon. 
The acceptance windows of the VFDs can be computed by performing scans in $(E_{loss},t)$ and
computing the probabilities of reaching the detectors. 
The Fig. \ref{RP_acceptance} shows the contour plots of the detectors acceptance, in this 
$(E_{loss},t)$ plane. It is obvious from these graphs that the VFD acceptances mostly depend on 
$E_{loss}$, and have a very small sensitivity in $t$, within a large $t$ range. For VFD at 
$220$ m, an area forbidden by kinematics is visible, for low $|t|$ and high $E_{loss}$. 
Corresponding profiles at fixed virtuality are shown at $220$ m and $420$ m (Fig. \ref{RP_acceptance_1D}).

\subsection{Chromaticity grids}

\begin{figure}[!ht] 
\centering
\begin{tabular}{cc}
\includegraphics[width=0.45\textwidth]{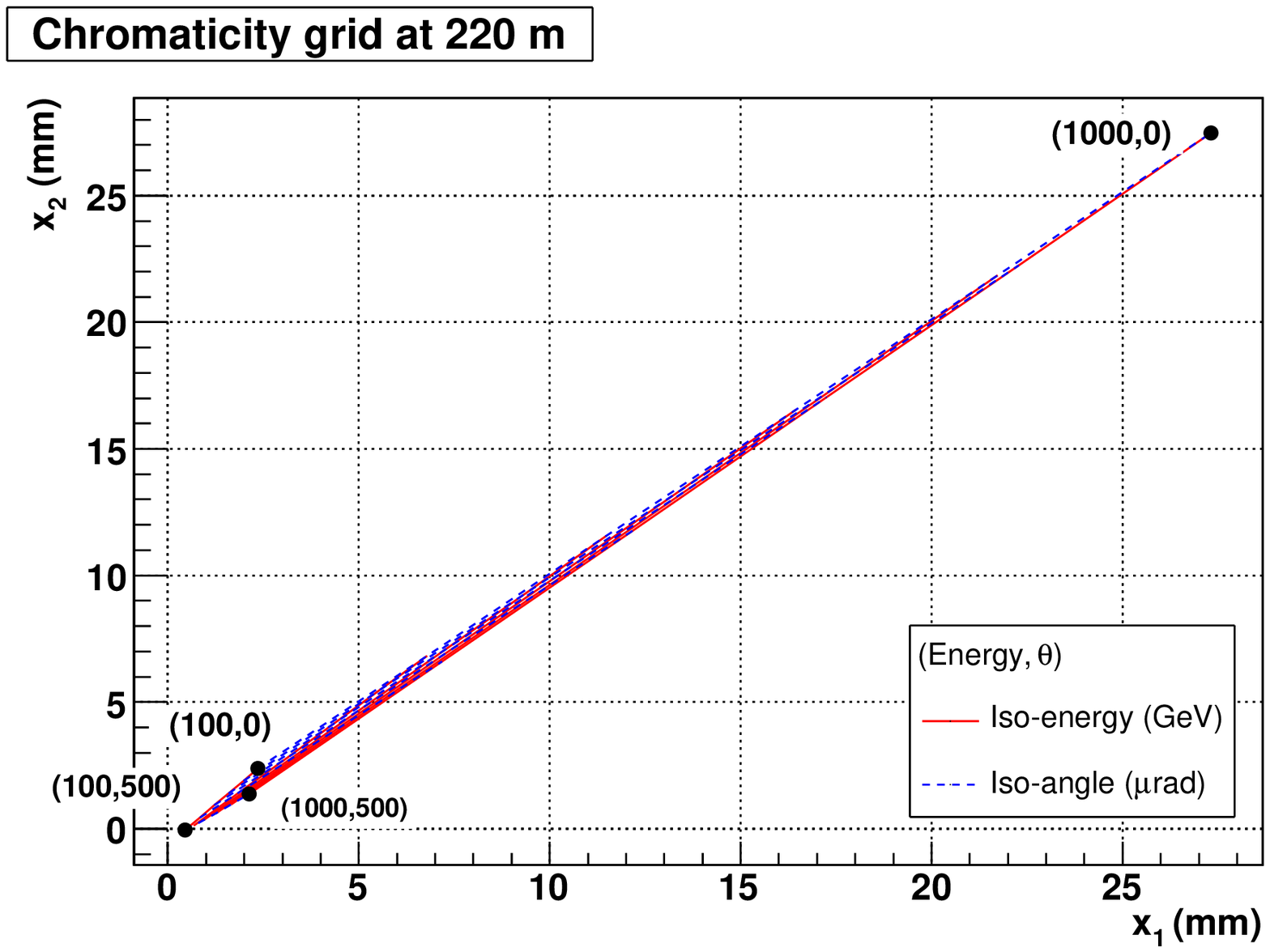}  & 
\includegraphics[width=0.45\textwidth]{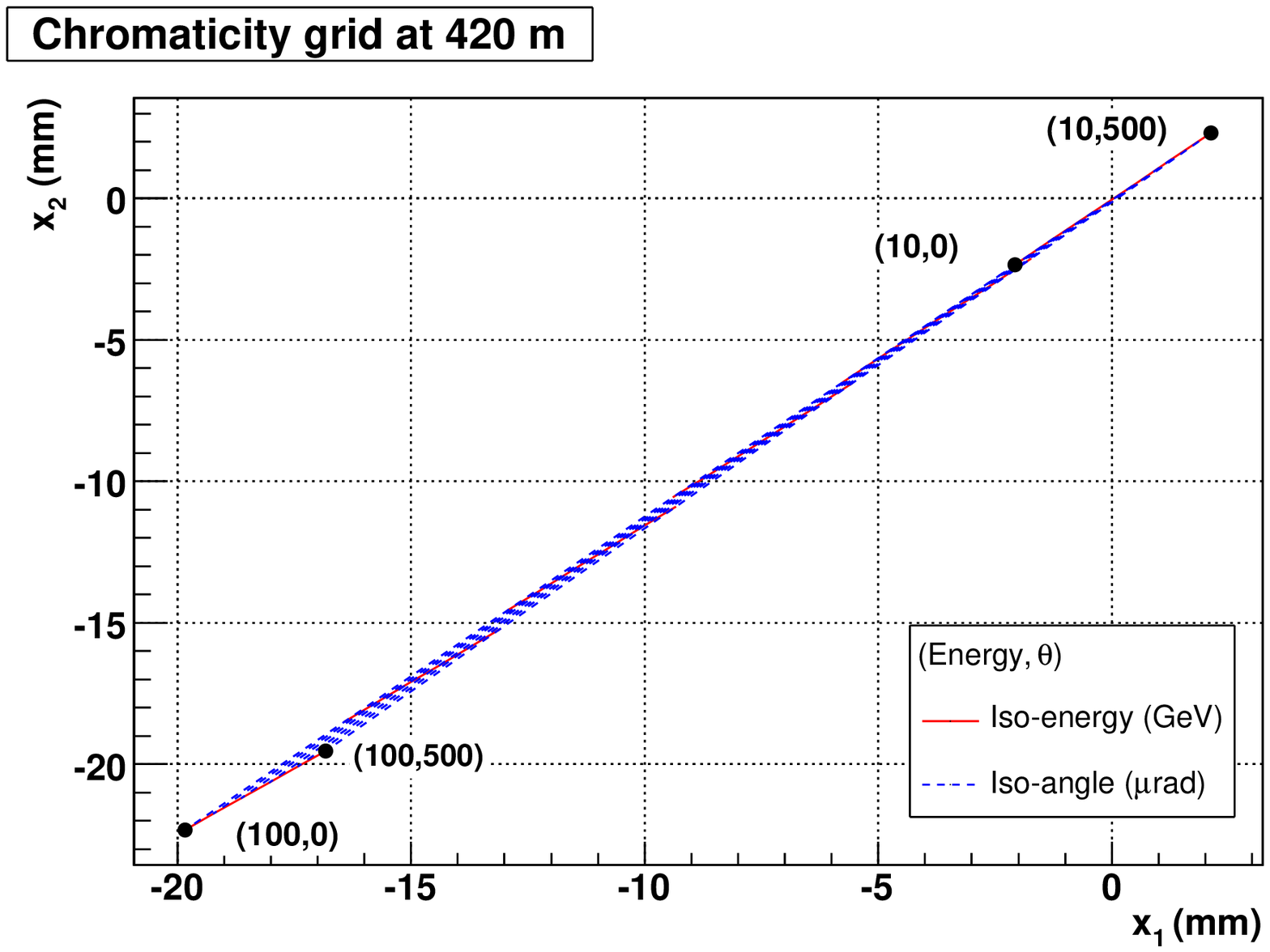}  \\
\includegraphics[width=0.45\textwidth]{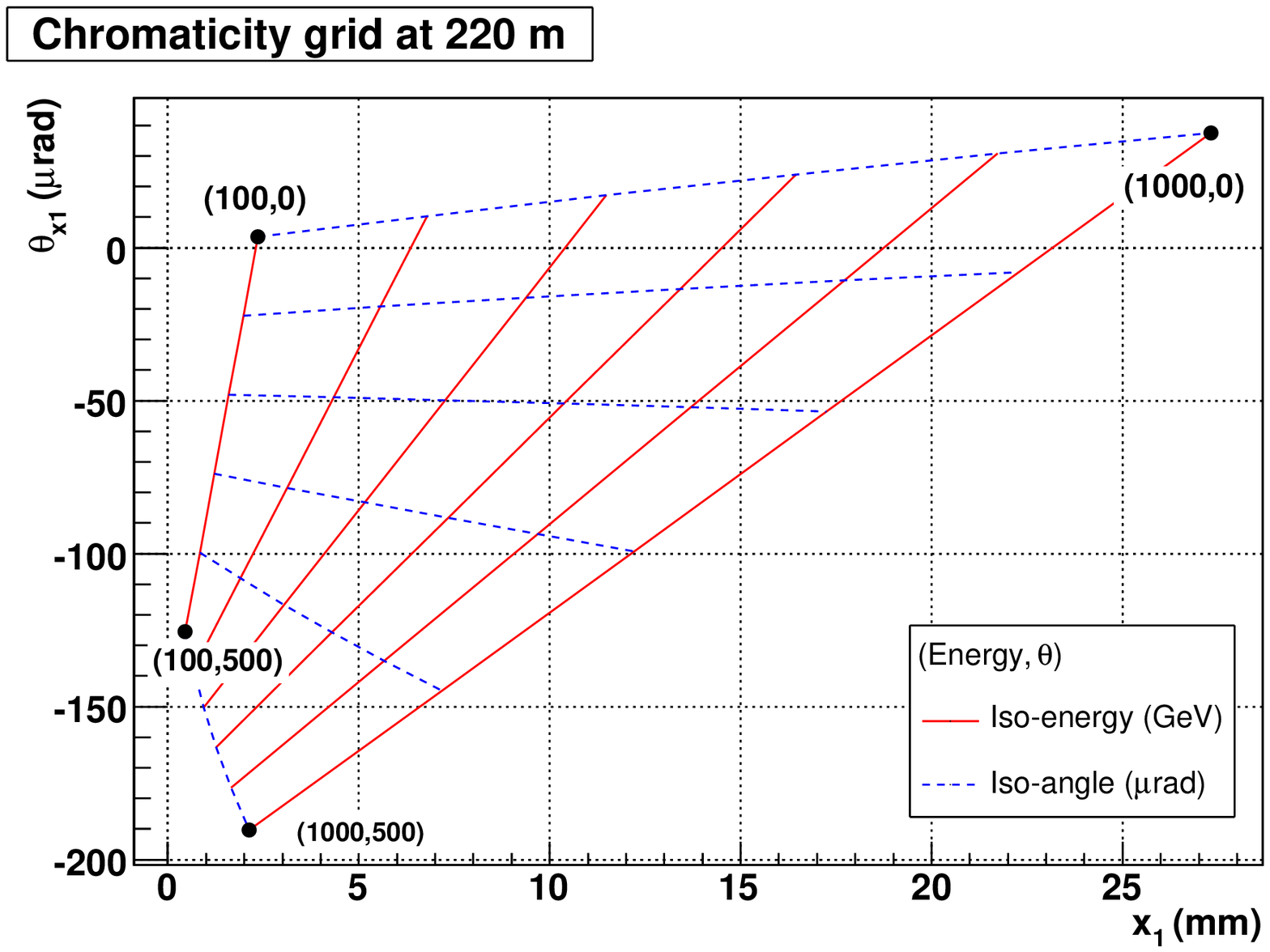}  & 
\includegraphics[width=0.45\textwidth]{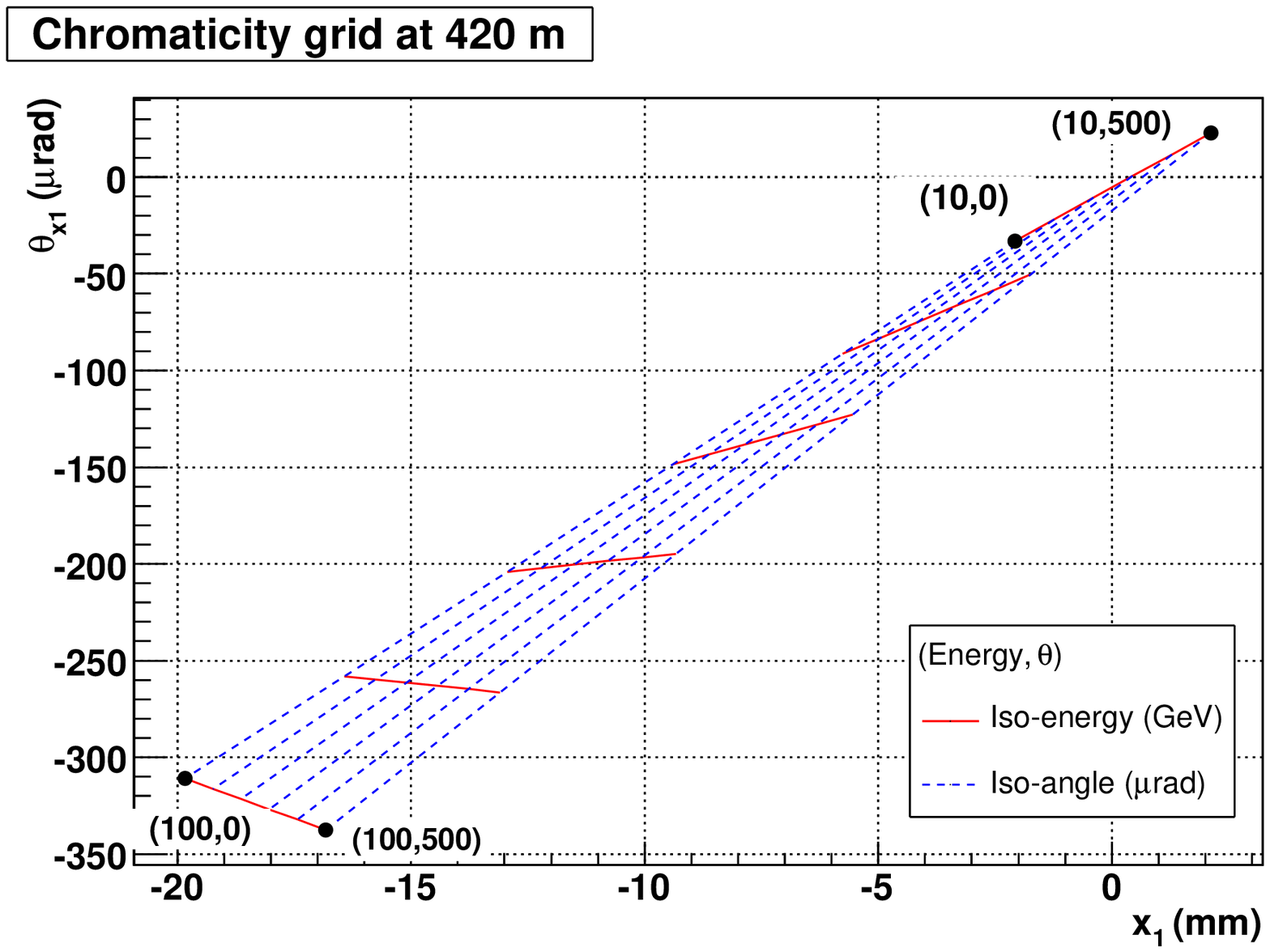}  \\
\end{tabular}
\caption[Chromaticity grids: Dependence on energy and virtuality loss at the IP]{Chromaticity grids: 
iso-energy and iso-angle lines for the VFDs at $220$ m (\emph{left}) and $420$ m (\emph{right}) away 
from the IP5, for the LHC beam 1. The graphs show the positions $x_1$ and $x_2$ of protons,  
given the energy loss $[0 ; 1000]$ GeV and $[0 ; 100]$ GeV, respectively, and the angular kick 
$[0 ; 500] ~ \mu$rad. The energy dependence of the transfer matrices induces a deformation of the grid, 
worsening the reconstruction power at higher angles. At $420$ m, the grid completely squeezes to a 
single line. The grids unfold themselves better in the $x$ - $x'$ plane.}
\label{Chromaticity_plots_x}
\end{figure}

\begin{figure}[!ht] 
\centering
\begin{tabular}{cc}
\includegraphics[width=0.45\textwidth]{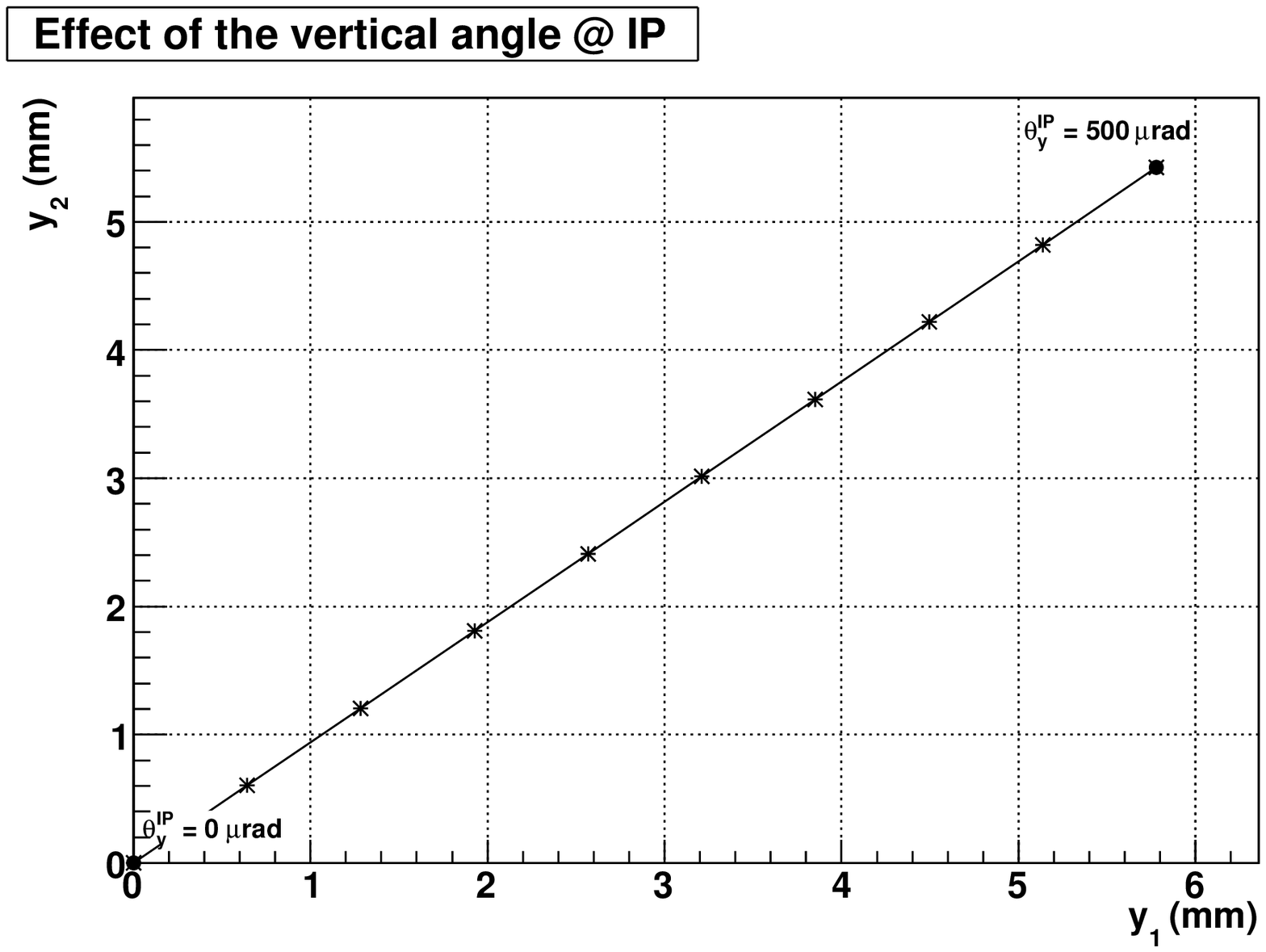}  & 
\includegraphics[width=0.45\textwidth]{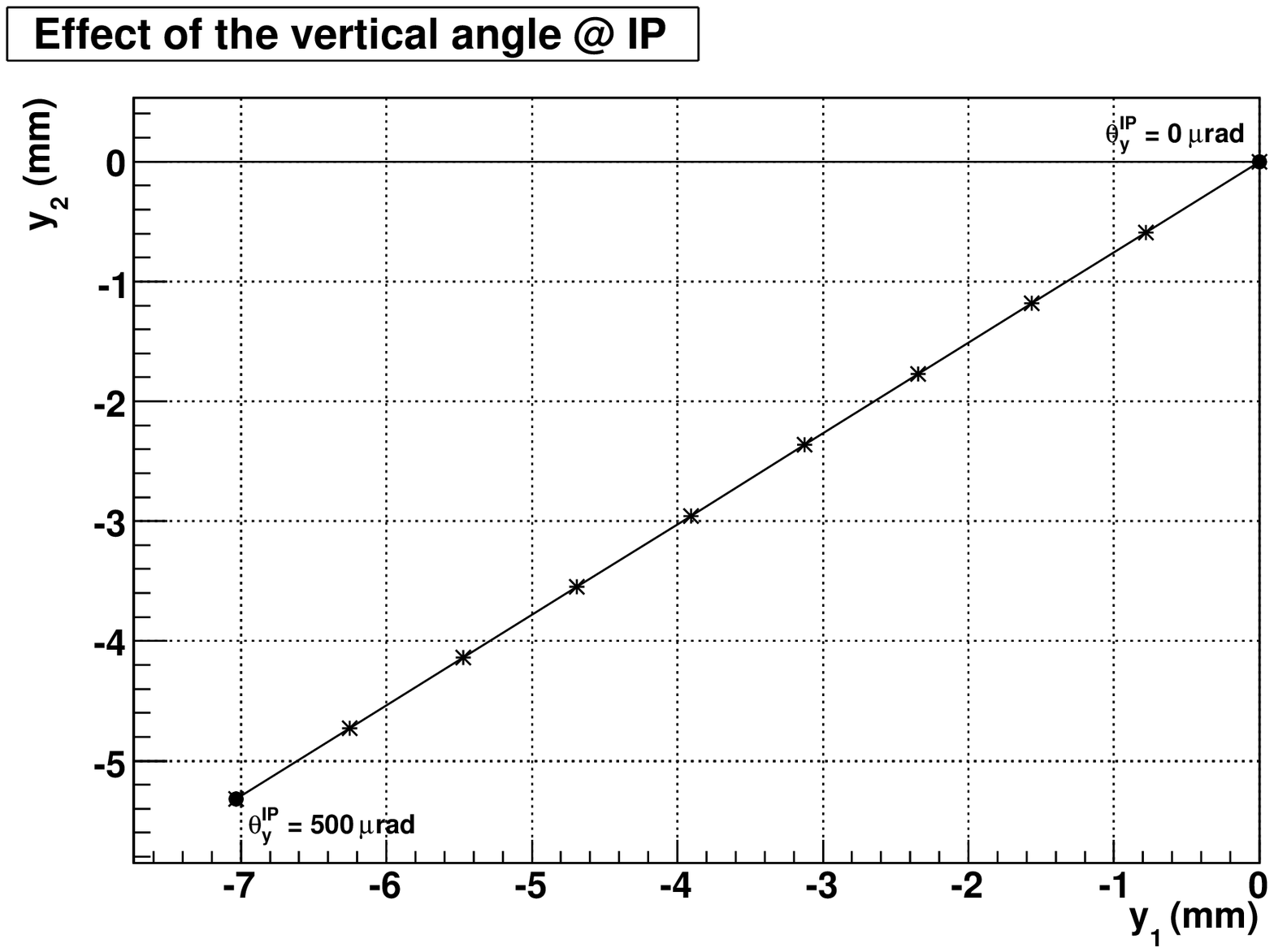}  \\
\end{tabular}
\caption[Dependence on vertical angle at IP]{Shift in vertical coordinate ($y_1$ and $y_2$) for VFDs at 
$220$ (left) and $420$ m (right) from the IP5, due to the vertical angle $\theta_y^*$ at the IP.}
\label{Chromaticity_plots_y}
\end{figure}

Once the acceptance windows of very forward tracking detectors are defined, it is interesting to see 
matching between the proton variables at the IP and those measured by VFDs. Depending on their energy 
and angle at the IP, forward protons will hit the VFDs at various positions. Drawing iso-energy and 
iso-angle curves for a set of sample protons produces a grid in the measurement related variables, 
$(x_1 , x_2)$ or $(x_1, \theta)$. Due to optics of the LHC beamlines, the grid unfolds itself in a 
much clearer way in the latter plane, and is almost invisible in the former one. The energy dependence 
of the transfer matrices implies a deformation of the grid -- without such a dependence, the grid 
would be a parallelogram. One should note, that uncertainty of the transverse position of the 
proton vertex at the IP results effectively in smearing the chromaticity grids. Anyway, these chromaticity 
grids provide a straightforward tool for unfolding the energy and angle at the IP of the measured 
particle. The grids in Fig \ref{Chromaticity_plots_x} were calculated in the energy range accessible 
to the VFDs. Finally, the correlation between the vertical angle at the IP and the vertical coordinate measurement 
is shown in Fig \ref{Chromaticity_plots_y}.
 
\subsection{Irradiation levels}
\begin{figure}[bp]                                                                                                                                      \centering
\includegraphics[width=0.8\textwidth]{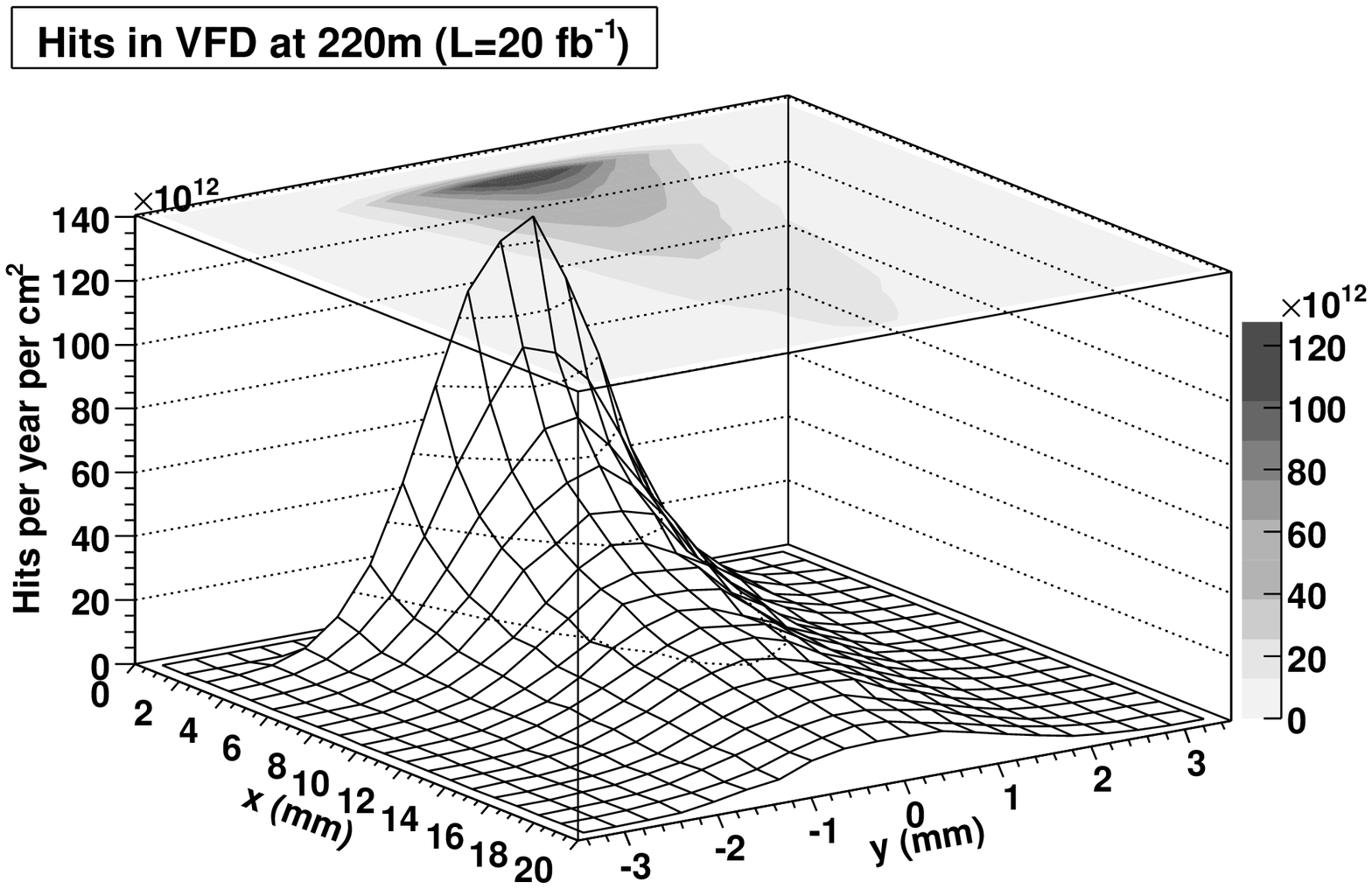}
\includegraphics[width=0.8\textwidth]{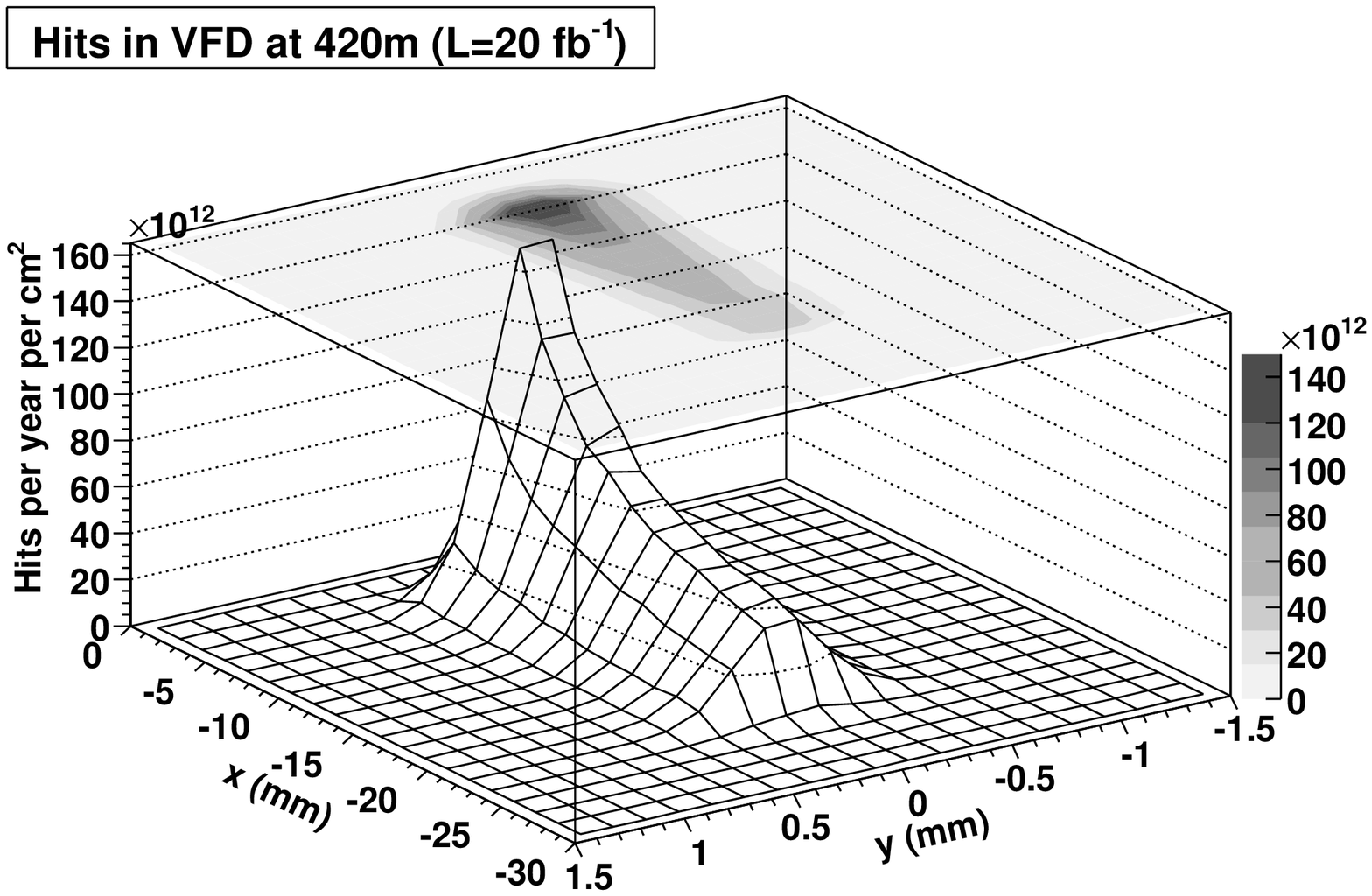}
\caption[Irradiation levels from $\mathrm{pp} \rightarrow \mathrm{pX}$ ]{Irradiation levels due to 
$\mathrm{pp} \rightarrow \mathrm{pX}$ processes, of VFDs located at $s=220 ~ \mathrm{m}$ (\emph{above}) 
and $s=420 ~ \mathrm{m}$ (\emph{below}) from the IP5. The horizontal position of the detector edge is 
respectively $x=2 ~ \mathrm{mm}$ and $x=4 ~ \mathrm{mm}$ from the nominal position of the center of 
the corresponding beam. The fluence is given per year and square centimeter. The integrated 
luminosity of $L = 20 ~\mathrm{fb}^{-1}$ was assumed.
For this analysis, \textsc{Hector} has propagated protons with 4-momentum generated by 
\textsc{Pythia} 6.2.10 (process number 93). The irradiation levels can locally exceed 
$1 \times 10^{14} $ protons per square centimeter.}
\label{irrad_level}
\end{figure}

The total diffractive cross-section at the LHC is very large, resulting in a high rate of 
diffractive protons hitting the VFDs. As a result, it causes extremely high irradiation levels, requiring
in turn usage of very radiation hard detectors. As an illustration, the number of 
hits and their respective map, due to the $\mathrm{pp} \rightarrow \mathrm{pX}$ processes has been 
investigated (Fig. \ref{irrad_level}). In this approach, we neglect other radiation sources, like 
beam halo or secondary particles coming from interactions in the beam pipe for instance.

The $\mathrm{pp} \rightarrow \mathrm{pX}$ processes are generated by \textsc{Pythia} \cite{PYTHIA} 
6.2.10 (single diffraction, process number 93). The number of hits are normalized by year by square centimeter, assuming 
an integrated luminosity of $L = 20~\mathrm{fb}^{-1} $. From the $10^6$ generated events, 
$17$ \% have been detected at $220 ~ \mathrm{m}$, and about $12$ \% at $420 ~ \mathrm{m}$. This 
simulation shows that the hottest spots in the detectors are localized with areas of only a few square
mm, reaching proton annual proton fluencies of more than $4\times 10^{14}$ protons/cm$^2$.

\clearpage

\subsection{Reconstruction}
One of the main physical motivations for \textsc{Hector} is the reconstruction of the event kinematics. 
For instance, if a beam particle has exchanged a photon at the IP, one could reconstruct photon's energy ($E$) 
and virtuality ($Q^2$). The particle energy at a given position in the beamline is obtained from the 
measured particle position and angle within the matrix formalism by solving these equations: 
$$ \bigg\{ \begin{array}{ll}
	x_s = a_s x_0 + b_s x'_0 + d_s E \\
	x'_s = \alpha_s x_0 + \beta_s x'_0 + \gamma_s E\\
  \end{array} 
$$

The transfer matrix of the beamline yields the coefficients $a$, $b$, etc. The introduction of an 
energy dependence on the strength of optical elements refines the transfer matrix, becoming a 
function of $E$: $a_s(E)$, $b_s(E)$,... This dependence will introduce \emph{non linearities}. 
Let us consider two reconstruction methods for the energy. In the first, so-called, \emph{trivial 
method} only the dispersion $D = d_s$ is considered.

\begin{figure}[!hbp]
\begin{tabular}{cc}
\includegraphics[width=0.49\textwidth]{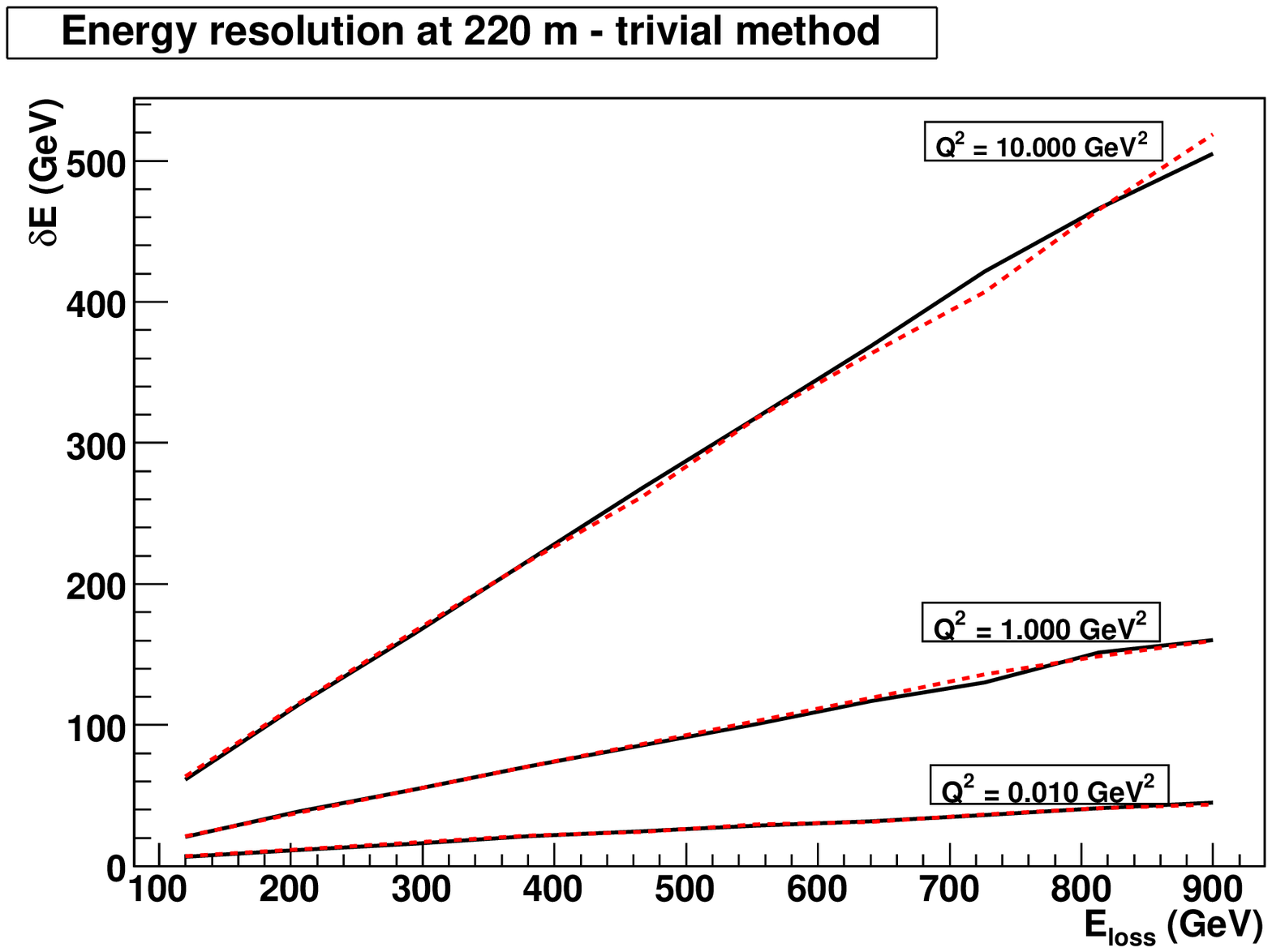} &
\includegraphics[width=0.49\textwidth]{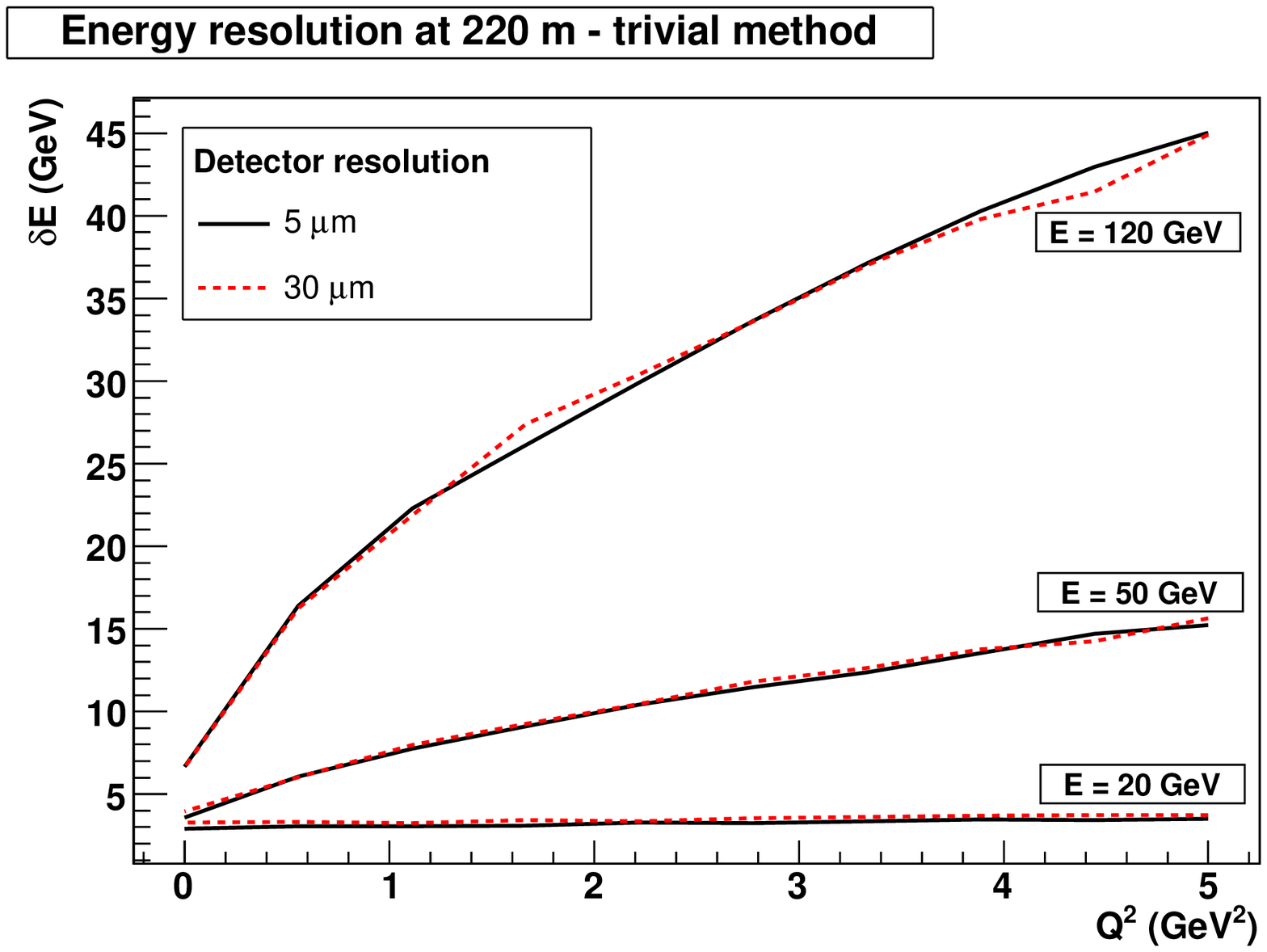} \\
\includegraphics[width=0.49\textwidth]{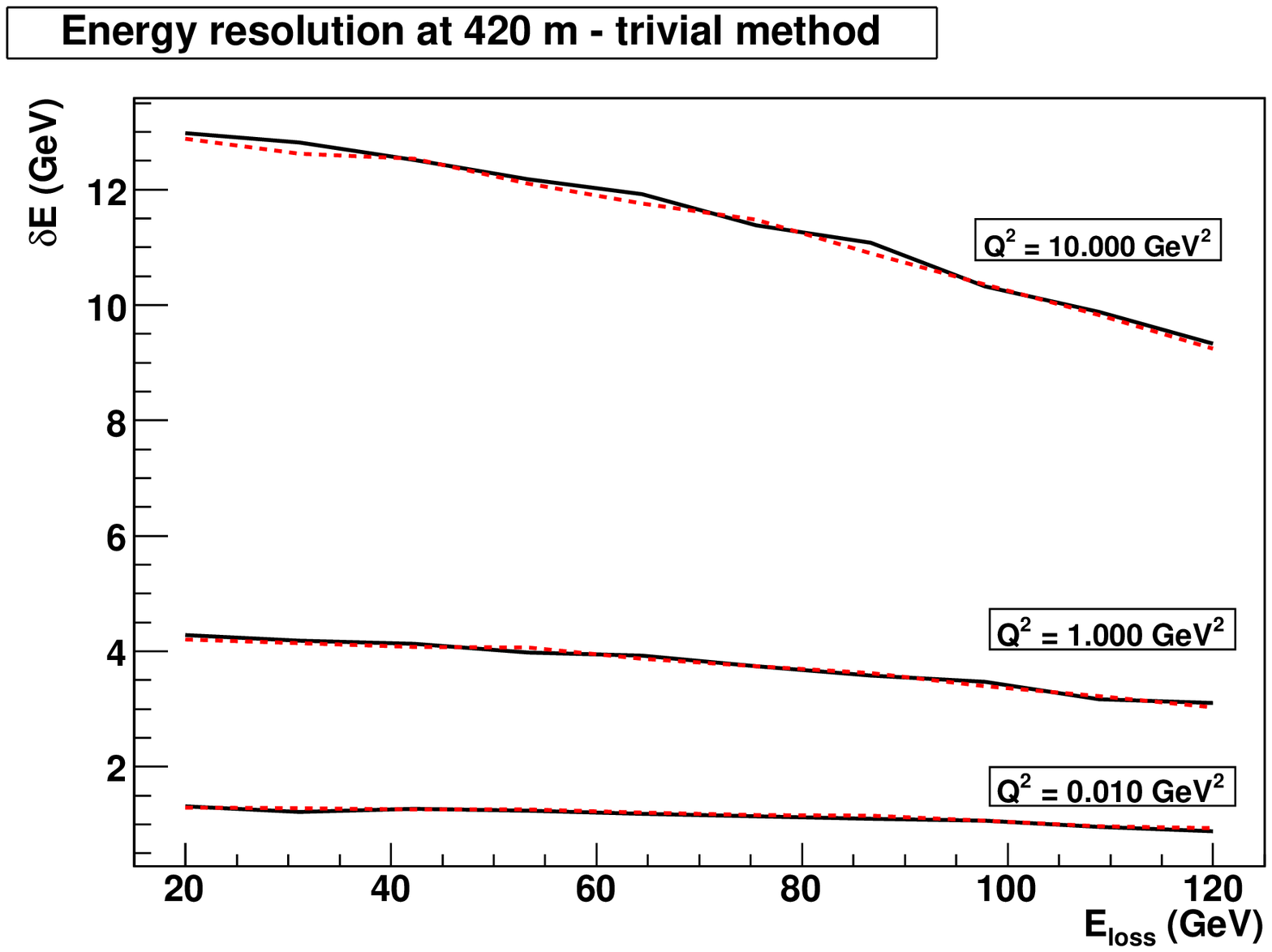} &
\includegraphics[width=0.49\textwidth]{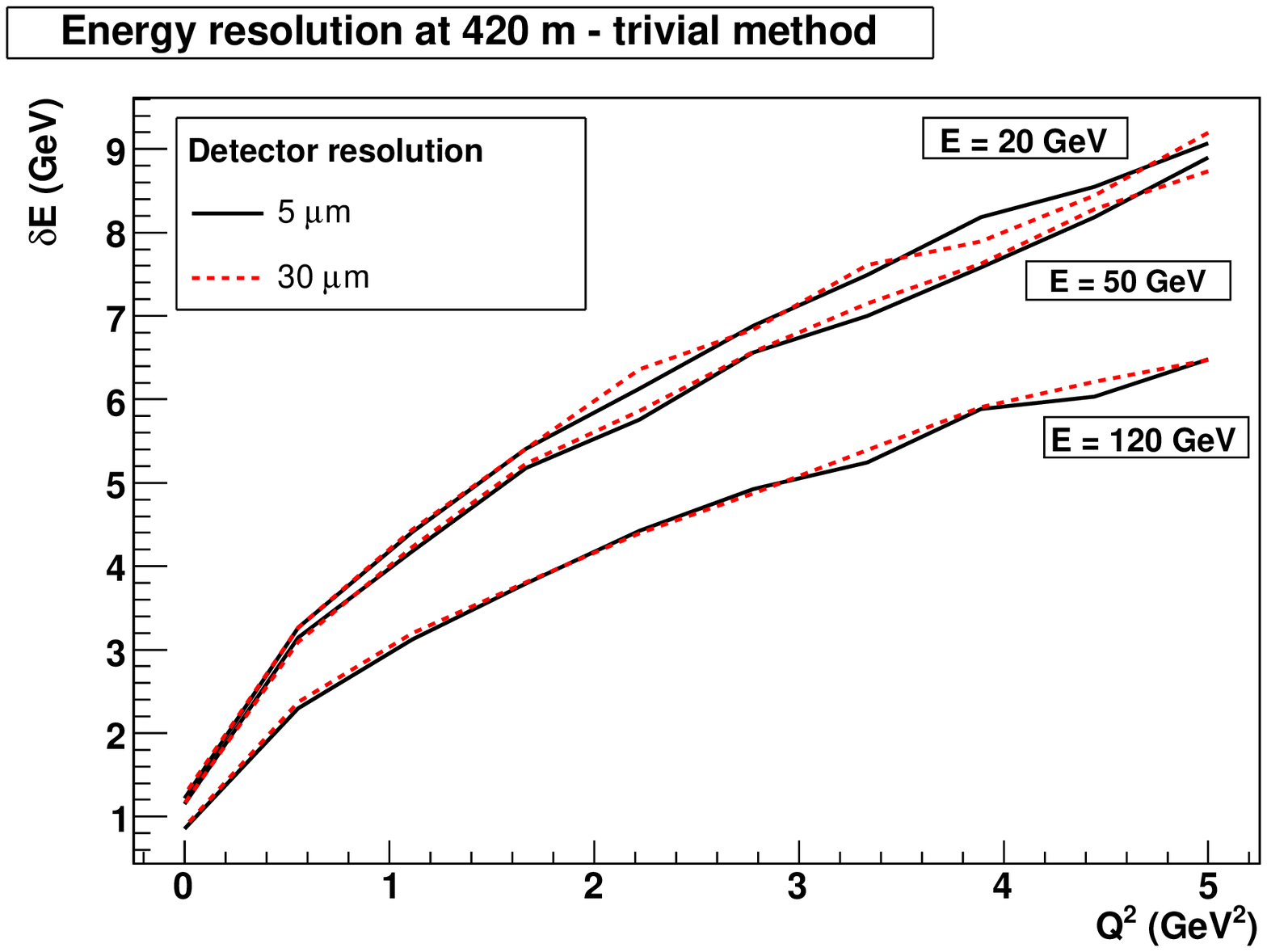} \\
\end{tabular}
\caption[Resolution of the reconstructed energy]{Energy resolution for the trivial reconstruction, 
for protons measured at $220$ m and $420$ m from the IP5. The protons have exchanged a particle 
taking away some energy, at a given virtuality. The resolution is shown as a function of the energy 
loss and the virtuality. The effects of some error on position measurement, due for instance to the spatial resolution of the detector stations 
($5 ~ \mu$m and $30 ~ \mu$m), are taken into account. The beam energy dispersion is not included.}
\label{reco_E_res}
\end{figure}
\begin{figure}[!hbp]
\centering
\begin{tabular}{cc}
\includegraphics[width=0.49\textwidth]{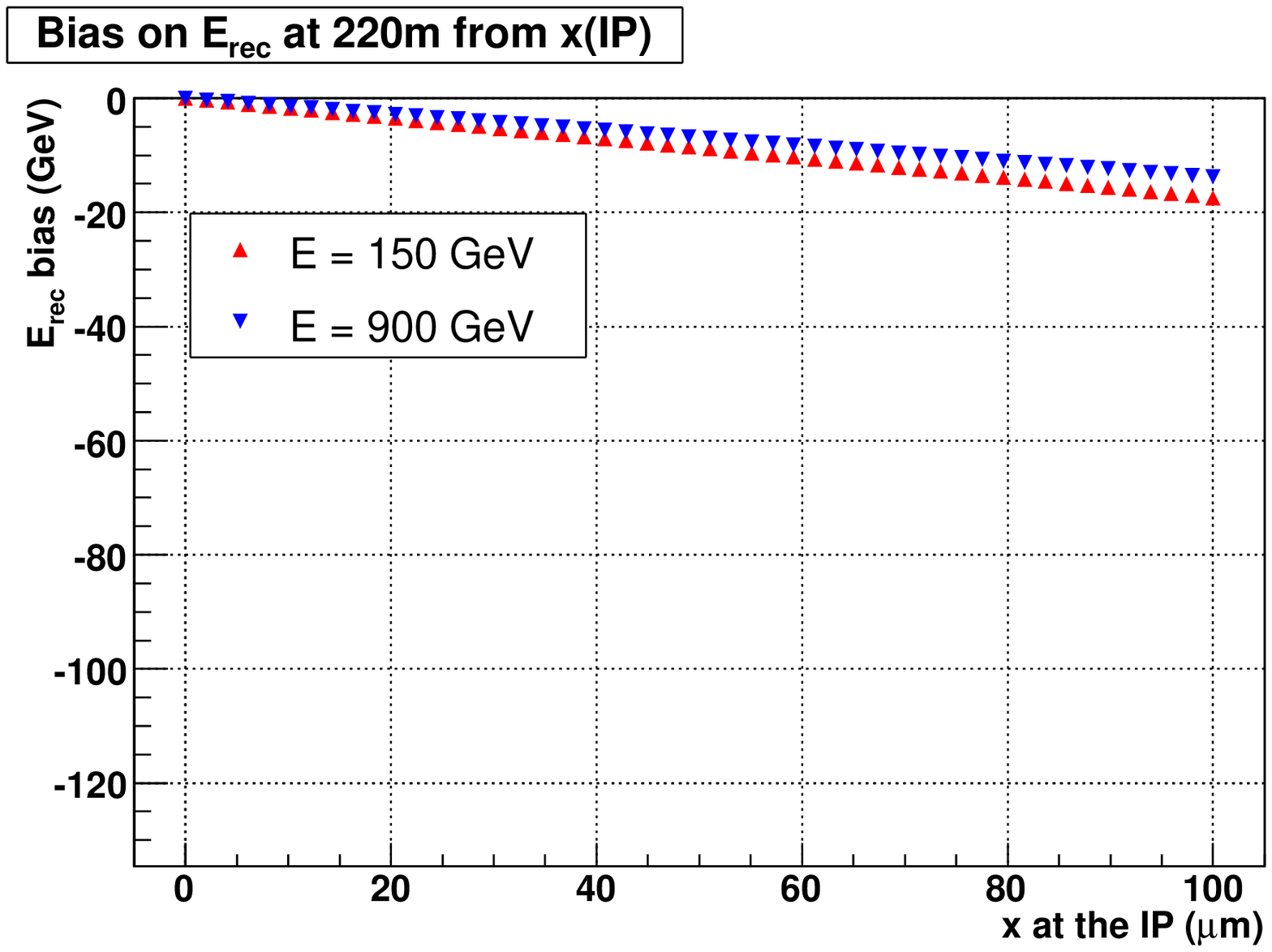} & 
\includegraphics[width=0.49\textwidth]{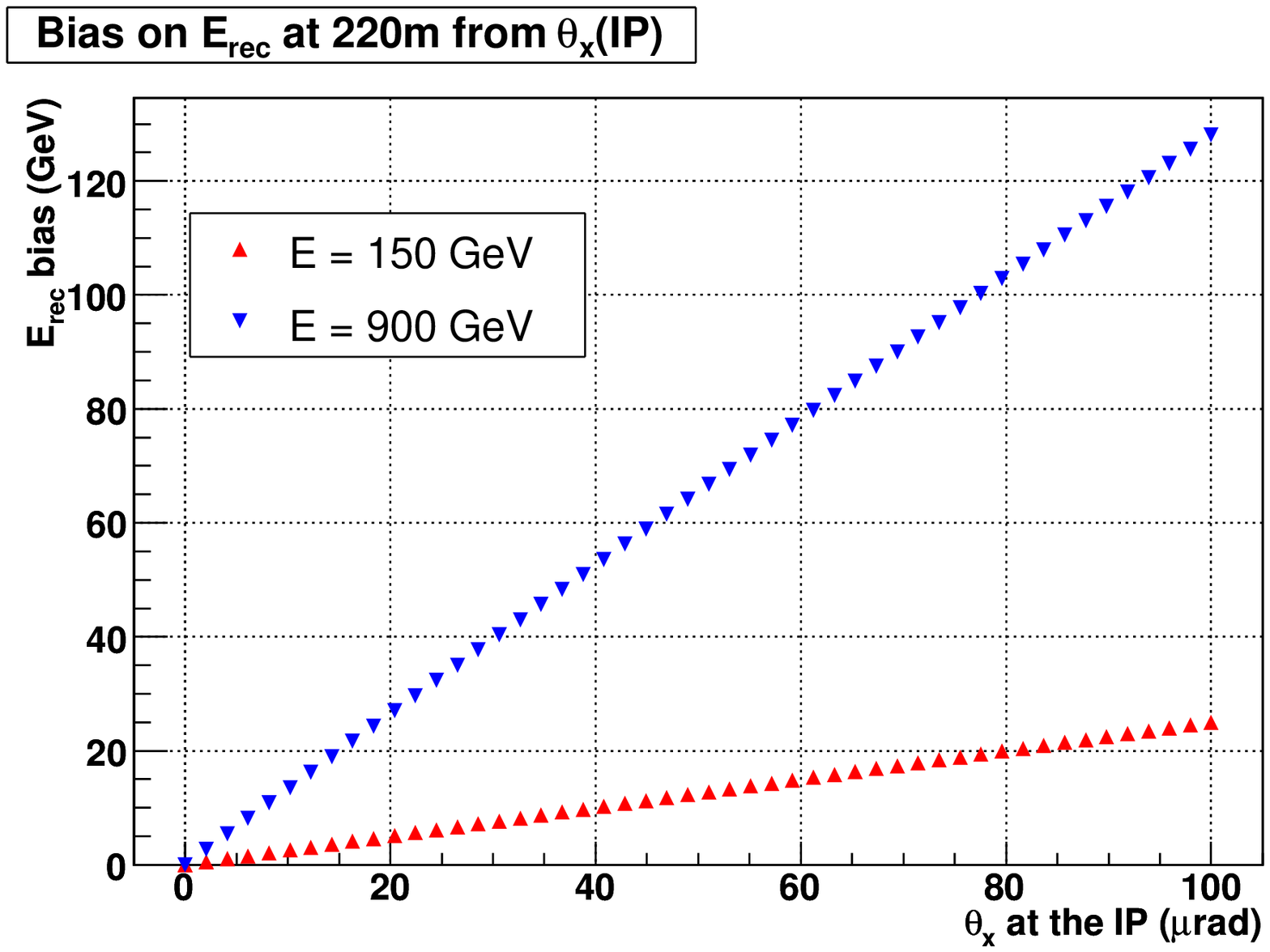} \\ 
\includegraphics[width=0.49\textwidth]{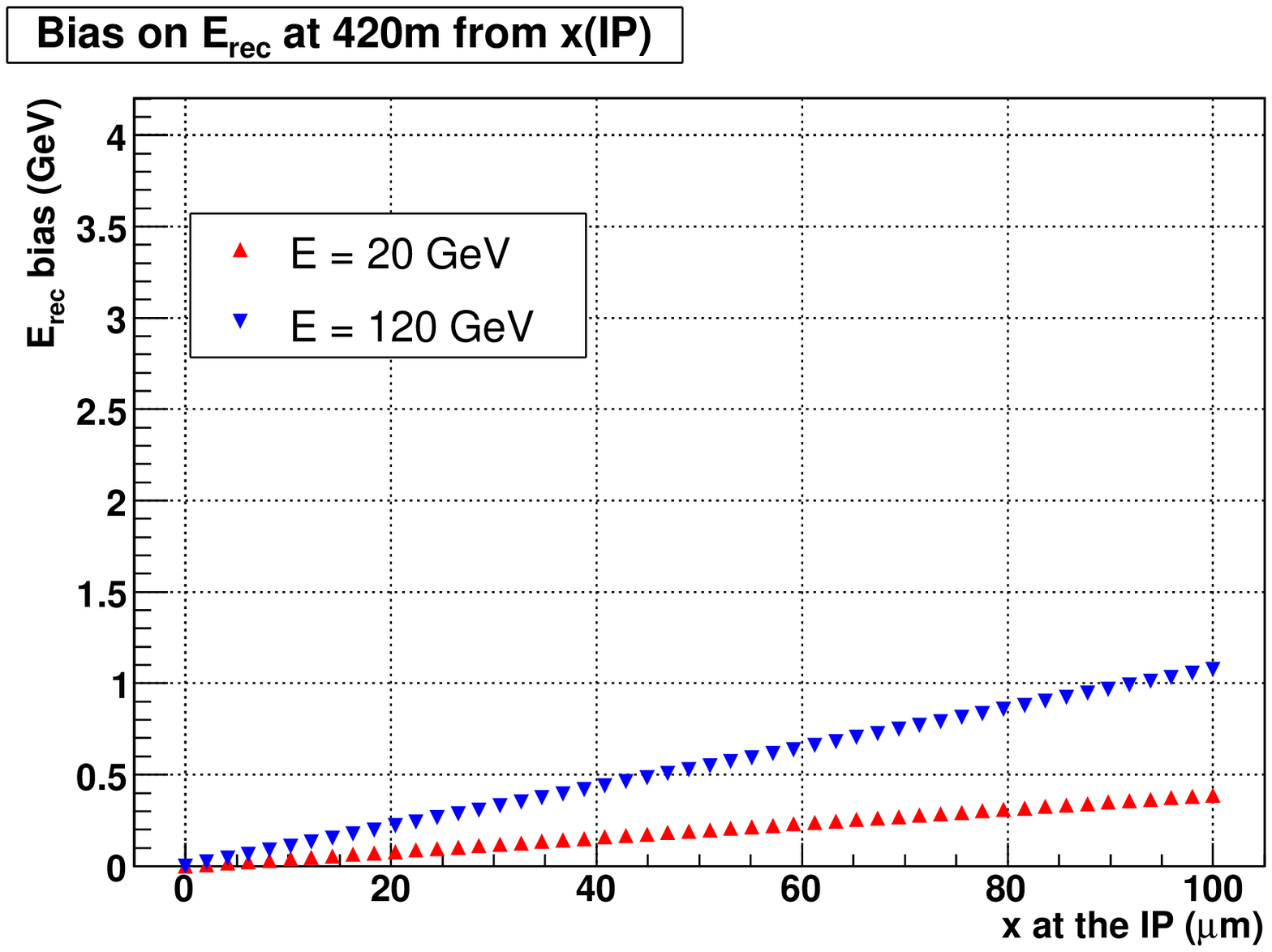} & 
\includegraphics[width=0.49\textwidth]{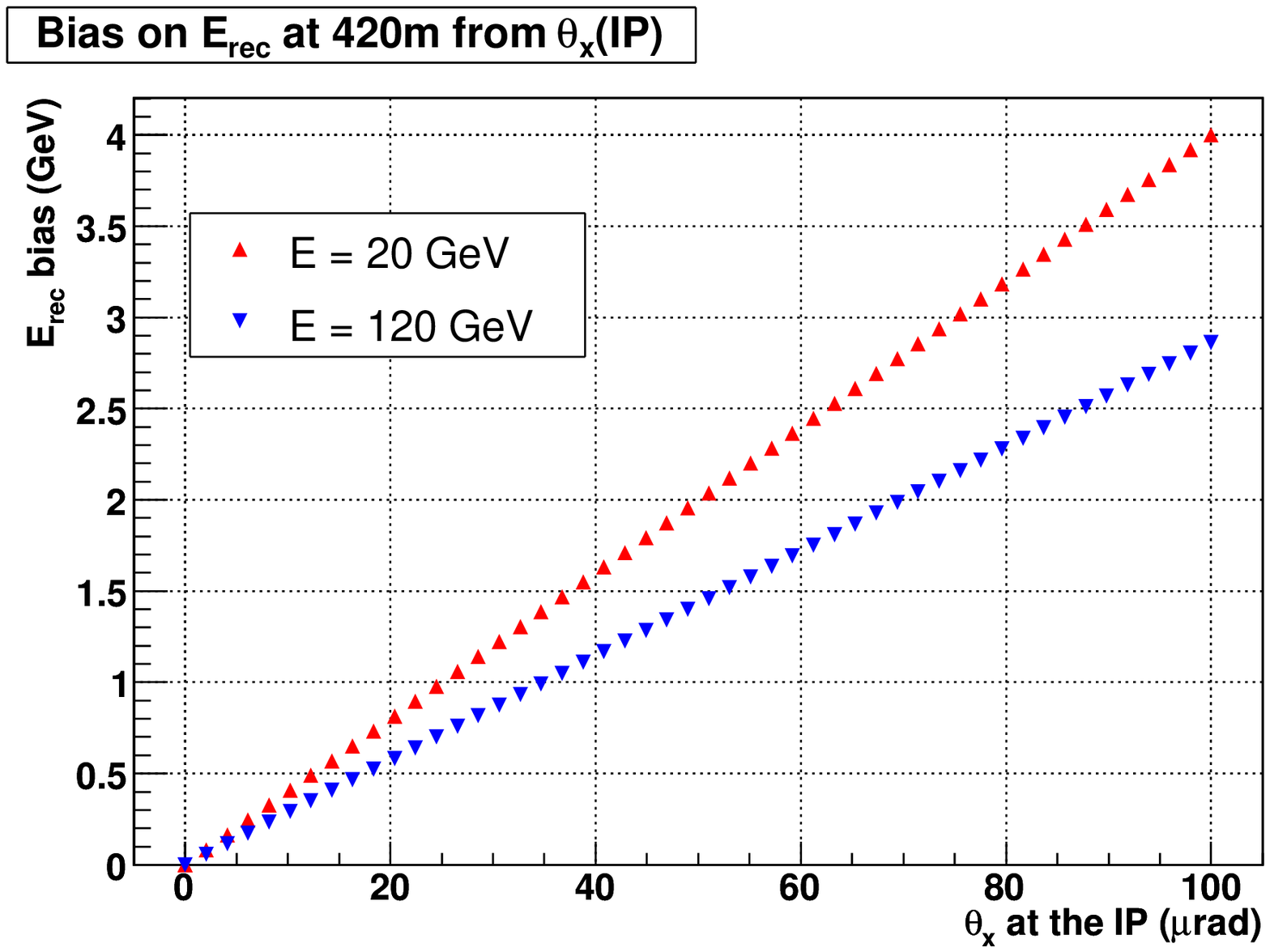} \\ 
\end{tabular}
\caption[Bias studies for the energy reconstruction]{Bias studies for the trivial energy reconstruction
for protons measured at $220$ m (\emph{above}) and $420$ m (\emph{below}). As the energy is computed 
using only the dispersion term of the global transport matrix, any non nominal values of the event 
vertex $x^*$ and the beam tilt $\theta_x^*$ at the IP, results in a biased energy reconstruction.}
\label{bias}
\end{figure}

The calculation of the particle energy using a beam transfer matrix corresponding to the nominal 
$7$ TeV energy (i.e. assuming constant $D$) obviously induces some errors. Therefore, in order to 
compensate for the energy dependence of $D$, a polynomial fit to the average (horizontal) position 
at a given detector location was done as a function of the particle energy. Then, for a fixed 
four-momentum transfer (or virtuality $Q^2$), the energy resolution $\delta E$ is almost independent 
of energy itself (see Fig. \ref{reco_E_res}), apart from the large $Q^2$ values for the detector at 
220 m. At the zero-degree angle the energy resolution is about 3 and 1 GeV at 220 m and 420 m, 
respectively, but the higher is the scattering angle (or the virtuality), the worse is the energy 
resolution, as shows Fig. \ref{reco_E_res}. 
So, if no correction is made due to non-zero scattering angle, the average energy resolution for a 
typical $Q^2$ of 0.3 GeV$^2$ is worse, about 5 and 2 GeV at 220 m and 420 m, respectively.
The trivial method is not much sensitive on the position resolution of VFDs - only if resolutions are 
close to the horizontal beam size at a given location (70 and 250 $\mu$m, respectively), they would 
change significantly the energy resolution. However, in this method neglecting the terms 
$a_s x_0$ and $b_s x'_0$ leads to significant sensitivity to the non-nominal values of the average 
vertex position and beam direction ($tilt$) at the IP, as shown in Fig. \ref{bias}. While the average 
vertex position can be very well measured using the central detectors, the beam tilt is more 
difficult to control -- it should be known to better than 10--20 $\mu$rad to avoid causing a 
significant bias. 

\begin{figure}[!hbp]
\centering
\begin{tabular}{c}  
\includegraphics[width=0.80\textwidth]{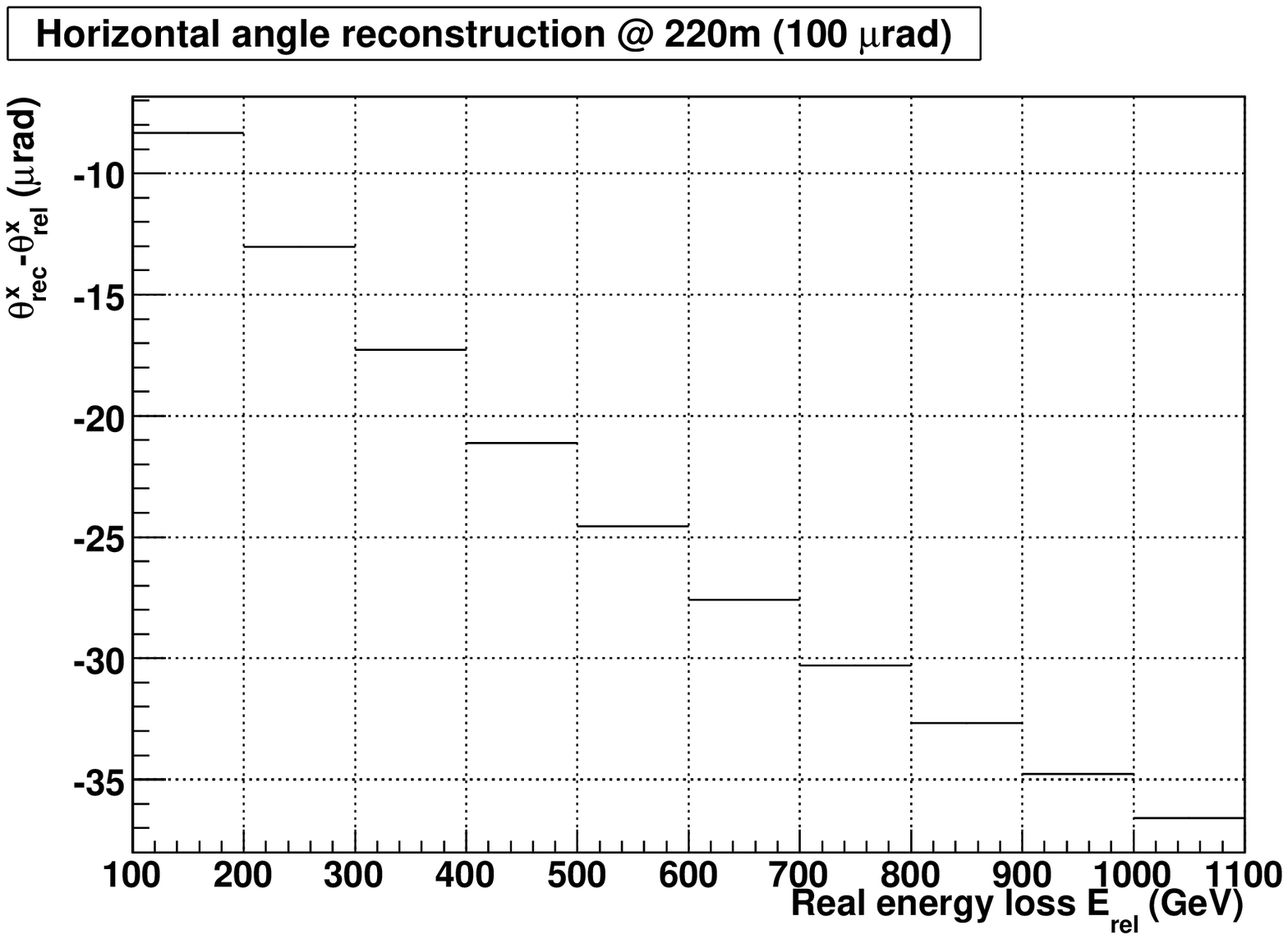}  \\
\includegraphics[width=0.80\textwidth]{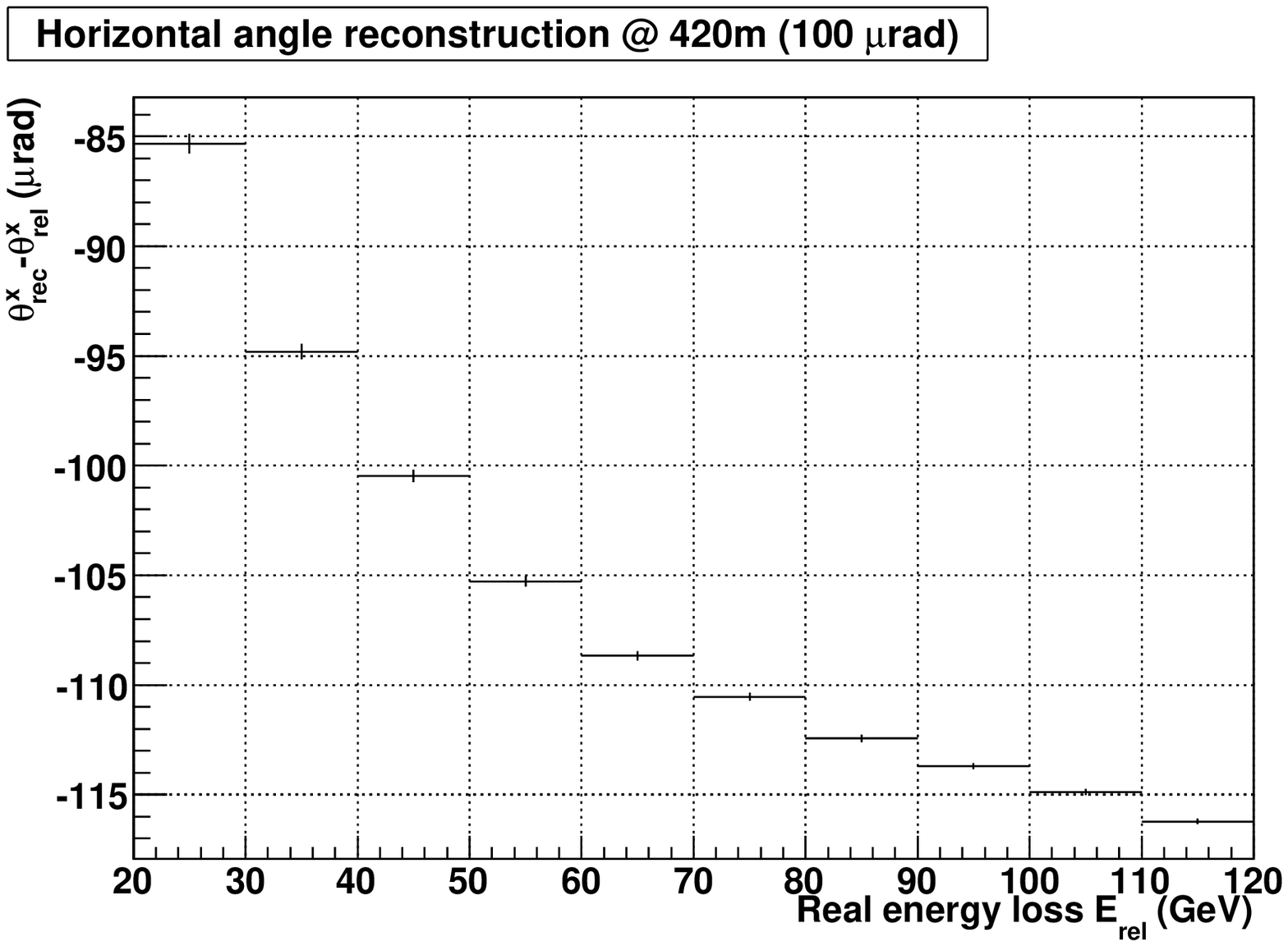}  
\end{tabular}
\caption[Reconstruction of the hor angle at IP]{Resolution of the reconstruction of the 
horizontal ($\theta_x$) scattering angles, for a 100 $\mu$rad scattering angle  
as a function of proton energy loss, for VFDs at $220$ (\emph{upper plot}) and $420$ m 
(\emph{lower plot}) from IP5. The trivial energy reconstruction method is used,
and the beam divergence at the IP is not taken into account.}
\label{h_angle}
\end{figure}

Once the energy has been reconstructed (Fig. \ref{reco_E_res} and \ref{bias}), the scattering angles at 
the IP can subsequently be computed, after applying all the usual smearings on other variables. 
As it can be seen in Fig.  \ref{Chromaticity_plots_y}, the reconstruction of vertical angle is 
really good both at 220 and 420 m. Even for very modest vertical position resolutions of 50 $\mu$m, the
reconstruction resolutions are already limited only by the vertical beam divergence of about 30 $\mu$rad.
This reflects the fact that all dipole fields are horizontal around the IP5, and it is effectively a
simple `geometrical' measurement.
Unfortunately, as expected, the situation is much worse in the horizontal plane, where the errors and biases 
are large and increase with energy loss (see Fig \ref{h_angle}). However, in the case of the 220 m detectors it 
seems possible to correct for the bias and recover the angular resolution close to the ultimate 30 $\mu$rad. It is
much more difficult at 420 m from the IP - this means that $\theta_x$ at the IP cannot be well reconstructed at 
420 m with the trivial energy reconstruction method, and possibly could be only used to apply selection cuts.



In the second (\textit{advanced}) method one reconstructs the energy and scattering angles using position measurements
at two detector stations at the same time. This requires to solve the equations for $x_s$ at both
detectors for E. The $a_s$, $b_s$ and $d_s$ coefficients of the transfer matrix depend on 
energy with rather complicated shapes, as they are products of many magnets matrices. One efficient 
way to get those is then simply to fit each coefficient as a function of energy. Various fitting 
functions were tried, but a quadratic fit proved to be sufficient to avoid any visible bias or 
resolution degradation.

Using these fitted coefficients, one can easily get a formula for $x_0$ as a function of $x$ at both 
detectors and of the energy. For each pair of detector $x$ coordinates, the used method is to numerically 
find the root ($x_0$ = 0) of the formula to get the energy corresponding to a $x_0$ = 0, and thus 
neglect the interaction point transverse extension.

This method allows to get reconstructed energy independently of the angle of the particle at the IP. 
Energy reconstruction resolutions does not degrade anymore if the particle transverse momentum $p_T$ rises, 
but the price to pay is that the detectors resolutions become critical. In particular, it means that the
uncertainty of the reconstructed angle at a given detector location should be better than the beam
angular divergences there (6 and 1.5 $\mu$rad, respectively). The transverse momentum can thus also be 
computed from the matrix coefficients once the energy has been reconstructed.

\begin{figure}[!hbp]
\centering
\begin{tabular}{c}
\includegraphics[width=0.80\textwidth]{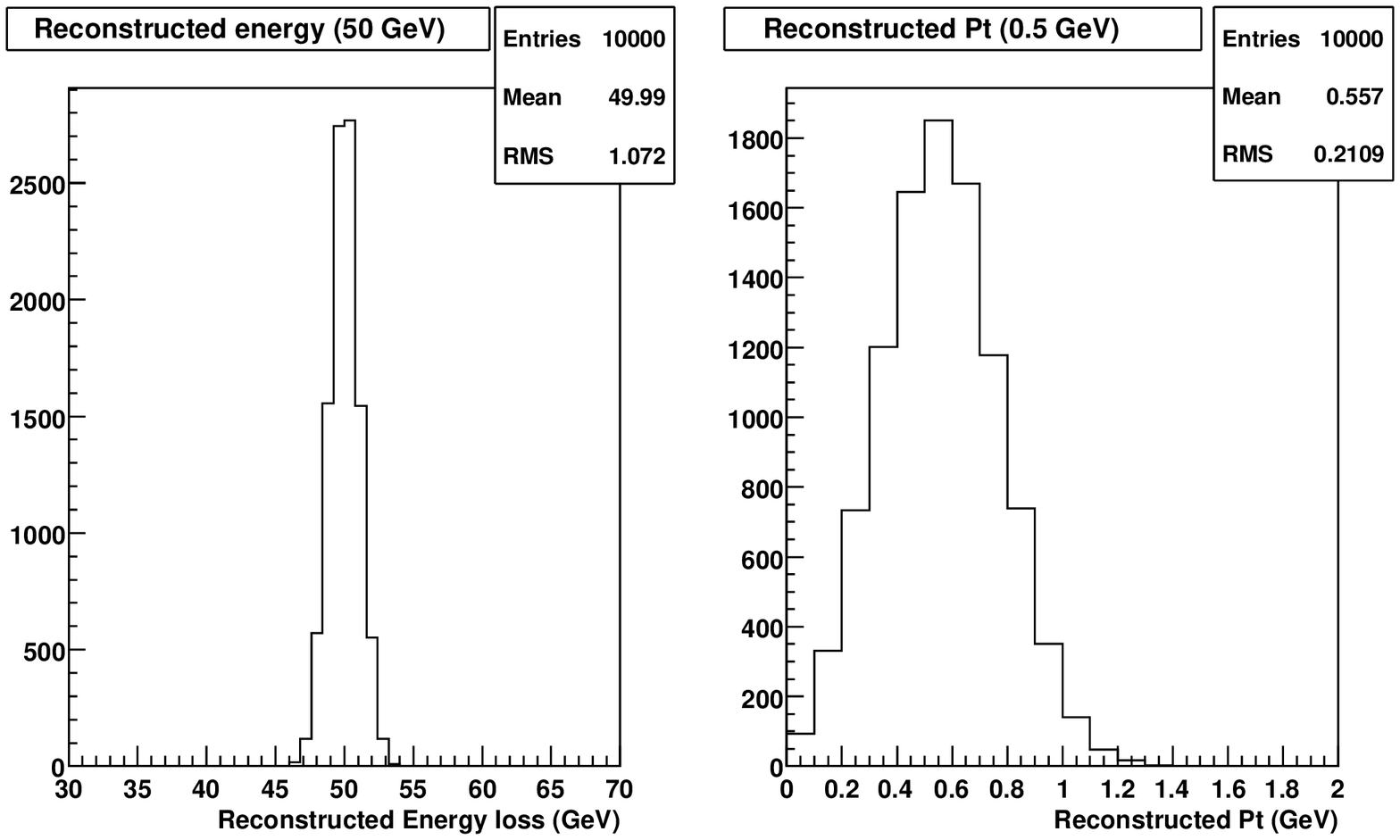}  \\
\includegraphics[width=0.80\textwidth]{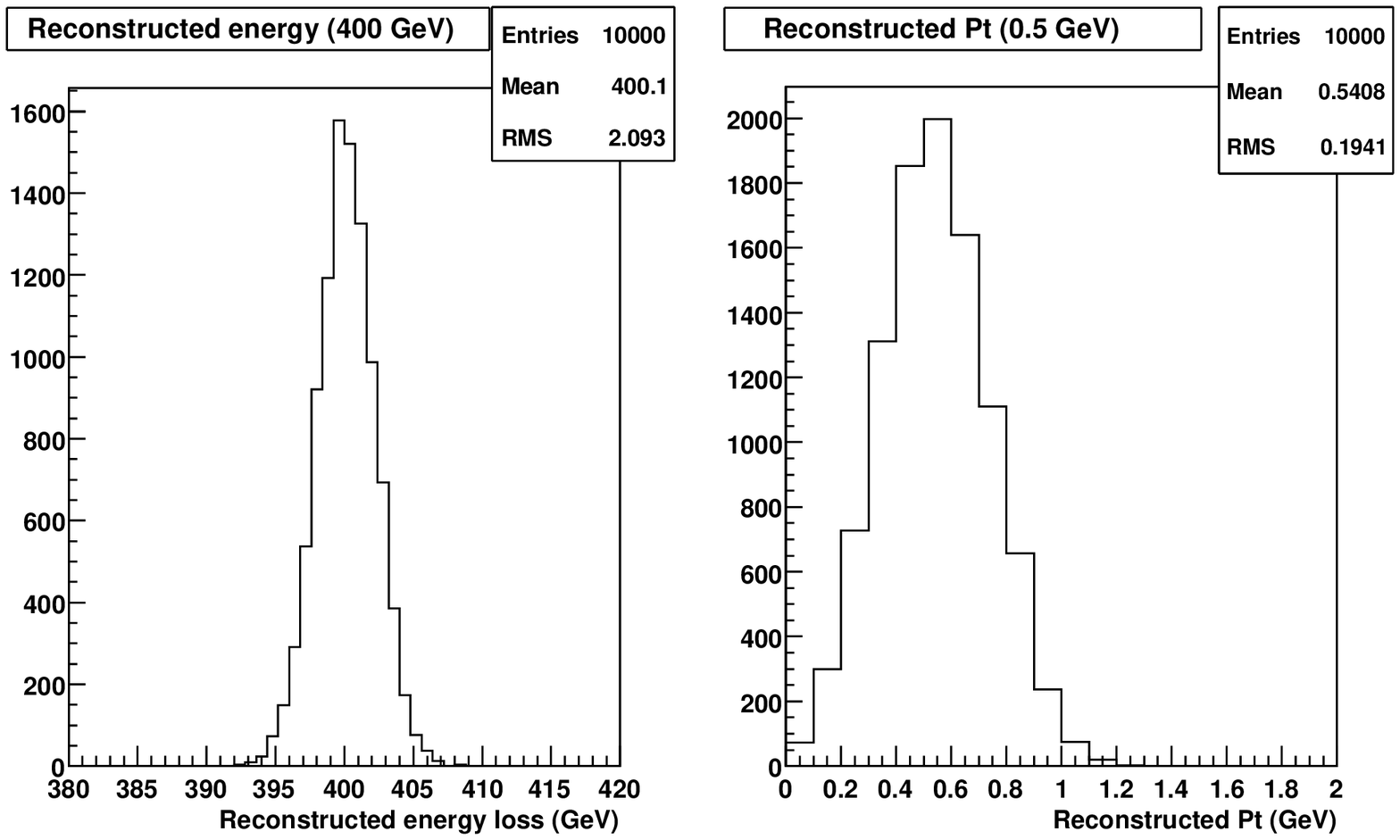}
\end{tabular}
\caption[Energy and $p_T$ reconstruction using the advanced method]{The so-called 
	\lq advanced method \rq allows to reconstruct at the same time the energy 
	and transverse momentum (or scattering angle). Here examples are given for detectors at $420$ m (\emph{upper plot}) 
	and $220$ m (\emph{lower plot}) from the IP, corresponding to energy loss of respectively $50$ and $400$ GeV, and transverse momentum 
	of $0.5$ GeV (uniformly distributed in azimuthal angle). Ideal detector is assumed.}
\label{amrecexample}
\end{figure}

\begin{figure}[!hbp]
\centering
\vspace{-1.5cm}
\includegraphics[width=0.90\textwidth]{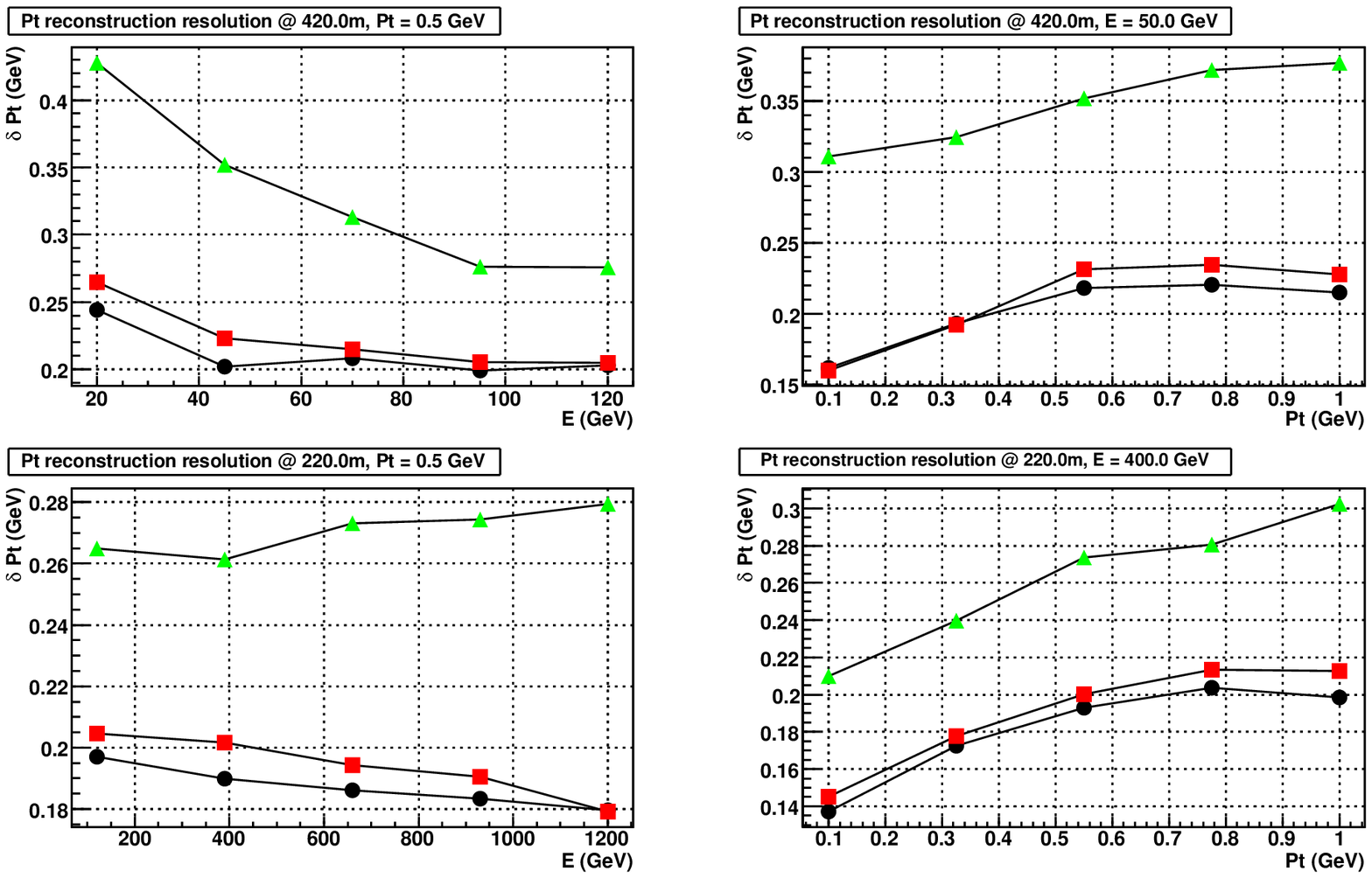}  
\caption[Reconstructed $p_T$ resolution using the advanced method]{Resolution of the 
	reconstruction of the particle transverse momentum $p_T$, as a function of the 
	energy loss (\emph{left}) and of the transverse momentum (\emph{right}), for 
	VFDs at $420$ (\emph{upper plot}) and $220$ m (\emph{lower plot}) from IP5.
	The advanced method, using full detector information, is used. Dots 
	correspond to different scenarios of detectors resolutions, namely perfect
	detectors (\emph{circles}), $5~\mu$m (\emph{squares}) and $30~\mu$m (\emph{triangles}) spatial 
	resolution.}
\label{amrecpt}

\includegraphics[width=0.90\textwidth]{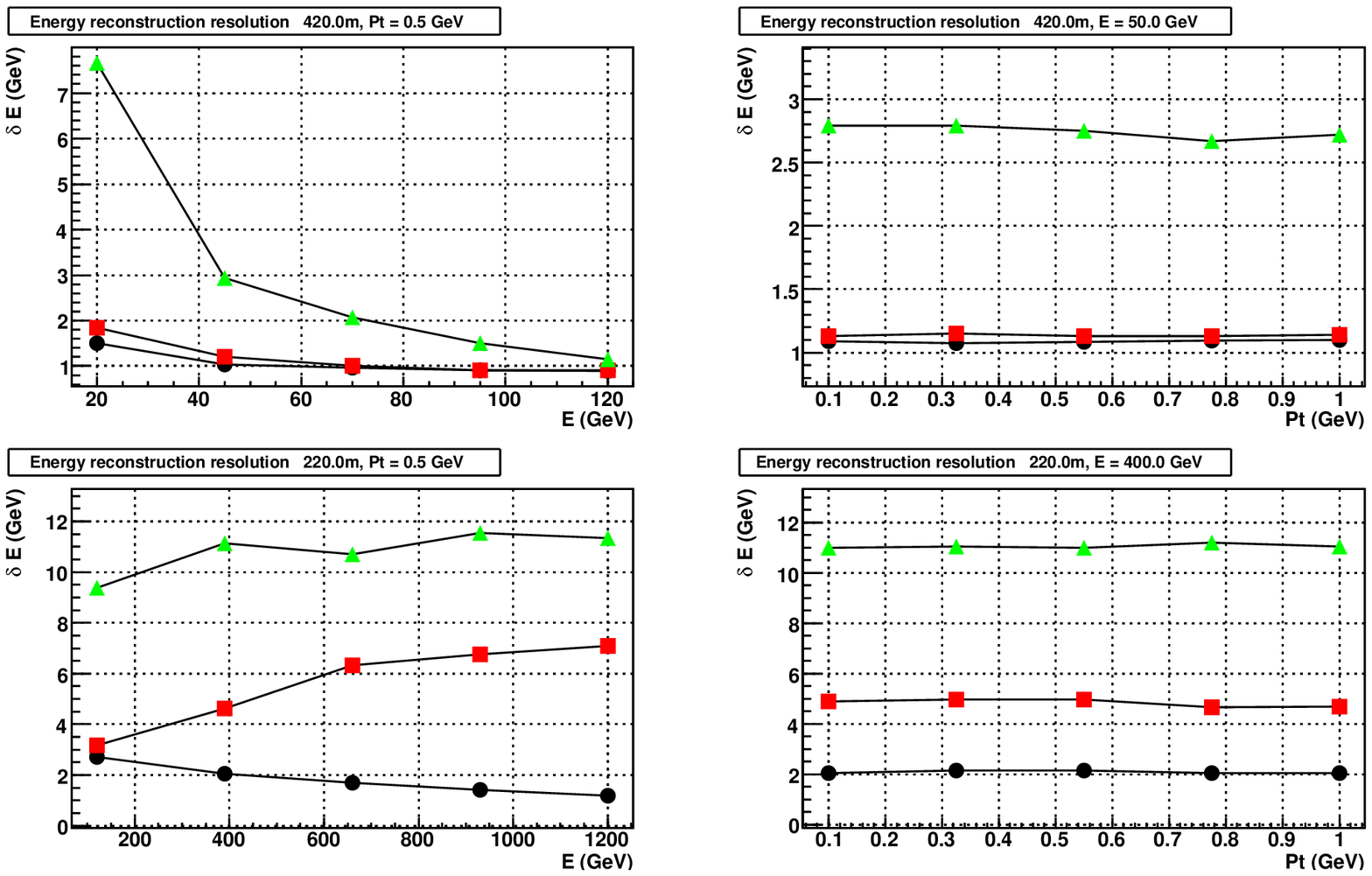}
\caption[Reconstructed Energy resolution using the advanced method]{Resolution 
	of the reconstruction of the particle energy loss $E$, as a function of the
    energy loss (\emph{left}) and of the transverse momentum (\emph{right}), for
    VFDs at $420$ (\emph{upper plot}) and $220$ m (\emph{lower plot}) from IP5.
    The advanced method, using full detector information, is used. Dots
    correspond to different scenarios of detector resolutions, namely perfect
    detectors (\emph{circles}), $5~\mu$m (\emph{squares}) and $30~\mu$m (\emph{triangles}) spatial
    resolution.}
\label{amrece}
\end{figure}

Distributions of reconstructed energy and transverse momentum for detectors at 220 m and 420 m are shown in Fig. 
\ref{amrecexample}. One can see that the reconstructed energy resolution stays very good even with
non-negligible initial particle $p_T$. Figures \ref{amrecpt} and \ref{amrece} show the $p_T$ and 
energy dependence of resolutions on the transverse momentum and energy, for various detectors resolutions. As expected, the
energy resolution is independent of $p_T$ but is sometimes sensitive to the energy as 
expected from the chromaticity grids of Fig. \ref{Chromaticity_plots_x}. The effect of the
detector resolution, which was absent for the simple reconstruction method described previously,
is now clearly visible even for an excellent resolution of 5~$\mu$m, especially for detectors 
at 220m from IP.


In summary, the proton angular distribution at the IP affects the most the energy 
reconstruction. In contrast, the vertex lateral distribution has negligible
impact. As a result, the most important beam parameter, which can change 
run-to-run, is the beam tilt at the IP. In principle, it can be indirectly controlled
by the beam position monitors at 220 and 420 m, but more direct tilt measurements
are favored, as by using BPMs next to the IP, or by monitoring the direction of
neutral particle (photons, neutrons, or neutral pions) production in the zero-degree 
detectors (ZDDs).

The energy resolution squared of the scattered proton $\sigma_E$ can be then
approximately decomposed into four terms:
$$ \sigma_E^2 = \sigma_0^2 + \sigma_{vtx}^2 + \sigma_{ang}^2 +  \sigma_{det}^2 , $$
where the nominal beam energy dispersion $\sigma_0\approx0.8$~GeV, the contribution due to the vertex spread 
$\sigma_{vtx}\approx0.7$~GeV at 420 m and 1--2~GeV at 220 m, the contribution due to the detector resolution, neglecting the angular effects, 
$\sigma_{det}^2$ is small for resolutions better than 50~$\mu$m, and the contribution due to the proton 
non-zero angle at the IP, $\sigma_{ang}$, which is very sensitive the angular reconstruction at the VFDs; 
even for a zero-degree scattering if one neglects effects due to the beam angular divergence at the IP,
$\sigma_{ang}\approx1$~GeV at 420 m and $\approx3$~GeV at 220 m.

\clearpage

\subsection{Calibration}
Usually, the calibration of the reconstructed variables at forward detectors can be well maintained using the physics processes and the central detectors. 
At HERA, for example, the elastic $\rho$ meson photoproduction was used where the momentum
of the scattered proton could be deduced from two decay charged pions using the central tracking. At
the LHC the two-photon exclusive production of dimuon pairs seems a good calibration process, at 
least for the detectors at $420 ~ \textrm{m}$. The visible cross section is large 
($\sigma_{\textrm{pp}} \leq 7 ~ \textrm{pb}$), including the acceptance of central detectors.
This should even allow for a run-by-run calibration of the scattered proton energy scale within a full acceptance
range. Finally, the expected reconstruction power of central detectors is excellent for such di-muon events,
giving the proton 
energy uncertainty about $10^{-4}$ per event \cite{opus}! One should note, however, that using this process 
it is not possible to check the angular reconstruction, and that it has much more limited 
statistics within acceptance of the detectors at 220 m. The high-energy bremsstrahlung $pp \rightarrow pp\gamma$ 
is a possible candidate for the energy scale calibration of the VFDs at 220 m. It has relatively large cross-section,
about 10 nb for $E_\gamma>$ 100 GeV, and photons are emitted into a very forward cone, so could be detected in the 
zero-degree detectors (ZDDs) at about 100 m from the IP, proposed for measurements of neutral particles produced at very small angles. 
A simultaneous measurement of the scattered proton in the VFD and of the bremsstrahlung photon in the ZDD would allow then for 
cross-calibration of these two devices. In addition, these photons are emitted at very small angles (typically, at
less than 150 $\mu$rad), so could be used to control the beam direction (tilt) at the IP. 

Finally, one should add that the angular measurements could be controlled by studying the reconstruction of the beam 
angular divergence at the IP, and comparing the obtained results with the direct measurements at the ZDDs, or with 
estimates obtained using the LHC instrumentation, or derived from the LHC luminosity.

\subsection{Misalignments}
The misalignment of LHC optical elements could have a significant impact on the measurements with very
forward detectors. As the deflection of the particle paths depends on their positions in quadrupoles, 
a misplacement of these optical elements implies a change in the nominal beam position.
In turn, as the accurate position measurement with the forward tracking detectors (as well as the 
information inferred from the segmentation of forward calorimeters) is referred to the ideal beam 
location, changing this reference results in a biased reconstruction of the measured particles.

\begin{figure}[!h]
\centering
\begin{tabular}{cc}
\includegraphics[width=0.5\textwidth]{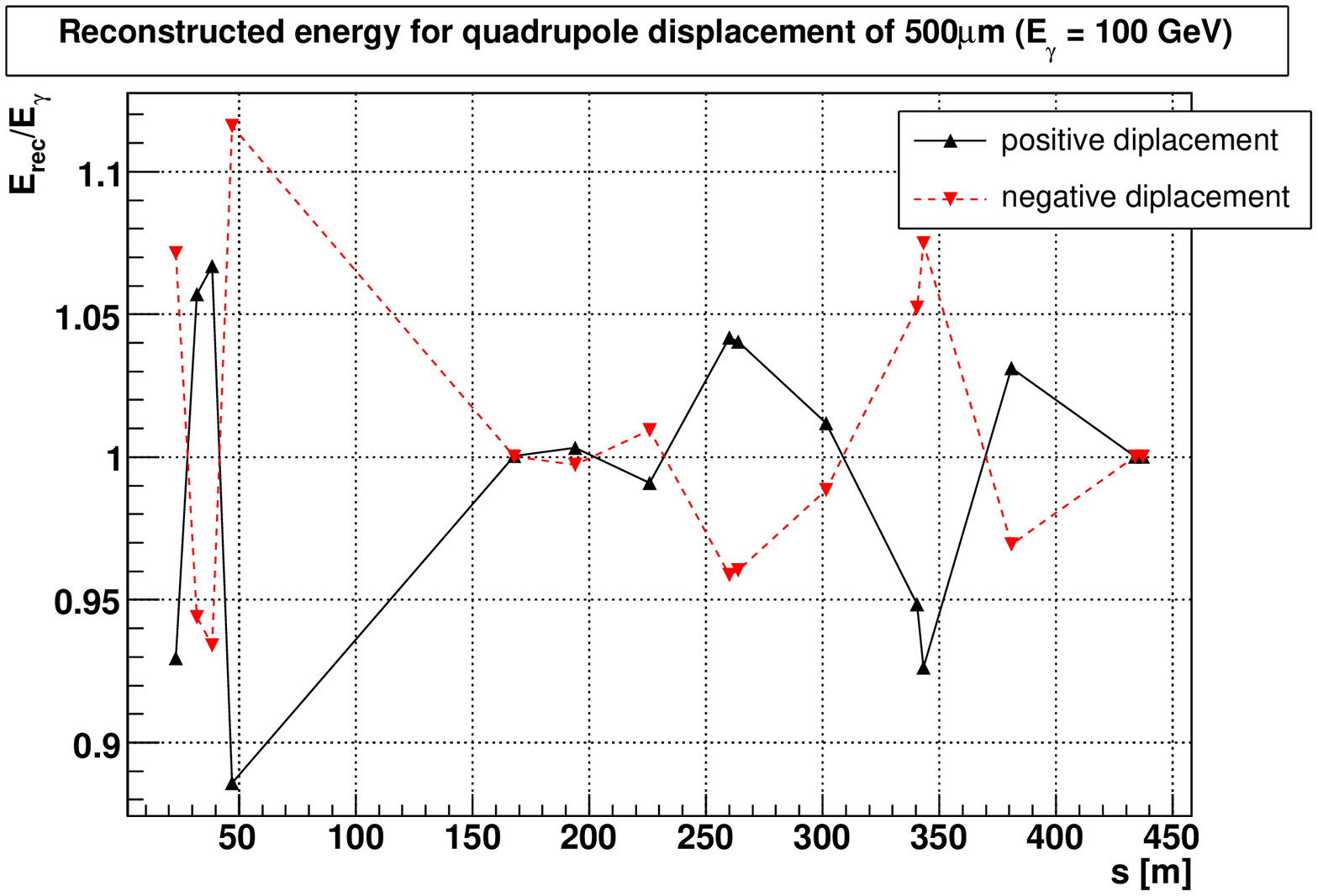} &
\includegraphics[width=0.5\textwidth]{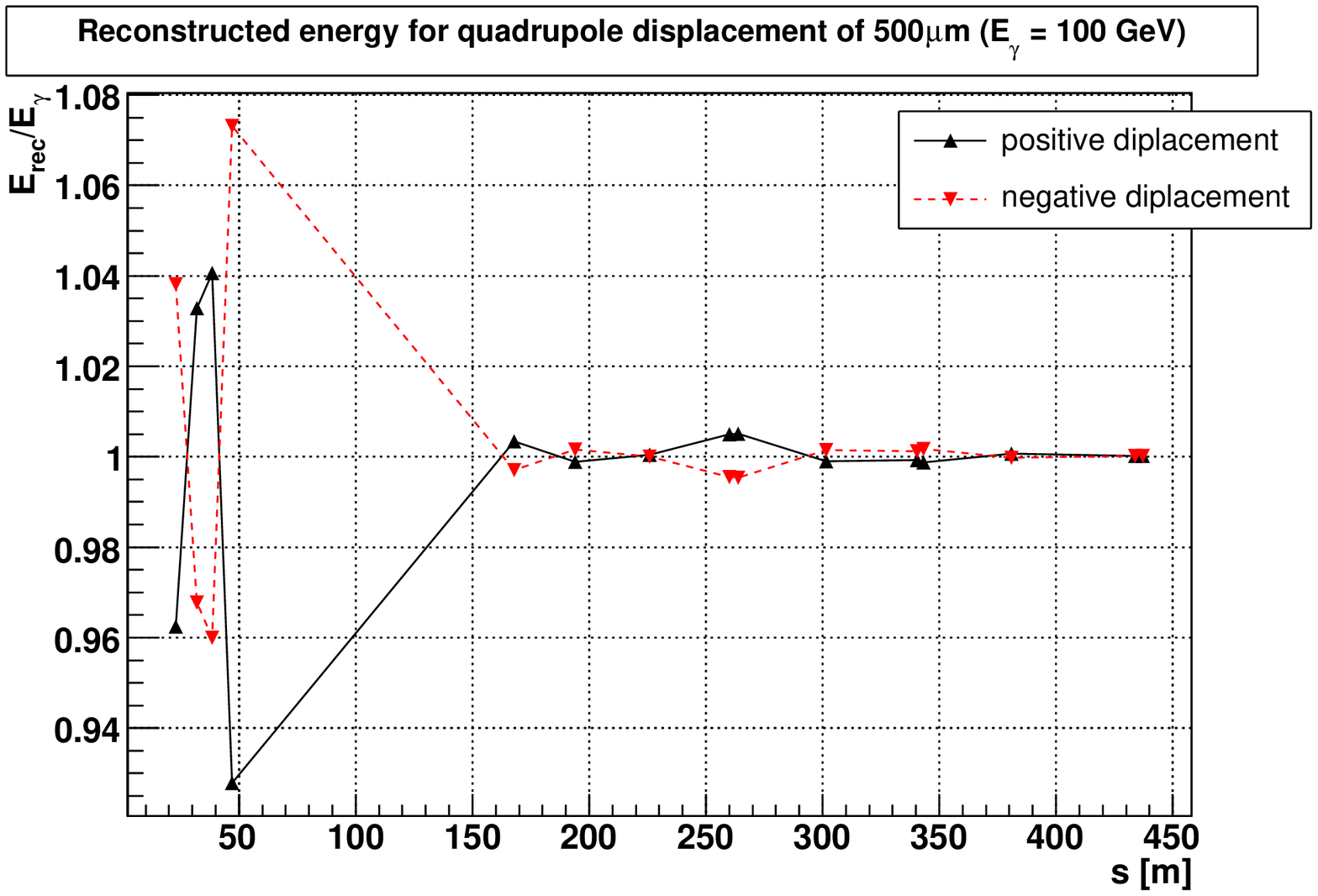} \\
\includegraphics[width=0.5\textwidth]{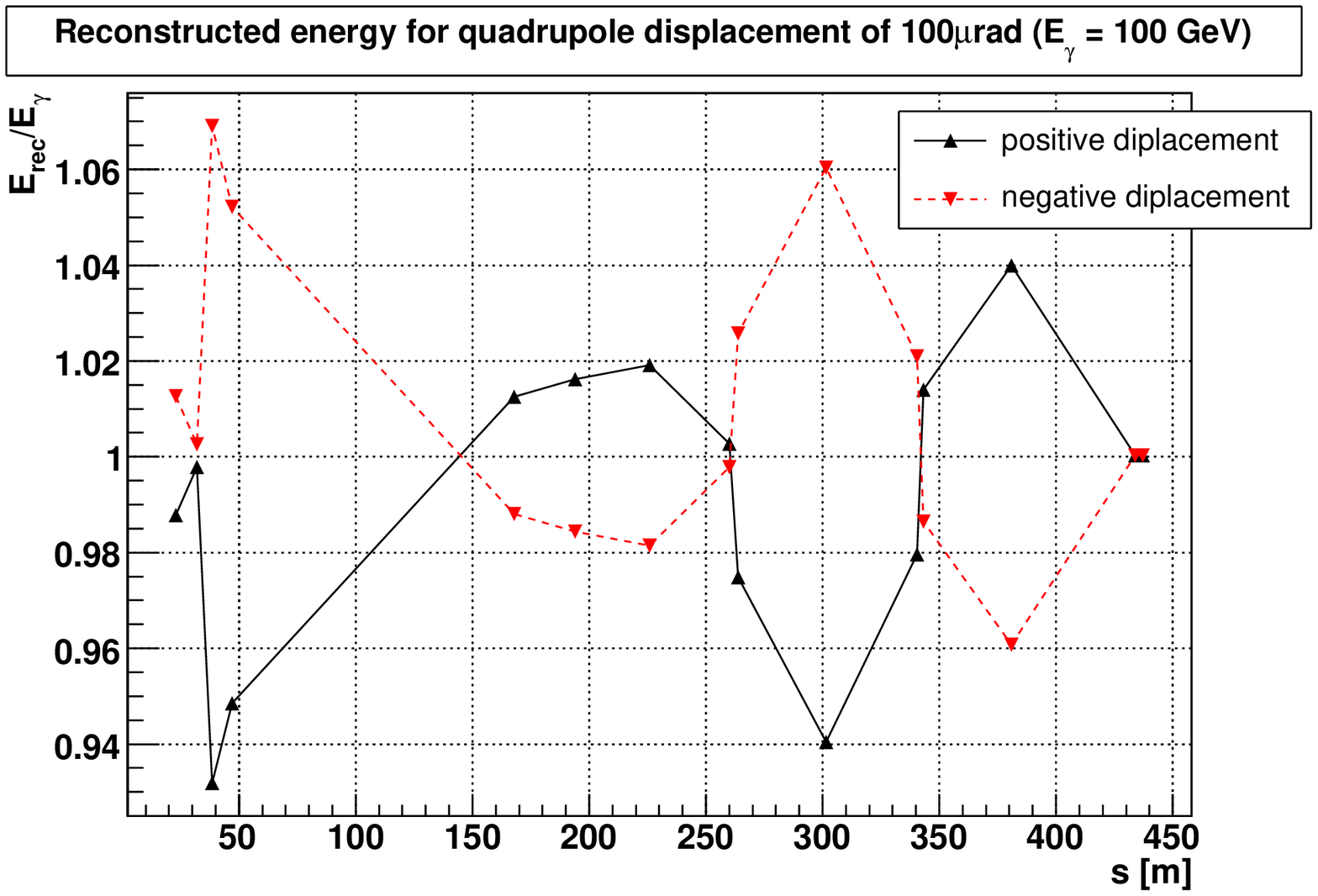} &
\includegraphics[width=0.5\textwidth]{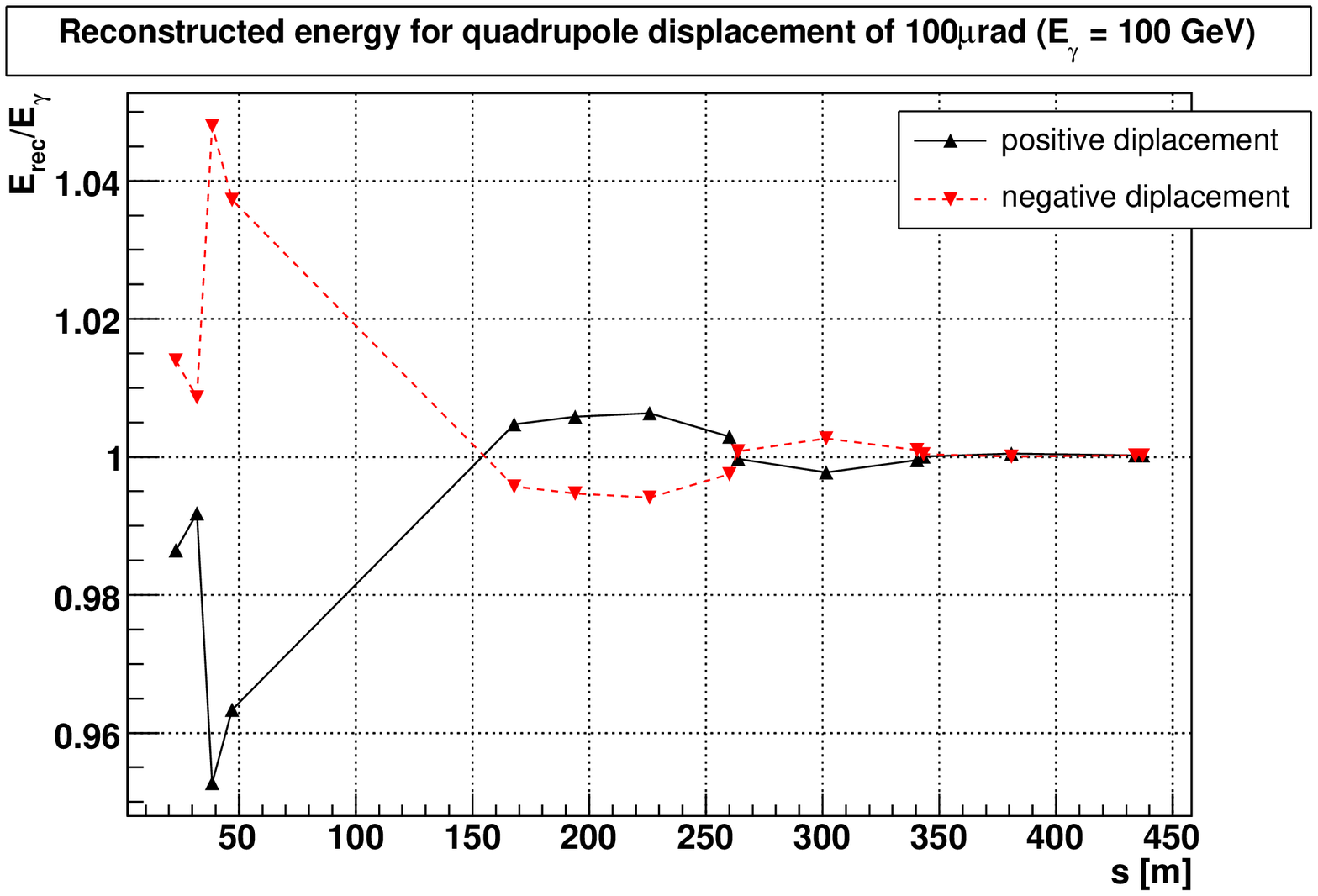} \\
\end{tabular}
\caption[Reconstruction error due to the misalignment of quadrupoles]{The misalignment of the LHC 
quadrupoles bias the energy reconstruction. 
The graphs show the bias for the reconstruction (with the trivial method) of a 100 GeV energy loss  
assuming the misaligned quadrupoles at various positions. Each element is separately shifted ($500~\mu$m) or tilted($100~\mu$rad), assuming a perfect alignment for the 
rest of the beamline. The impact of the misalignment can be important. Even a perfect knowledge of the actual beam position at the VFD (right) does not compensate for this bias, depending on the position of the misplaced quadrupole.}
\label{misalignment}
\end{figure}

Fig. \ref{misalignment} shows the impact of possible shifts ($0.5 ~\textrm{mm}$) and tilts 
($0.1 ~ \textrm{mrad}$) of the beamline quadrupoles on the energy reconstruction with VFDs at 
$420 ~ \textrm{m}$. The reconstruction assumes ideal beamline in which only one quadrupole at a time 
is separately moved. Effects even higher than $10 ~ \%$ could be expected.
The closer to the IP (i.e. the further from the VFDs), the more insensitive the correction. 
One can partially correct for these effects using information from the beam position monitors, but better results
are obtained using a physics calibration process like the two-photon muon pair exclusive production, 
at least in case of the VFDs at 420 m.

\begin{figure}[!h]
\centering
\begin{tabular}{c}
\includegraphics[height=8.5cm]{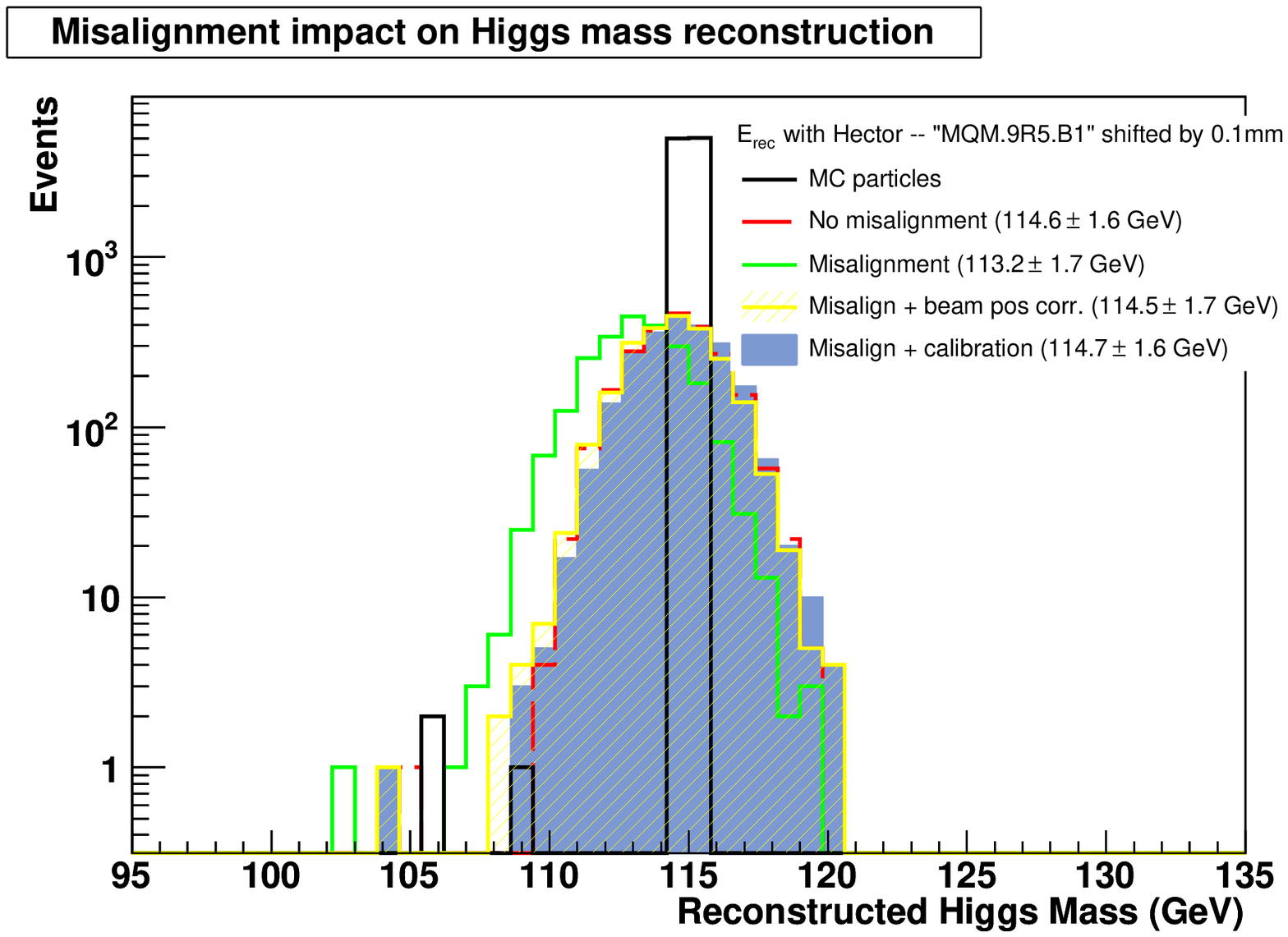} \\ 
\includegraphics[height=8.5cm]{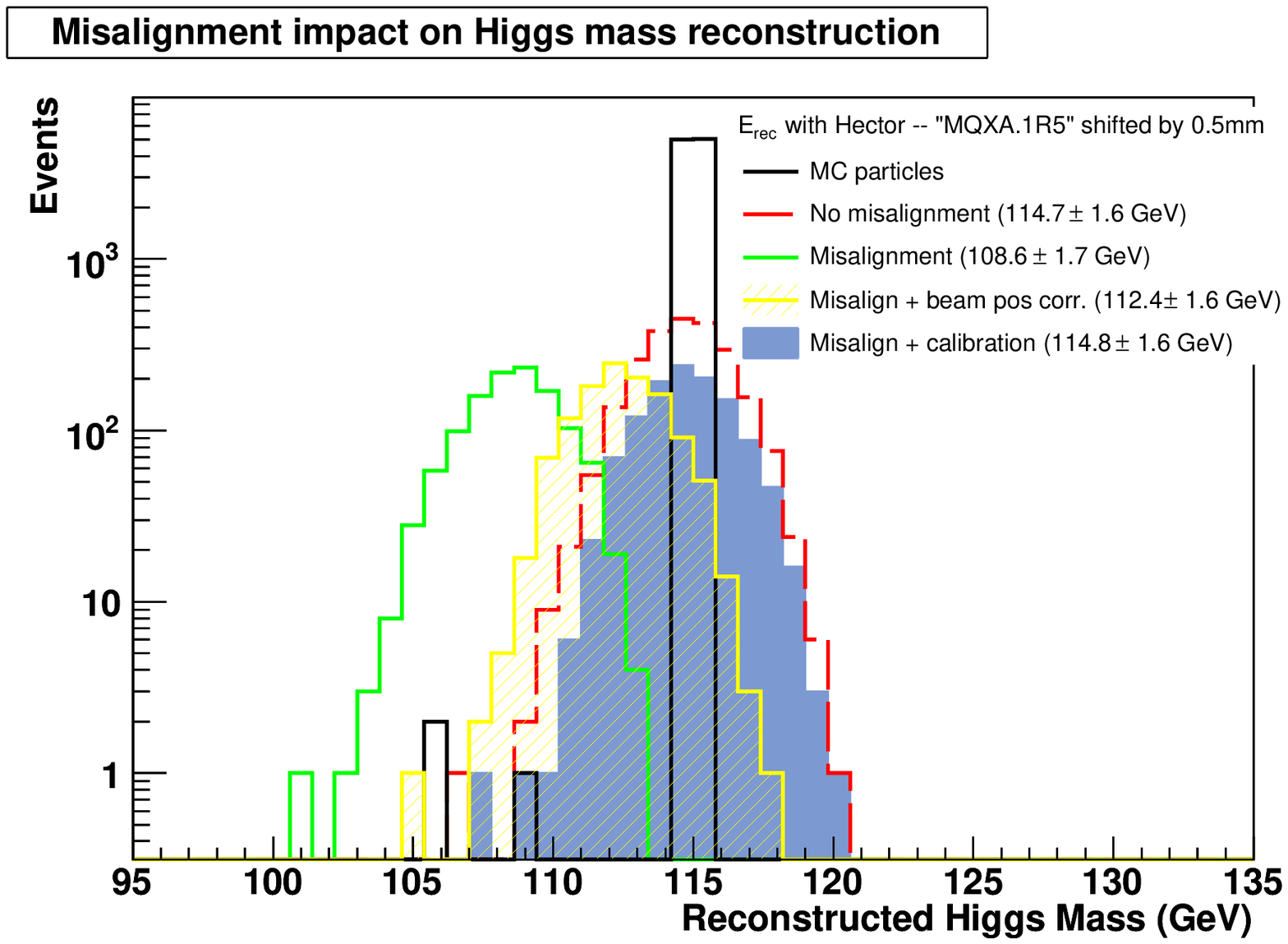} \\
\end{tabular}
\caption[Higgs two-photon exclusive production: error on reconstructed mass from misalignment]
{Illustration of the effects in the energy reconstruction due to the misalignment of LHC quadrupoles. 
The graphs show the reconstructed Higgs boson mass in the two-photon exclusive production, using 
energy of two forward scattered protons. In the upper plot, a quadrupole 
(MQM9R5, $s= 347$ m) close to the detector has been shifted by $100~\mu \textrm{m}$. Misaligning an optical 
element (MQXA1R5, $s=29$ m) close to the IP leads to a loss of acceptance (lower plot). 
The reconstructed values including the correction due to the dimuon calibration 
is also plotted. In brackets, the average reconstructed mass and its resolution are given, 
without including the beam energy dispersion.}
\label{misalignment_higgs}
\end{figure}

These misalignment effects and the corrections are illustrated (Fig. \ref{misalignment_higgs}) by 
the study of two-photon exclusive production of the SM Higgs boson 
($\textrm{pp} \rightarrow \textrm{pp}(\gamma \gamma) \textrm{H}$) with $\textrm{M}_{\textrm{H}} = 115 ~ \textrm{GeV}$.
The measurement of the energy of two scattered protons straightforwardly yields the boson mass, by means of the
so-called \textit{missing mass method}. 
As a consequence, an uncorrected measurement with misalignment leads to a bad mass calculation.
A quadrupole (MQM9R5, $s= 347$ m) close to the detector has been shifted by $100~\mu \textrm{m}$, and another 
quadrupole (MQXA1R5, $s=29$ m) close to the IP by  $500~\mu \textrm{m}$; in the latter case, the 
misalignment-induced change in the VFD acceptance is visible.
The limitations of the beam-position-based corrections are clearly visible, even assuming no systematic errors, 
while the muon-calibration stays unbiased (though only a relatively small sample of 700 dimuon events was used to
get the correction factors).

\section{Summary and outlook}

A new tool for the transport of particles in beamlines was presented, exhibiting high performances thanks to the transfer matrix model. The tool was validated with the \textsc{Mad-X} software, on the LHC beamlines. Much information is produced from the raw transport of particles and study of the transport up to forward detectors is subsequently possible. The reconstruction of variables at IP is tested as well as the impact of the misalignment of beamline optical elements on this reconstruction.

\textsc{Hector} is being ported in the software of the CMS experiment. Finally, further studies on the reconstruction methods, effects of the beamline misalignments and the related corrections will be led.

\section*{Acknowledgments}
The authors are grateful to Monika Grothe, Pierre Van Mechelen, Valentina Avati and Samim Erhan for the useful discussions and comments.

\clearpage
\appendix
\section*{Appendix}
\section{Transfer matrices}

We define the transport matrices of optical elements as the following ones. The length of the optical element is $l$ and $k$ represents its magnetic field. We added the 6th line and row in order to take the kickers into account. The energy dependence of the magnetic deflection of the particle is set by rescaling the $k$ factor by $\frac{E}{E-\Delta E}$.Note : Our convention is transposed compared to \cite{Klaus_Wille} : $x_0.M = x_1$ instead of $x_1 = M.x_0$ so the matrices should be transposed. The matrices should have the following units :

      $$
      \mathbf{M_{units}} =
      \left(
      \begin{array}{cccccc}
          1 & 1/m & 1 & 1/m & GeV/m & 1 \\
          m & 1 & m & 1 & GeV & 1 \\
          1 & 1/m & 1 & 1/m & GeV/m & 1\\
          m & 1 & m & 1 & GeV & 1 \\
          m/GeV & 1/GeV & m/GeV & 1/GeV & 1 & 1\\
          1 & 1 & 1 & 1 & 1  & 1 \\
      \end{array}
      \right)
      $$

For a vertically focusing quadrupole (\textit{H\_VerticalQuadrupole}), where $\omega(k) = l \sqrt{k}$ :

\begin{equation}
\label{v_quad}
 \mathbf{M_{vertical-quadrupole}} =
 \left(
 \begin{array}{cccccc}
 \cosh(\omega) & \sqrt{k}\sinh(\omega) & 0 & 0 & 0 & 0\\
 (1/\sqrt{k})sinh(\omega) & \cosh(\omega) & 0 & 0 & 0 &0\\
 0 & 0 & \cos(\omega) & -\sqrt{k}sin(\omega) & 0 &0\\
 0 & 0 & (1/\sqrt{k})*sin(\omega) & \cos(\omega) & 0 &0\\
 0 & 0 & 0 & 0 & 1 &0\\
 0 & 0 & 0 & 0 & 0 &1 \\
 \end{array}
 \right)
\end{equation}

For a horizontally focusing quadrupole (\textit{H\_HorizontalQuadrupole}), where $\omega(k) = l \sqrt{k}$ : 

\begin{equation}
\label{h_quad}
 \mathbf{M_{horizontal-quadrupole}} =
 \left(
 \begin{array}{cccccc}
 \cos(\omega) & -\sqrt{k}\sin(\omega) & 0 & 0 & 0 & 0\\
 (1/\sqrt{k})sin(\omega) & \cos(\omega) & 0 & 0 & 0 & 0\\
 0 & 0 & \cosh(\omega) & \sqrt{k}sinh(\omega) & 0 & 0\\
 0 & 0 & (1/\sqrt{k})sinh(\omega) & \cosh(\omega) & 0 & 0\\
 0 & 0 & 0 & 0 & 1 & 0\\
 0 & 0 & 0 & 0 & 0 & 1 \\
 \end{array}
 \right)
\end{equation}

For a rectangle dipole (\textit{H\_RectangularDipole}), where $r(k) = 1/k$ :

\begin{equation}
\label{r_dip}
 \mathbf{M_{rectangular-dipole}} =
 \left(
 \begin{array}{cccccc}
 \cos(l/r) & \frac{-1}{r} \sin(l/r) & 0 & 0 & 0 & 0\\
 r \sin(l/r) & \cos(l/r) & 0 & 0 & 0 & 0\\
 0 & 0 & 1 & 0 & 0 &0\\
 0 & 0 & l & 1 & 0 &0\\
 2r \sin^2(l/2r)/BE & \sin(l/r)/BE & 0 & 0 & 1 &0\\
 0 & 0 & 0 & 0 & 0 & 1\\
 \end{array}
 \right)
\end{equation}

Note : $r*(1-\cos(l/r))$ causes a numerical sensitivity. We use $ 2\sin^2(x/2) = 1-\cos(x)$ instead.
The rectangular shape implies an edge focusing of the dipole. In the beam frame, we apply the edge focusing as described in \cite{Klaus_Wille} :
$$ \mathbf{M_{rdip-edge-focusing}} = \mathbf{M_{edge}} \mathbf{M_{rdip}} \mathbf{M_{edge}} $$ 
where $\Psi = l /2r$ and
$$ \mathbf{M_{edge}} =      
   \left(
	\begin{array}{cccccc}
	1 & tan(\Psi)/r & 0 & 0 & 0 & 0\\
	0 & 1 & 0 & 0 & 0 & 0\\
	0 & 0 & 1 & -tan(\Psi)/r & 0 & 0\\
	0 & 0 & 0 & 1 & 0 & 0\\
	0 & 0 & 0 & 0 & 1 & 0\\
	0 & 0 & 0 & 0 & 0 & 1\\
	\end{array}
  \right)
$$

For a sector dipole (\textit{H\_SectorDipole}), either with their bending effect on or off, where $r(k) = 1/k$ :

\begin{equation}
\label{s_dip_off}
 \mathbf{M_{sector-dip-bending-off}} =
 \left(
 \begin{array}{cccccc}
 \cos(l/r) & \frac{-1}{r} \sin(l/r) & 0 & 0 & 0 & 0\\
 r \sin(l/r) & \cos(l/r) & 0 & 0 & 0 & 0\\
 0 & 0 & 1 & 0 & 0 & 0\\
 0 & 0 & l & 1 & 0 & 0\\
 0 & 0 & 0 & 0 & 1 & 0\\
 0 & 0 & 0 & 0 & 0 & 1\\
 \end{array}
 \right)
\end{equation}

\begin{equation}
\label{s_dip}
 \mathbf{M_{sector-dip-bending-on}} =
 \left(
 \begin{array}{cccccc}
 \cos(l/r) & \frac{-1}{r} \sin(l/r) & 0 & 0 & 0 & 0\\
 r \sin(l/r) & \cos(l/r) & 0 & 0 & 0 & 0\\
 0 & 0 & 1 & 0 & 0 &0\\
 0 & 0 & l & 1 & 0 &0\\
 2r \sin^2(l/2r)/BE & \sin(l/r)/BE & 0 & 0 & 1 & 0\\
 0 & 0 & 0 & 0 & 0 & 1\\
 \end{array}
 \right)
\end{equation}

The bending can be switched off when working in the detector frame. Obviously, this should not be used for non nominal energy beam particles.

For a drift (\textit{H\_Drift}) :

\begin{equation}
\label{drift}
 \mathbf{M_{drift}} =
 \left(
 \begin{array}{cccccc}
 1 & 0 & 0 & 0 & 0 & 0\\
 l & 1 & 0 & 0 & 0 & 0\\
 0 & 0 & 1 & 0 & 0 & 0\\
 0 & 0 & l & 1 & 0 & 0\\
 0 & 0 & 0 & 0 & 1 & 0\\
 0 & 0 & 0 & 0 & 0 & 1\\
 \end{array}
 \right)
\end{equation}

For a horizontal kicker (\textit{H\_HorizontalKicker}), where $k$ is the angular kick in rad :

\begin{equation}
\label{h_kick}
 \mathbf{M_{horizontal-kicker}} =
 \left(
 \begin{array}{cccccc}
 1 & 0 & 0 & 0 & 0 & 0\\
 l & 1 & 0 & 0 & 0 & 0 \\
 0 & 0 & 1 & 0 & 0 & 0\\
 0 & 0 & l & 1 & 0 & 0\\
 0 & 0 & 0 & 0 & 1 & 0\\
 l \tan(k) /2 & k & 0 & 0 & 0 & 1\\
 \end{array}
 \right)
\end{equation}

For a vertical kicker (\textit{H\_VerticalKicker}), where $k$ is the angular kick in rad :

\begin{equation}
\label{v_kick}
 \mathbf{M_{vertical-kicker}} =
 \left(
 \begin{array}{cccccc}
 1 & 0 & 0 & 0 & 0 & 0\\
 l & 1 & 0 & 0 & 0 & 0 \\
 0 & 0 & 1 & 0 & 0 & 0\\
 0 & 0 & l & 1 & 0 & 0\\
 0 & 0 & 0 & 0 & 1 & 0\\
 0 & 0 & l \tan(k) /2 & k & 0 & 1\\
 \end{array}
 \right)
\end{equation}

\section{User guide}
\subsection{The usual "getting started" chapter}

In order to run \textsc{Hector} on your system, first download its sources and compile the \emph{libHector} library\footnote{The Makefile is compatible for a compilation on Linux and Windows/Cygwin systems. For such Windows, just modify the library extension name to LEXT=dll in the Makefile.} :

\begin{quote}
\begin{verbatim}
me@mylaptop:~$ wget http://www.fynu.ucl.ac.be/perso/xrouby/files/hector_1_3_2.tgz
me@mylaptop:~$ tar jxvf hector_1_3_2.tgz
me@mylaptop:~$ cd Hector
\end{verbatim}
\end{quote}

\textsc{Hector} has been tested with \textsc{Root} versions from 4.xx. When launched from \textsc{Hector}'s main directory, \textsc{Root} will automatically load libHector (see rootlogon.C) and display a welcome prompt. In case you are not using Pythia, one can simply ignore it by commenting \texttt{$\sharp$define \_include\_pythia\_} in ./include/H\_Parameters.h\footnote{If your \textsc{Root} version does not contain the \textsc{Pythia} libraries, the make command will return an error. In that case, you will have to remove by hand the references to libPythia and libEGPythia in the Makefile, then run "make clean" and "make".}.

\begin{quote}
\begin{verbatim}
me@mylaptop:~/Hector$ make
me@mylaptop:~/Hector$ root
Ready for Hector -- enjoy !
root [0]
\end{verbatim}
\end{quote}

As a general comment, don't forget that every class you use in your routine should be included in the routine file, for instance :

\begin{quote}
\begin{verbatim}
#include "H_BeamLine.h"
#include "H_BeamParticle.h"
\end{verbatim}
\end{quote}

\subsection{The everyday life with \textsc{Hector}}

In this chapter, we will tell you how to use \textsc{Hector} to perform basics tasks such as using predefined beamlines and designing yours, propagating particles along those, and computing some beam properties. Generally speaking, the detailed constructors and methods arguments for \textsc{Hector} classes will not be detailed here, as it can be easily found in the Internet reference manual\footnote{This online manual, as well as all information on Hector, can be found here : \url{http://www.fynu.ucl.ac.be/hector.html} }.

\subsubsection{Genesis 1 : the beamline}

The first step to use \textsc{Hector} is to define the list of optical elements that the particles will cross while in the beam pipe. This can be done in two ways : 

\begin{itemize}
\item Using (existing) magnets tables
\item Building the line yourself
\end{itemize}

\subsubsection*{Tables of optical elements}

This way is the simplest. You only need a text file containing columns with the magnet name, position, length, strength. Apertures can also be specified in this file. The order of the columns is meaningless but you should use column headers with the following "codenames" : 

\begin{itemize}
\item NAME : name
\item KEYWORD : type
\item S : position along the line
\item L : length
\item K0L : dipole strength (horizontal)
\item K1L : quadrupole strength
\item HKICK : horizontal kick angle
\item VKICK : vertical kick angle
\item APERTYPE : aperture type
\item APER\_X : aperture size (X runs from 1 to 4)
\end{itemize}

The "KEYWORD" code defining the element type can take the following values : 

\begin{itemize}
\item "DRIFT" : no-field zone\footnote{The drift spaces lines in the tables are not read by \textsc{Hector}, because it automatically fills the gaps between other elements with drifts of the right length.}
\item "QUADRUPOLE" : regular quadrupole
\item "RBEND" : rectangular dipole
\item "SBEND" : sector dipole
\item "VKICKER" : vertical kicker
\item "HKICKER" : horizontal kicker
\item "MARKER" : dummy element\footnote{This can be used to tag special places such as Interaction points.}
\item "RCOLLIMATOR" : rectangular collimator
\end{itemize}

The "APERTYPE" can be : 

\begin{itemize}
\item "NONE" : no aperture limitation
\item "CIRCLE" : circular aperture. the radius is given by APER\_1.
\item "RECTANGLE" : rectangular aperture. The $x$ and $y$ sizes are given by APER\_1 and APER\_2.
\item "ELLIPSE" : elliptic aperture. Same size definition as the rectangular aperture.
\item "RECTELLIPSE" : intersection between a rectangle and an ellipse. The APER\_X parameters give the rectangle then the ellipse size.
\end{itemize}

When your table is properly written, the method to create your beamline is the following one\footnote{As all the methods/functions using a H\_BeamLine as argument require it to have the "pointer" type, it is heavily recommended to declare it with that type in all the cases.} : 

\begin{quote}
\begin{verbatim}
float length = 500;
int direction = 1;
H_BeamLine* mybeamline = new H_BeamLine(direction,length);
mybeamline->fill("table.txt",1,"starting point")
\end{verbatim}
\end{quote}

The two arguments of the $H\_BeamLine$ creator are the line length and the direction in which the table will be read (forwards (1) or backwards (-1) ). The $fill$ Method requires the name of the table file, the propagation direction of the particles from the starting point and the name of the starting point marker.

The meaning of these two direction parameters can seem unclear, so let's suppose you use the following table : 

\begin{quote}
\begin{verbatim}
NAME              KEYWORD          S   ...

"element1"        "rbend"          0   ...
"element2"        "vquadrupole"   10   ...
"starting point"  "marker"        20   ...
"element3"        "vquadrupole"   30   ...
"element4"        "rbend"         40   ...
\end{verbatim}
\end{quote}

If the $H\_BeamLine$ creator argument is 1, we read only the last three elements. Has it been -1, we would have read the first three. The $fill$ argument being 1, the particles will go from top to bottom. If it is -1, it goes from bottom to top\footnote{It can be useful to notice that the order of the lines does not matter as the ordering is done from the S parameter. However, the elements have to be on their real position (before or after) with respect to the starting point.}.

Examples of elements tables - taken from the LHC IP5 region - can be found in the "data" directory of \textsc{Hector} as a working example.

\subsubsection*{Element-by-element beam building}

Another way to build the beamline is to add each element separately in the beamline. This is done by this kind of code : 

\begin{quote}
\begin{verbatim}
float length = 500;
int direction = 1;
H_BeamLine* mybeamline = new H_BeamLine(direction,length);
float position = 10;
float strength = -0.001;
float length = 1;
H_VerticalQuadrupole hvq("myquadrupole",position,strength,length);
mybeamline->add(hvq);
\end{verbatim}
\end{quote}

This way is not recommended, as the other one is more convenient in most of the cases. But it still can be used to add markers or user-defined elements useful only in specific routines, or to design a new beamline automatically from Monte Carlo techniques.

It is important to note that the $H\_Dipole$ and $H\_Quadrupole$ classes are purely abstract and should then never be used directly. 

\subsubsection*{Alignment effects}

After building the beamline, one can move elements around their nominal position using the following method : 

\begin{quote}
\begin{verbatim}
string name = "MCBXA.1R5"
double delta_x = 50;
double delta_y = 30;
mybeamline->alignElement(name,delta_x,delta_y);
\end{verbatim}
\end{quote}

This will displace the element called $MCBXA.1R5$\footnote{Please note that in LHC tables the names include the double quotes, meaning that the name should be something like "\textbackslash"MCBXA.1R5\textbackslash"".} by the specified amounts in the x and y directions. Corresponding angles can be changed using the similar method : 

\begin{quote}
\begin{verbatim}
mybeamline->alignElement(name,delta_theta_x,delta_theta_y);
\end{verbatim}
\end{quote}

\subsubsection{Genesis 2 : The particles}

Using the lines : 

\begin{quote}
\begin{verbatim}
double mass = MP;
double charge = 1;
H_BeamParticle myparticle(mass, charge);
\end{verbatim}
\end{quote}

one particle with given mass and charge\footnote{If not specified, the default particle used is a proton.} is created at the "starting point" (IP) with energy defined by the BE (beam energy) given in the $H\_Parameters.h$ file. The other main variables in this file are : 

\begin{itemize}
\item SBE : beam energy dispersion
\item PX : $x$ position at IP
\item PY : $y$ position at IP
\item TX : $x$ angle at IP 
\item TY : $y$ angle at IP
\item SX : $x$ dispersion of the beam at IP
\item SY : $y$ dispersion of the beam at IP
\item STX : $x$ angle dispersion of the beam at IP
\item STY : $y$ angle dispersion of the beam at IP
\end{itemize}

The $x$ and $y$ positions are always given in $\mu$m and the corresponding angles are in $\mu \mathrm{rad}$. The $s$ coordinate is conveniently expressed in meters.

The particle can then be smeared\footnote{These methods use a simple Gaussian smearing.} using the SBE, SX, SY, STX, STY parameters by the following methods : 

\begin{quote}
\begin{verbatim}
myparticle.smearE();
myparticle.smearPos();
myparticle.smearAng();
\end{verbatim}
\end{quote}

Other useful methods for the particles at the starting point include :  

\begin{quote}
\begin{verbatim}
H_BeamParticle::emitGamma(float Energy, float virtuality);
\end{verbatim}
\end{quote}

which allows to simulate the emission of a virtual particle with given properties, causing an energy loss and an angle displacement. Another way to simulate an energy change is to use the

\begin{quote}
\begin{verbatim}
H_BeamParticle::setE(float Energy);
\end{verbatim}
\end{quote}

method which only changes the energy parameter of the particle without any other effects. The effect of the energy loss is obviously simulated by \textsc{Hector} \footnote{The energies and virtualities are always given in GeV and GeV$^2$. The virtuality is by definition a negative number.}.

\subsubsection{Exodus : the particle propagation}

Once the beamline has been set and the particle has all the desired properties, the latter can be propagated along the former using the following : 

\begin{quote}
\begin{verbatim}
myparticle.computePath(mybeamline);
\end{verbatim}
\end{quote}

This computes the positions and angles of the particle at the entrance of each element and allows to interpolate its position in all the drift spaces. This method is the most time-consuming in \textsc{Hector} and thus defines the speed of the program. The typical timescale is less than $4 ~\mu\mathrm{s}$ per particle and per element (including drifts)\footnote{Tested on a $1.7 ~ \mathrm{GHz}$ Centrino processor running Linux.}.

\subsubsection*{Aperture check}

If one wants to check if the particle has been stopped by any aperture, the method following 

\begin{quote}
\begin{verbatim}
bool stopped = myparticle.stopped(mybeamline); 
\end{verbatim}
\end{quote}

should be used. It checks all the element apertures and stores the closest intercepting element from the IP. This element can then be obtained using :

\begin{quote}
\begin{verbatim} 
const H_OpticalElement* myopticalelement = myparticle.getStoppingElement();
cout<<myopticalelement->getName()<<endl;
\end{verbatim}
\end{quote}

for instance to show the name of the element.

\subsubsection*{Getting the particle information}

Once the path has been computed, it is quite legitimate to be interested on the particle transverse position\footnote{The $x$ and $y$ coordinates are respectively the transverse horizontal and vertical displacement of the particle.} and angle at some position in s. This is achieved the following methods : 

\begin{quote}
\begin{verbatim}
float position = 100;
myparticle.propagate(position);
float x = myparticle.getX();
float y = myparticle.getY();
float theta_x = myparticle.getTX();
float theta_y = myparticle.getTY();
\end{verbatim}
\end{quote}

The particle can then be "propagated" to any other other position and the variables obtained at that new positions by the same methods. The complete path can be obtained via the \textsc{Root} $TGraph$ class using the following :

\begin{quote}
\begin{verbatim}
TGraph* ppath_x;
TGraph* ppath_y;
ppath_x = p1.getPath(0,color_x);
ppath_y = p1.getPath(1,color_y);
ppath_x->Draw("AL");
ppath_y->Draw("AL");
\end{verbatim}
\end{quote}

A convenient way to deal with big numbers of particles is to use the $H\_Beam$ class. It builds the beam in the following way : 

\begin{quote}
\begin{verbatim}
int number_of_particles = 1000;
H_Beam mybeam;
mybeam.createBeamParticles(number_of_particles);
\end{verbatim}
\end{quote}

where the $createBeamParticles$ method smears all the particle variables using the pre-defined parameters. Most of the single-particle methods can then be applied to the full beam, such as $computePath$ and $propagate$. One can access one particle of the beam using the simple

\begin{quote}
\begin{verbatim}
int particle_i = 10;
H_BeamParticle* myparticle = mybeam.getBeamParticle(particle_i);
\end{verbatim}
\end{quote}

for instance to get it's properties or to use methods such as $emitGamma$ on it. Using a $H\_Beam$ is especially useful when the user is interested in global properties of the beam, such as its spatial extension, given by the $\beta$ functions. This is achieved in a straightforward way using the method : 

\begin{quote}
\begin{verbatim}
float position = 100;
float beta_x_error, beta_y_error;
float beta_x = mybeam.getBetaX(position, beta_x_error);
float beta_y = mybeam.getBetaY(position, beta_y_error);
\end{verbatim}
\end{quote}

which returns the value of the $\beta$ function at the given position and the error on this value.

\subsubsection{Kings : Remarks for the courageous user}

In this section we will detail some features of \textsc{Hector} that most people should not use, some problems that will appear to only few of the users, and other exceptionally boring things that could help you in some very seldom cases.

\subsubsection*{Absolute frame}

First of all, \textsc{Hector} is designed to work in "relative frame". This means the normal transverse position for a particle with no angle or displacement at the IP is always 0, and then particles with nominal beam energy will not be deflected by the dipoles.

However, in some case it can be interesting to check the absolute trajectory of the beam. This can be achieved with \textsc{Hector}, if and only if you can consider all your beam elements as parallel. In that case, it suffices to add two lines at the beginning of your routine : 

\begin{quote}
\begin{verbatim}
extern bool relative_energy;
relative_energy = false;
\end{verbatim}
\end{quote}

You can also displace all beam elements after a given point laterally using the following :

\begin{quote}
\begin{verbatim}
float s_start = 100;
float x_offset = -0.1:
mybeamline->offsetElements(s_start,x_offset);
\end{verbatim}
\end{quote}

This will displace horizontally all elements from 100m onwards from the IP, by 10cm in the negative $x$ direction.

\subsubsection*{Kickers}

One exception to the "no-deflection" rule of the previous section is obviously the presence of an initial angle of the beam compared to its natural propagation direction, for instance a crossing angle. This kind of case is usually coped with using "kickers" which deflect the beam by a given angle. These kickers are obviously included in \textsc{Hector} and are the only elements that doesn't respect the "no-deflection" law. If one feels more comfortable to switch this effect off, it is easily be done by including : 

\begin{quote}
\begin{verbatim}
extern int kickers_on;
kickers_on = 0;
\end{verbatim}
\end{quote}

at the beginning of the routine.

\subsubsection{Revelation : compiling and running your code}

Once you have written a routine and made sure you put it in the right place - for instance the very convenient \textsc{Hector}/routines directory - you only need to compile it. This is most easily done by running \textsc{Root} and doing the following\footnote{Here we suppose that \textsc{Hector} is already properly installed on your computer.} : 

\begin{quote}
\begin{verbatim}
me@mylaptop:~/Hector$ root -l
Ready for Hector -- enjoy !
root [0] .L routines/myroutine.cpp++
root [1] myprogram()
\end{verbatim}
\end{quote}

\subsection{A simple example}

As a reward for your patience, here's a little working example of \textsc{Hector} plotting a transverse view of the beam at $220 ~ \mathrm{m}$ from the IP.

\begin{quote}
\begin{verbatim}

// C++ #includes
#include <iostream>

// ROOT #includes
#include "TH2F.h"

// Hector #includes
#include "H_BeamLine.h"
#include "H_BeamParticle.h"

using namespace std;

void drawProfile() {

  const int NParticle = 1000;
  H_BeamLine* beamline = new H_BeamLine(1,500);
  beamline->fill("data/twiss_ip5_b1_v6.5.txt",1,"IP5");

  TH2F* hp = new TH2F("Positions","",100,-2.5,2.5,100,-2.5,2.5);

  for (unsigned int i=0; i<NParticle ; i++) {
        H_BeamParticle p1;
        p1.smearPos();
        p1.smearAng();
        p1.computePath(beamline);
        p1.propagate(220);
        hp->Fill(p1.getX()/1000.,p1.getY()/1000.);
  }
  hp->Draw();
}

\end{verbatim}
\end{quote}

\newpage
\tableofcontents
\newpage
\listoffigures
\end{document}